\documentclass[11pt]{amsart}
\usepackage[T1]{fontenc}
\usepackage[utf8]{inputenc}
\IfFileExists{lmodern.sty}{\usepackage{lmodern}\usepackage[nopatch=footnote]{microtype}}{\usepackage[expansion=false,nopatch=footnote]{microtype}}
\usepackage{a4wide}
\usepackage{amsmath,amssymb,amsfonts,mathtools}
\usepackage{amsthm}
\usepackage{enumitem}
\usepackage{booktabs}
\usepackage{array}
\usepackage[numbers,sort&compress]{natbib}
\usepackage{xcolor}

\usepackage{graphicx}
\usepackage[hyperfootnotes=false]{hyperref}
\hypersetup{hidelinks}

\usepackage{calc}
\usepackage{ragged2e}
\usepackage{setspace}
\newcommand{\FundingLogos}{%
  \raisebox{0pt}{\includegraphics[height=1.5cm]{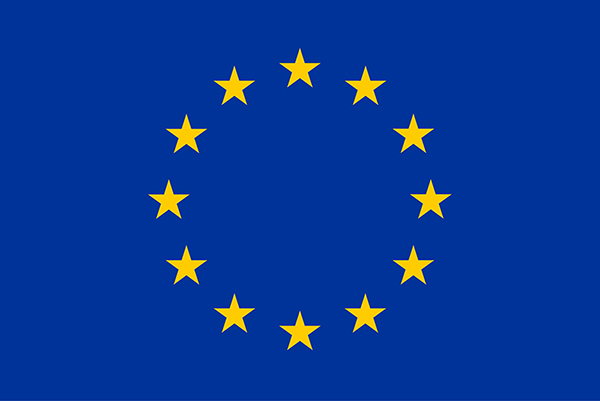}}%
  \hspace{1em}%
  \raisebox{0pt}{\includegraphics[height=1.5cm]{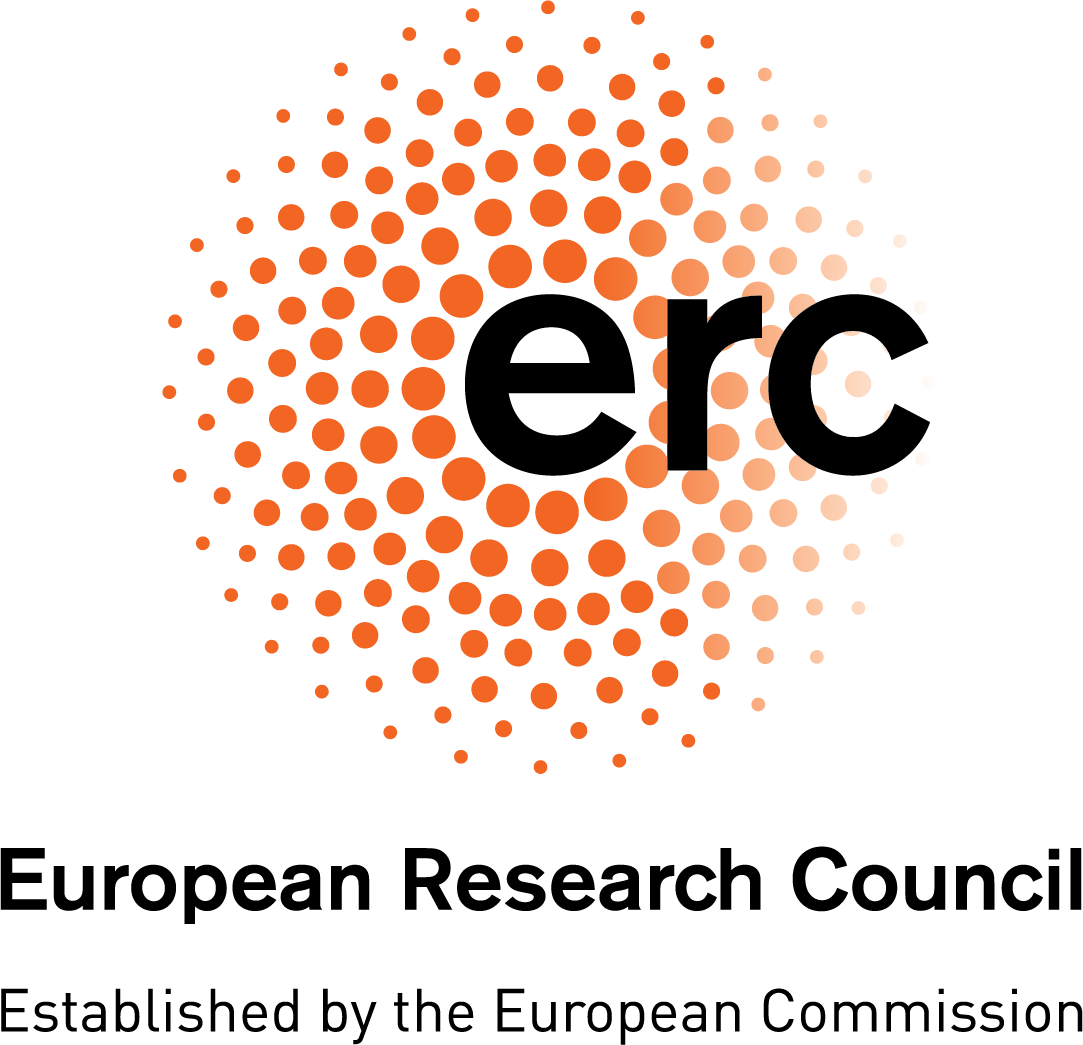}}%
}

\usepackage{tcolorbox}
\tcbset{colback=white,boxrule=0.5pt,arc=2pt,left=4pt,right=4pt,top=4pt,bottom=4pt}

\setlist[enumerate]{leftmargin=2.25em,itemsep=0.34em,topsep=0.42em,label=(\roman*)}
\setlist[itemize]{leftmargin=2.15em,itemsep=0.30em,topsep=0.40em}

\numberwithin{equation}{section}

\theoremstyle{plain}
\newtheorem{theorem}{Theorem}[section]
\newtheorem{proposition}[theorem]{Proposition}
\newtheorem{lemma}[theorem]{Lemma}
\newtheorem{corollary}[theorem]{Corollary}
\theoremstyle{definition}
\newtheorem{definition}[theorem]{Definition}
\newtheorem{remark}[theorem]{Remark}
\newtheorem{example}[theorem]{Example}

\title[Adapted Wasserstein Distance for Credit Risk]{Adapted Wasserstein Distance for Credit Risk:\\
 Rating Information and Model-Transfer Hedging}
\author{Luca Di Persio}
\address{Department of Computer Science, University of Verona, Strada le Grazie 15, 37134 Verona, Italy}
\email{luca.dipersio@univr.it}

\author{Francesco Mattesini}
\address{Department of Mathematics, Technical University of Munich, Boltzmannstra{\ss}e 3, 85748 Garching bei M{\"u}nchen, Germany, and Munich Center for Machine Learning (MCML)}
\email{francesco.mattesini@tum.de}

\date{30 October 2026}

\subjclass[2020]{Primary 91G40; Secondary 49Q22, 60G55, 91G20}
\keywords{adapted Wasserstein distance, bicausal optimal transport, credit risk}

\begin{document}

%

\begin{abstract}
Two credit models can quote the same bond prices today and still disagree on how information about the issuer unfolds over time. We quantify this disagreement by means of the adapted Wasserstein distance, an optimal transport distance between laws of processes consistent with the flow of information associated with them. For defaultable bond price processes, the squared distance is the smallest probability by which the two default indicators must disagree. This yields closed-form distances for rating reviews and a dynamic programming principle for multi-period rating migration. For a single review, the distance can be estimated from observed posterior survival probabilities at the parametric rate without losing calibration.
For intensity models, we obtain an explicit formula for the adapted distance between bond prices that controls the error of bootstrap hedges along a quoted term structure, determine its sharp range over hazard schedules calibrated to the same prices, and show that it is stable under perturbations of the hazard rates. Finally, we complement the theoretical findings with applications to S\&P transition data and Monte Carlo experiments to illustrate the size of these effects.
\end{abstract}

\maketitle

\section{Introduction}\label{sec:introduction}

A credit model is calibrated to the prices at which the bonds and credit default swaps of an issuer trade today. 
With zero recovery and deterministic interest rates, these prices fix the survival probabilities of the issuer at the quoted maturities. However, they determine neither the survival curve between these maturities nor, more importantly, how these probabilities are revised as information about the issuer reaches the market.
 Two models can therefore agree on every quoted price and still describe very different markets: in one of them the credit quality of the issuer is revealed gradually, in the other it is revealed in a single rating review. Consequently, in one of them a desk that hedges a long bond by rolling short bonds never has to inject cash; in the other it does. In this paper, we quantify this kind of disagreement, computing it explicitly in the models used in credit practice, and we relate it to the hedging errors it produces.

\medskip
The basic object is the price of a defaultable zero-coupon bond with zero recovery. With zero interest rates, its price at a date $t$ before maturity $N$, on survival of the issuer, is the probability of surviving until $N$ given the information available at $t$. We refer to it as the \emph{defaultable bond price process}. 
It is a martingale starting at price $p$ quoted today, and at maturity it equals $1$ if the issuer has survived and $0$ otherwise. A pricing model is the law of this process, together with the information it carries. If two models quote the same price $p$, their bond price processes have the same law at maturity, namely a Bernoulli law with parameter $p$, and they are distinguished solely by
the way their prices move in between.

\medskip
As a guiding example, consider a bond maturing at date $2$ and two models with the same price $p$. In the \emph{coarse} model, no information is unveiled before maturity, and the bond price stays equal to $p$ until it jumps to $0$ or $1$. In the \emph{rating} model, a review at date $1$ occurs and moves the bond price to a random level $Q$, the posterior survival probability given the outcome of the review, with $\mathbb{E}Q=p$ so that the price today is still $p$. For instance, we may consider $Q=p\pm\delta$ with probability one half each. The two models quote the same price and predict the same probability of default, but a desk that hedges in the rating model must trade at date $1$, while in the coarse model it need not.

\medskip
A natural way of comparing two pricing models is provided by optimal transport. Optimal transport compares two probability laws by constructing both random objects on a common space, a \emph{coupling}, and measuring how far apart they are on average. For two pricing models, this means generating a price path of each model jointly and comparing the paths. However, the ordinary Wasserstein distance allows every coupling and hence is agnostic of the time information. In the guiding example, it may first draw the whole path of the rating model, outcome of the review included, and then let the coarse model default exactly when the rating model does. The two default indicators then always agree. The only discrepancy is the price gap $|Q-p|$ at date $1$, and the distance is of order $\delta$. This coupling uses information the coarse model does not have: at date $1$ the coarse model knows nothing, yet its future default has been chosen to match the outcome of the review.

\medskip
To overcome this flaw, we admit only couplings that respect the flow of information: at every date, the part of the coupling constructed for one model may depend on the past of the other model, but not on its future. Such couplings are called \emph{bicausal} \cite{Lassalle2018,BackhoffBeiglbockLinZalashko2017}, and the \emph{adapted Wasserstein distance} is the optimal transport cost computed over them \cite{PflugPichler2012,BackhoffBartlBeiglbockEder2020Finance,BartlBeiglbockPammer2024}. As cost we take the squared differences of the price increments, since this is the quantity that controls the profit and loss of a common trading strategy.\footnote{The canonical definition of \cite{BartlBeiglbockPammer2024} uses the squared differences of the path values. We follow instead \cite[Section~3.3]{BackhoffBartlBeiglbockEder2020Finance} and \cite[Remark~6.5]{BartlBeiglbockPammer2024}, where the increment cost is argued to be the natural one for problems involving stochastic integrals, see Remark~\ref{rem:level-convention} for details. In discrete time, the two costs are equivalent up to constants, and all qualitative statements of the paper hold for both.}
In the guiding example, bicausality forces the default of the coarse model to be independent of the outcome of the review. If the review has moved the survival probability to $p+\delta$, say, the coarse model still survives with probability $p$, so the two defaults must disagree with probability at least $\delta$, and each disagreement costs a jump of size one. The squared adapted distance is therefore of order $\delta$, and the distance itself of order $\sqrt\delta$, much larger than $\delta$ for small revisions. Since the two models have the same initial price and the same terminal survival law, the whole discrepancy is produced by information.


\medskip
The same mechanism is at work in the intensity models that are standard in credit practice, where default can occur at any time and arrives with a \emph{hazard rate}, the instantaneous probability of default per unit of time given survival \cite{Lando1998,DuffieSingleton1999}. For two hazard paths, we show that the squared adapted distance between the two bond price processes is again the smallest probability that their survival indicators disagree at maturity, when both are constructed on a common flow of information. We call it the \emph{adapted default mismatch}, and we show that it has an explicit formula and is attained by a common-shock construction. It also has a direct meaning for hedging, which makes precise the situation described at the beginning of the introduction. A natural strategy is to fit a \emph{bootstrap} model, that is, a credit curve with piecewise constant hazard rate that reprices the quoted bonds \cite{OKane2008}, and to hedge a long bond by rolling short ones, sized by the forward survival factors of the fitted curve. We call this the \emph{bootstrap hedge}. It replicates the long bond exactly under the bootstrap model. Under a market model with rating reviews, however, it requires cash injections, and the adapted default mismatch controls their size.

\subsection{Main results}\label{subsec:original-results}

Our results on defaultable bond prices can be summarized as follows.

\begin{theorem}[Information revealed by credit ratings]\label{thm:intro-ratings}
Let $X$ and $Y$ be the defaultable bond price processes of two models with the same initial price $p$.
\begin{enumerate}
\item\label{it:terminal} The squared adapted Wasserstein distance between the two models is the smallest probability, over bicausal couplings, that their default indicators disagree at maturity, see Proposition~\ref{prop:terminal-representation} for a precise statement.
\item\label{it:one-review} If one model reveals nothing before maturity and the other reveals the posterior $Q$ at an intermediate date, then $\mathcal{AW}_2^2=\mathbb{E}|Q-p|$, while the ordinary Wasserstein distance satisfies $\mathbb{E}|Q-p|^2\le\mathcal{W}_2^2\le2\,\mathbb{E}|Q-p|^2$, see Theorems~\ref{thm:exact-aw-one-rating} and~\ref{thm:exact-w2-one-rating} for a precise statement.
\item\label{it:timing} If both models reveal the same posterior, but at different dates, then $\mathcal{AW}_2^2=\mathbb{E}|Q-\widetilde Q|$, where $\widetilde Q$ is an independent copy of $Q$ (see Theorem~\ref{thm:early-late-aw} for a precise statement).
\item\label{it:chains} If both defaultable bond prices are driven by finite-state rating chains, the distance is the value of a backward recursion over pairs of current ratings. If migration is monotone in credit quality, the recursion is solved by driving both chains with the same uniform random variables, see Theorems~\ref{thm:dpp} and~\ref{thm:synchronous} for a precise statement.
\end{enumerate}
\end{theorem}


\medskip
In particular, close to the coarse model, that is, when the review revises the survival probability only slightly, the two distances behave very differently. For a symmetric revision $Q=p\pm\delta$, part~\ref{it:one-review} gives $\mathcal{W}_2$ of order $\delta$ but $\mathcal{AW}_2$ of order $\sqrt\delta$, so that the ratio $\mathcal{AW}_2/\mathcal{W}_2$ is of order $\delta^{-1/2}$ and blows up as $\delta\to0$. A small rating revision is therefore almost invisible to the ordinary Wasserstein distance, while the adapted distance registers it at a much larger scale.

\medskip
Parts~\ref{it:terminal} and~\ref{it:chains} build on known results. That, for martingale laws, the adapted cost reduces under bicausal couplings to the squared difference of the terminal values is used by Backhoff-Veraguas, Bartl, Beiglb\"ock and Eder in their computation for geometric Brownian motions \cite[Proposition~3.6]{BackhoffBartlBeiglbockEder2020Finance}. What part~\ref{it:terminal} adds is that for defaultable bond prices the terminal values are default indicators, so that the terminal cost is a probability of disagreement, which can be reduced once more to the prices at the last date before maturity. Part~\ref{it:chains} is a consequence of part~\ref{it:terminal}, of the dynamic programming principle of \cite{BackhoffBeiglbockLinZalashko2017}, and of the optimality of the synchronous, or Knothe-Rosenblatt, coupling between stochastically monotone Markov processes, which goes back to R\"uschendorf \cite{Ruschendorf1985} and is proved in the bicausal setting in \cite[Proposition~5.3]{BackhoffBeiglbockLinZalashko2017}. Parts~\ref{it:one-review} and~\ref{it:timing}, and more generally the explicit formulas for the information structures of rating-based models, follow from the second reduction in part~\ref{it:terminal}. Together with part~\ref{it:chains}, they turn the adapted distance into an exactly computable measure of how differently two calibrated rating systems release information, for instance, a point-in-time and a through-the-cycle rating philosophy.

\medskip
In the model with one review, the distance can also be estimated from data. If the law of the posterior is replaced by the empirical law of observed posteriors, rescaled so that it still reproduces the price $p$, the squared adapted distance is estimated with expected error at most $n^{-1/2}$, see Theorem~\ref{thm:posterior-estimation}.

\medskip
For intensity models, our results read as follows.

\begin{theorem}[Default mismatch and model-transfer hedging]\label{thm:intro-cox}
Let $a$ and $b$ be hazard paths with cumulative hazards $\mathsf{A}_t=\int_0^ta_s\,\mathrm{d}s$, $\mathsf{B}_t=\int_0^tb_s\,\mathrm{d}s$ and $\mathsf{M}_t=\int_0^t\max\{a_s,b_s\}\,\mathrm{d}s$.
\begin{enumerate}
\item\label{it:mismatch} If $a$ and $b$ are deterministic, the squared adapted Wasserstein distance between the price processes of the zero-recovery bonds maturing at $t$ in the two Cox models is the smallest probability, over bicausal couplings, that their survival indicators disagree at $t$, and it equals
\[
\Psi_t(a,b)=e^{-\mathsf{A}_t}+e^{-\mathsf{B}_t}-2e^{-\mathsf{M}_t}.
\]
It is attained for all maturities simultaneously by a common shock with intensity $\min\{a,b\}$. For random intensities, among couplings that preserve the doubly stochastic structure of both models, the problem reduces to an adapted transport problem between the factor laws with cost $\Psi$, see Theorem~\ref{thm:aw-cox} and Theorem~\ref{thm:default-layer} for a precise statement.
\item\label{it:roll} The expected absolute cash residual of the bootstrap roll, implemented under a rating-driven market model, equals the expected absolute difference of the two survival factors at maturity, and is bounded by $\mathbb{E}\Psi$. The gap is produced by crossings of the two hazard paths inside the bucket and is given explicitly, see Theorem~\ref{thm:cox-hedge-error} for a precise statement.
\item\label{it:term-structure} Along a quoted term structure, the expected absolute cash residuals of the rolling bootstrap are bounded bucket by bucket by the adapted default mismatch between the market model and a model that switches to the bootstrap one bucket ahead, with equality when ratings are reviewed at the calibration dates and select flat hazards for the following buckets, see Theorem~\ref{thm:term-structure} for a precise statement.
\end{enumerate}
\end{theorem}

Part~\ref{it:mismatch} is the continuous-time counterpart of Theorem~\ref{thm:intro-ratings}\ref{it:terminal}. In both settings, the squared adapted distance between two defaultable bond prices is the minimal bicausal probability that the two defaults disagree, and for deterministic hazards this probability is available in closed form. The bounds in parts~\ref{it:roll} and~\ref{it:term-structure} are therefore bounds by adapted Wasserstein distances between bond prices.

\medskip
Calibration to the quoted prices leaves the adapted default mismatch largely undetermined. If the hazard rates are bounded, we compute its exact range over all hazard schedules with prescribed survival probabilities at the ends of a bucket, see Theorem~\ref{thm:hazard-envelope}. In particular, two schedules with the same survival probability can have a mismatch as large as the intensity bound allows, see Corollary~\ref{cor:invisible-calibrated-mismatch}. The mismatch is moreover Lipschitz in the hazard rates with respect to the $L^1$ norm, with sharp constant one, which controls the error made when hazard paths are approximated, see Theorem~\ref{thm:psi-L1-stability}.

\medskip
Finally, a desk that knows a rating-driven market model and can trade bonds of several maturities can remove its hedging error altogether. Indeed, in the time-homogeneous rating-driven market considered below \cite{Norberg2003}, finitely many consecutive maturities span the price of the long bond, see Proposition~\ref{prop:spanning}. We use this in Section~\ref{subsec:empirical-multi} to separate two sources of hedging error that are usually confounded: missing instruments and a misspecified model. We also show that, in the model with one review, an adapted Wasserstein ball around the coarse model is exactly a bound on the mean absolute deviation of the posterior, which yields explicit robust prices for options on the bond price at the review date. Moreover, under calibrated perturbations of rating-dependent hazard rates, the ordinary distance grows linearly in the size of the perturbation, while the adapted one grows like its square root, see Theorem~\ref{thm:hazard-asymptotics} and Corollary~\ref{cor:asymptotic-separation}.



\medskip
In Section~\ref{sec:examples} we apply these results to the rating transition data published by S\&P Global Ratings \cite{SP2025}. We compare rating systems that agree on today's bond price but release information differently: U.S.\ against European rating histories, the long-run average against a single recent year, and a point-in-time system that reacts to the business cycle against a through-the-cycle one. The adapted distance between such systems turns out to be a sizeable fraction of their distance from a model that reveals no information at all. The way in which information is released therefore matters even when the initial price of the bond being compared agrees, while static measures of rating mobility can hardly tell some of these systems apart. Agency matrices are not exactly monotone, but the synchronous coupling of Theorem~\ref{thm:intro-ratings}\ref{it:chains} remains optimal up to a negligible error so that the distances can be computed in closed form. We then simulate the bootstrap hedge in a market calibrated to the same data. The bound of Theorem~\ref{thm:intro-cox}\ref{it:term-structure} is exact when ratings change only at the calibration dates and still captures most of the realised error when they change in between. Finally, when the desk can trade bonds of several maturities, a few bonds suffice to remove the hedging error if the market model is known, whereas the bootstrap hedge improves only slowly. With a rich hedging universe, most of the remaining error therefore comes from the model rather than from missing instruments.

\medskip
\textbf{Comments on the proofs.} All explicit results on defaultable bond prices rest on a single observation. Under a bicausal coupling, neither model learns anything about the other's future, so both price processes remain martingales with respect to the joint flow of information, and so does their difference. The martingale isometry then turns the sum of squared increment differences into the squared difference of the terminal values, which for defaultable bond prices is the indicator that the two defaults disagree. Conditioning on the information at the last date before maturity, the two defaults are Bernoulli variables with parameters equal to the current prices, and the elementary fact that two Bernoulli variables with parameters $a$ and $b$ disagree with probability at least $|a-b|$ reduces the problem further to the distance between the prices at that date (Proposition~\ref{prop:terminal-representation}). The explicit formulas of Section~\ref{sec:ratings} follow by identifying which variables bicausality forces to be independent, typically because the history of one model is deterministic up to the relevant date.

\medskip
For rating chains, causality and the Markov property imply that, given the current pair of ratings, the next pair is distributed according to a coupling of the two one-step transition laws, which yields the dynamic programming principle of \cite{BackhoffBeiglbockLinZalashko2017} as a backward recursion over pairs of current states. When migration is monotone, every value function of the recursion has the Monge property, under which the comonotone coupling is optimal at each step, and composing the comonotone couplings gives the synchronous coupling. We include this short argument to keep the paper self-contained.

\medskip
For intensity models with deterministic hazards, the bond price is a martingale of finite variation that jumps only at default, and its filtration is that of the default indicator. The quadratic variation of the difference of two such prices is the sum of its squared jumps, and the martingale isometry again turns the cost into the probability that the two defaults disagree. Bicausality forces each default indicator to keep its hazard rate in the joint flow of information. The lower bound is then obtained by following the probability that both names survive. Integration by parts expresses its evolution through the two hazard rates and the rate of simultaneous default, which cannot exceed the smaller hazard. A differential inequality and Gr\"onwall's lemma then bound the joint survival probability by $e^{-\mathsf{M}_t}$, and three independent Poisson processes, one common and two idiosyncratic, attain the bound. That this construction is bicausal follows from the martingale representation of single-jump processes. For random intensities, conditioning on the factor paths reduces the problem to deterministic hazards. The hedging results all follow the same pattern: by the Cox pricing formula, each cash residual is the conditional expectation, given the current information, of a difference of survival factors under the market model and under a bootstrap or hybrid model. Conditional Jensen's inequality and the explicit form of $\Psi$ then separate the realised error, the information arriving inside a bucket, and the adaptation slack. 


\subsection{Related literature and what is new}\label{subsec:position}
In this section we place our results in the context of the literature on adapted transport, stability in mathematical finance, explicit solutions of transport problems and credit risk, and we highlight where our main contributions lie.

\medskip
\textbf{Adapted transport.} 
The idea of comparing stochastic processes through couplings that respect the flow of information appears independently in several communities.
A first construction in this direction is Vershik's iterated Kantorovich distance, developed for the classification of filtrations \cite{Vershik1994}, and the bicausality condition already appears in R\"uschendorf's work on couplings of Markov processes \cite{Ruschendorf1985}. Independently, Pflug and Pichler introduced nested distributions and the nested distance, a metric on scenario trees under which the values of multistage programs are continuous, and were the first to exploit it systematically in multistage stochastic optimization \cite{Pflug2009, PflugPichler2012,PflugPichler2014}. In optimal transport, causal and bicausal couplings were formulated by Lassalle and by Backhoff-Veraguas, Beiglb\"ock, Lin and Zalashko \cite{Lassalle2018,BackhoffBeiglbockLinZalashko2017}. The latter also derived a dynamic programming principle for bicausal transport in discrete time and showed that the Knothe-Rosenblatt rearrangement is optimal for co-monotone processes. The relation between causal couplings and enlargement of filtrations was clarified in \cite{AcciaioBackhoffZalashko2020}. Backhoff-Veraguas, Bartl, Beiglb\"ock and Eder showed that the resulting topologies all coincide, and that they agree with older notions such as Aldous' extended weak topology and Hellwig's information topology \cite{BackhoffBartlBeiglbockEder2020Top}, while fundamental metric properties of these distances were established in \cite{BackhoffBeiglbockEderPichler2020}. Bartl, Beiglb\"ock and Pammer identified the completion of the space of laws under the adapted distance with the space of filtered processes, and showed that it is a geodesic space in which martingales form a closed convex subset \cite{BartlBeiglbockPammer2024}. Statistical estimation, which requires modified empirical measures because the ordinary empirical measure does not converge in adapted topologies, and numerical methods are treated in \cite{BackhoffBartlBeiglbockWiesel2022,EcksteinPammer2024}. Continuous-time versions of the theory are developed in \cite{BionNadalTalay2019,BackhoffKallbladRobinson2025,BartlBeiglbockPammerSchrottZhang2025,BeiglbockPfluglSchrott2026}.

\medskip
\textbf{Stability in mathematical finance.} The paper closest to ours in spirit is \cite{BackhoffBartlBeiglbockEder2020Finance}. There, Backhoff-Veraguas, Bartl, Beiglb\"ock and Eder define the adapted distance between semimartingale laws through the quadratic variation of the difference of the martingale parts and the total variation of the difference of the drifts, and explain why this cost, rather than the nested distance, which uses the cost on the path values, has the right scaling for a continuous-time limit \cite[Section~3.3]{BackhoffBartlBeiglbockEder2020Finance}. They prove that under bicausal couplings the coordinate processes keep their semimartingale decompositions \cite[Lemma~3.1]{BackhoffBartlBeiglbockEder2020Finance}, which is the fact behind our terminal representation, and they bound the difference of the stochastic integrals of a common bounded strategy \cite[Lemma~3.9]{BackhoffBartlBeiglbockEder2020Finance}. From these estimates, they derive Lipschitz stability of superhedging errors, both when the hedge is adapted to the new model and when the same Lipschitz strategy is used in both \cite[Theorem~1.5]{BackhoffBartlBeiglbockEder2020Finance}, of average value at risk, of utility maximization and of utility indifference prices, with bounds that are sharp for Brownian models and call options. They also compute the distance explicitly for diffusions with suitably simple coefficients, where the synchronous coupling is optimal, and for geometric Brownian motions, and show by example that the synchronous coupling is not optimal in general \cite[Section~3.2]{BackhoffBartlBeiglbockEder2020Finance}. Earlier, Glanzer, Pflug and Pichler bounded acceptability prices of contingent claims through the nested distance \cite{GlanzerPflugPichler2019}. Bartl and Wiesel computed the first-order sensitivity of multiperiod optimization problems with respect to adapted perturbations \cite{BartlWiesel2023}, and Acciaio, Backhoff-Veraguas and Pammer quantified the fundamental theorem of asset pricing under adapted perturbations \cite{AcciaioBackhoffPammer2025}. These results are bounds of the form $|V(\mathbb{P})-V(\mathbb{Q})|\le C\,\mathcal{AW}(\mathbb{P},\mathbb{Q})$ for broad classes of value functionals. We use them rather than reprove them: the stability of prices and of a common hedge for defaultable bond prices, Remark~\ref{rem:price-stability} and Proposition~\ref{prop:hedge-stability}, is a direct instance of \cite{BackhoffBartlBeiglbockEder2020Finance}, recorded with the constants we need. Robust pricing over ordinary Wasserstein balls is studied in \cite{MohajerinKuhn2018, BlanchetMurthy2019,BartlDrapeauOblojWiesel2021}, model uncertainty more broadly in \cite{Cont2006,GlassermanXu2014}, and the performance of hedges under a misspecified model in \cite{ElKarouiJeanblancShreve1998,FollmerSchweizer1991}.

\medskip
\textbf{Explicit solutions.} Explicit solutions of adapted transport problems are available only in a few cases. Besides the diffusion examples of \cite{BackhoffBartlBeiglbockEder2020Finance} and their extension to stochastic differential equations in \cite{BackhoffKallbladRobinson2025}, where the synchronous coupling is optimal, the most developed case is the Gaussian one. Gunasingam and Wong computed the adapted Wasserstein distance between Gaussian processes in discrete time in closed form \cite{GunasingamWong2025} and extended the analysis to filtered Gaussian processes \cite{GunasingamWong2026}. Acciaio, Hou and Pammer obtained closed forms for multidimensional Gaussian processes and for the entropic regularization of the problem \cite{AcciaioHouPammer2025}, and the geometry of the resulting adapted Bures-Wasserstein space is studied in \cite{AcciaioBartlGrassHouPammer2026}. Barycenters of Gaussian processes were characterized in \cite{GunasingamMattesiniWieselWong2026}, using a decomposition of the squared adapted distance into classical squared Bures-Wasserstein distances between covariance contributions of successive innovations. For Markov processes, R\"uschendorf identified the optimal coupling between stochastically monotone chains \cite{Ruschendorf1985}, and for stationary finite-state Markov chains O'Connor, McGoff and Nobel compute optimal transition couplings by policy iteration \cite{OConnorMcGoffNobel2022}. The present paper adds explicit solutions of a different kind. The laws involved are martingales with binary terminal values, which are very different from Gaussian laws, and the closed forms come from the terminal representation rather than from linear algebra. In continuous time, Theorem~\ref{thm:aw-cox} gives the adapted distance between the price processes of two Cox bonds with deterministic hazards in closed form, with a common-shock optimal coupling. 
We present this as an explicit solution for a particular class of single-jump credit martingales, without claiming priority over the broader coupling literature on jump processes.

\medskip
\textbf{Credit risk.} Intensity-based models represent default as the first jump of a process with a given hazard rate, possibly driven by economic factors, and are the standard framework for pricing and hedging defaultable claims \cite{Lando1998,DuffieSingleton1999,ElliottJeanblancYor2000,BieleckiRutkowski2002,Schonbucher2003,Lando2004,BieleckiJeanblancRutkowski2004,JeanblancYorChesney2009}. Rating-migration models describe credit quality by a Markov chain on rating classes \cite{JarrowLandoTurnbull1997,KalkbrenerPackham2026}, and the role of the information available to investors, which is the object of this paper, has been emphasized in models with incomplete or delayed information \cite{DuffieLando2001,JarrowProtter2004,FreyRoeslerLu2019}. Wasserstein distances have been used to proxy credit curves from static default distributions \cite{MichielonKhedherSpreij2024}, a setting in which information plays no role. Relative entropy, the usual alternative for comparing or pooling models \cite{Meucci2008}, is infinite in the direction from a nondegenerate one-review model to the coarse benchmark, because the former is not absolutely continuous with respect to the latter (Remark~\ref{rem:entropy-singular}). The same obstruction motivates the transport criteria of \cite{Vecer2026} for diffusion models. The optimal coupling in Theorem~\ref{thm:aw-cox} is a Marshall-Olkin common shock \cite{MarshallOlkin1967}, and couplings of killing times of this type are classical in probability, where they appear as optimal co-adapted couplings \cite{Lindvall1992,BurdzyKendall2000,ConnorJacka2008}. In credit risk, however, adapted transport does not seem to have been used so far to compare models or to quantify the hedging errors of credit strategies, and this is the gap that the present paper addresses.

\medskip
\textbf{What is new.} In view of the above, the contributions of this paper are the following. First, the identification of the squared adapted distance between defaultable bond prices with a minimal probability of default disagreement, and its reduction to the last pre-maturity prices, which yield closed forms for one review, for a change of its date and for the no-information benchmark. Second, the combination of this representation with the classical synchronous coupling, which turns the comparison of calibrated rating systems into an exact computation, carried out on agency data. Third, in continuous time, the closed-form adapted Wasserstein distance between the price processes of Cox bonds with deterministic hazards, attained by a common shock, and the reduction of factor-driven intensities to an adapted transport problem between factor laws. Fourth, exact identities and bounds for the hedging errors of bootstrap strategies, in a single bucket, along a term structure and with several instruments, together with their empirical evaluation. The spanning of rating-driven markets by finitely many bonds is a discrete-time instance of known completeness results \cite{Norberg2003,BieleckiRutkowski2002}, and what is new there is the quantitative comparison of bootstrap and market-based strategies on the same hedging universe. The stability bounds of \cite{BackhoffBartlBeiglbockEder2020Finance} are used, not reproved, and serve as a benchmark against which the exact identities can be compared.

\subsection{Open problems}\label{subsec:open}

We conclude this section with some questions that arise in view of our results.
\begin{enumerate}[label=(\arabic*)]
\item \emph{Nonlinear credit claims.} Our exact identities exploit the fact that the value of a zero-coupon bond, and more generally of a claim with deterministic cash flows such as a coupon bond or a leg of a credit default swap, is a linear function of the survival curve. The cash residual of a hedge is then a difference of survival factors, and its size is controlled by the adapted default mismatch. Many traded credit claims are not of this form. Callable and putable bonds contain exercise decisions. Step-up coupons and rating triggers make payments depend on future ratings, and options on bonds or on credit spreads depend nonlinearly on the conditional survival curve. For such claims, the hedging error is no longer determined by the default mismatch alone, since it also depends on how the two models revise the value of the claim over time. A natural goal is to bound, or ideally to compute, the largest model-transfer error over a class of such claims by a transport problem. Because the value of a nonlinear claim depends on conditional laws of the future, the relevant cost should be of weak type, that is, a function of conditional distributions rather than of paths. The open question is whether this problem admits a dynamic programming principle that makes it as tractable as the one used here for linear claims.
\item \emph{Continuous information.} In our rating models information arrives at finitely many dates, typically rating reviews. In practice, credit information is also revealed continuously, through the observation of spreads, accounting data or the issuer's asset value, and in filtering models the price of a defaultable bond evolves in continuous time \cite{DuffieLando2001,FreyRoeslerLu2019}. The terminal representation only uses the martingale property, and Theorem~\ref{thm:aw-cox} extends it to Cox bond prices with deterministic hazards, but computing the distance is harder when prices move continuously before default. A natural route is to discretize the observation dates and pass the multi-date formulas to the limit. This requires tightness of almost optimal bicausal couplings, a proof that the bicausality constraint survives in the limit, and control of the coupling of the two default times at maturity. In filtering models, the natural state variable is the conditional law of the unobserved credit factor, so that the problem becomes a transport problem between laws of measure-valued processes, in the continuous-time framework of \cite{BartlBeiglbockPammerSchrottZhang2025,BackhoffKallbladRobinson2025}. Explicit formulas in simple filtering models, comparable to those obtained here for rating reviews, would already be of interest.
%
\item \emph{Estimation.} The rating systems compared in Section~\ref{sec:examples} are estimated from finite samples of rating histories. Transitions out of investment-grade classes, and defaults in particular, are rare, histories are censored when ratings are withdrawn, and the estimated matrices are historical rather than risk neutral \cite{LandoSkodeberg2002,JafrySchuermann2004,FrydmanSchuermann2008}. Part of the adapted distance between two estimated systems may therefore reflect sampling noise rather than a genuine difference in the way information is released. Separating the two requires nonasymptotic bounds, or confidence intervals, for the adapted distance between a Markov chain and its estimate. In the finite-state case, such bounds should follow from the concentration of transition counts, in the spirit of the estimation results of \cite{BackhoffBartlBeiglbockWiesel2022}. A related question is how sensitive the distances are to the choice of the adjustment from historical to risk-neutral migration.
%
\item \emph{Portfolios and aggregation.} Our results compare a single name under two models, whereas credit indices, tranches and counterparty valuation adjustments depend on dependent defaults across many names. Comparing two portfolio models requires coupling vectors of default indicators, and it is not clear whether the multivariate Marshall-Olkin construction \cite{MarshallOlkin1967} remains optimal when defaults are dependent within each model. A related question is the aggregation of several credit models, such as different rating systems, into a consensus model. Candidates are the arithmetic pooling of experts in the spirit of \cite{Vecer2026} and adapted barycenters \cite{BartlBeiglbockPammer2024,GunasingamMattesiniWieselWong2026}. Since the adapted distance controls hedging errors, the distance of the consensus to each expert would bound the error of hedging with it when one of the experts is correct.
\item \emph{Couplings beyond the doubly stochastic class.} For factor-driven intensities, Theorem~\ref{thm:default-layer} minimizes the default mismatch over couplings that preserve the Cox structure of both models. Bicausal couplings of the full processes may also let the joint coupling of future factor innovations depend on the observed default history, while preserving each marginal conditional law, as explained in Remark~\ref{rem:feedback}. Whether such feedback can strictly reduce the mismatch is open.
\end{enumerate}

\subsection{Outline of the paper}\label{subsec:organization}

Section~\ref{sec:framework} develops the filtered price-law setting and the terminal transport reduction. Section~\ref{sec:ratings} treats one-review, timing and migration models, including calibration-preserving statistical estimation and dynamic hedging. Section~\ref{sec:cox-hedging} establishes the Cox mismatch formulas, sharp hazard envelopes and perturbation bounds, and analyzes model-transfer hedging, robust pricing and term-structure effects. Section~\ref{sec:examples} contains the empirical illustrations.

\section{Defaultable bond prices and adapted distances}\label{sec:framework}

\subsection{Standing assumptions}\label{subsec:standing-assumptions}

Throughout the paper, unless specified differently, we assume that the time is discrete, the interest rates are zero, and the recovery rate is zero. Moreover, we assume that all probability spaces are standard Borel spaces, so that regular conditional distributions exist.

\subsection{Defaultable bond prices}\label{subsec:survival-martingales}

First, we make precise the objects described in the introduction. Fix an integer $N\ge2$ and index time by $\{0,1,\ldots,N\}$. Let $(\Omega,\mathcal{F},\mathbb{F},\mathbb{Q})$ be a filtered probability space with filtration $\mathbb{F}=(\mathcal{F}_t)_{t=0}^N$ with $\mathcal{F}_0 = \{\emptyset, \Omega\}$, where $\mathbb{Q}$ is a pricing measure and $\mathcal{F}_t$ represents the information available to the market at date $t$. We take the money market account as numeraire and normalize it to one, so that interest rates are zero and all prices are discounted. The default time of the issuer 
is an $\mathbb{F}$-stopping time, with $\tau=+\infty$ if the issuer never defaults. 

\medskip
Consider a zero-recovery defaultable zero-coupon bond maturing at $N$, with terminal payoff $D=\mathbf{1}_{\{\tau>N\}}$. Its price process is
\begin{equation}\label{eq:bond-price}
S_t=\mathbb{E}_{\mathbb{Q}}[D\mid\mathcal{F}_t],\qquad 0\le t\le N.
\end{equation}
Then $S$ is a bounded $\mathbb{Q}$-martingale with values in $[0,1]$, and $S_N=D\in\{0,1\}$. 
The initial price is $p=S_0=\mathbb{Q}[\tau>N]$. Throughout we assume $p\in(0,1)$, since the degenerate values $p\in\{0,1\}$ carry no default uncertainty.

\medskip
Let $\Omega_N=[0,1]^{N+1}$ with canonical process $X_t(x)=x_t$ and canonical filtration $\mathcal{F}_t^X=\sigma(X_0,\ldots,X_t)$. For $p\in(0,1)$ define
\begin{equation}\label{eq:martingale-class}
\mathcal{M}_N(p)=
\bigl\{
\mathbb{P}\in\mathcal P(\Omega_N):
X_0=p\ \mathbb{P}\text{-a.s.},\ X\text{ is a }\mathbb{P}\text{-martingale},\ X_N\in\{0,1\}\ \mathbb{P}\text{-a.s.}
\bigr\}.
\end{equation}
Every price process of a zero-recovery defaultable zero-coupon bond induces a law in $\mathcal{M}_N(p)$. The following lemma shows that all these laws coincide at maturity, so that the initial price alone determines the law of the survival indicator at maturity.

\begin{lemma}[Terminal law]\label{lem:terminal-law}
Every $\mathbb{P}\in\mathcal{M}_N(p)$ satisfies $\mathbb{P}[X_N=1]=p$ and $\mathbb{P}[X_N=0]=1-p$.
\end{lemma}

\begin{proof}
By the martingale property, $\mathbb{E}_{\mathbb{P}}[X_N]=\mathbb{E}_{\mathbb{P}}[X_0]=p$. Since $X_N\in\{0,1\}$, $\mathbb{E}_{\mathbb{P}}[X_N]=\mathbb{P}[X_N=1]$.
\end{proof}

Lemma~\ref{lem:terminal-law} formalizes the calibration issue: once the time-zero bond price $p$ is fixed, all pricing laws in $\mathcal{M}_N(p)$ share the same terminal Bernoulli law. At the same time, their intermediate conditional survival probabilities and filtrations may differ.

\subsection{Bicausal couplings}\label{subsec:bicausal}

Let $\mathbb{P},\mathbb{Q}\in\mathcal P(\Omega_N)$. A coupling $\pi\in\operatorname{Cpl}(\mathbb{P},\mathbb{Q})$ is a probability measure on $\Omega_N\times\Omega_N$ with marginals $\mathbb{P}$ and $\mathbb{Q}$. We denote the canonical coordinates on the product by $X$ and $Y$.

\medskip
We now specify the causality constraint we will consider in the following. Informally, a coupling is causal from $X$ to $Y$ if, at every date, the part of the path $Y$ constructed so far depends on the path $X$ only through its past. It is bicausal if the same holds with the roles of $X$ and $Y$ exchanged.

\begin{definition}[Causal and bicausal couplings]\label{def:bicausal}
A coupling $\pi\in\operatorname{Cpl}(\mathbb{P},\mathbb{Q})$ is \emph{causal from $X$ to $Y$} if, for each $t\in\{0,\ldots,N\}$ and each bounded Borel function $f$ of $(Y_0,\ldots,Y_t)$, the conditional expectation $\mathbb{E}_\pi[f(Y_0,\ldots,Y_t)\mid X_0,\ldots,X_N]$ admits a $\sigma(X_0,\ldots,X_t)$-measurable version. It is \emph{bicausal} if it is causal from $X$ to $Y$ and from $Y$ to $X$. The set of bicausal couplings is denoted by $\operatorname{Cpl}_{\mathrm{bc}}(\mathbb{P},\mathbb{Q})$.
\end{definition}

Equivalently following \cite{BackhoffBeiglbockLinZalashko2017}, $\pi$ is causal from $X$ to $Y$ if and only if $X_{t+1:N}$ and $Y_{0:t}$ are conditionally independent given $X_{0:t}$ under $\pi$, for every $t$. The explicit couplings of this paper are built step by step, and the following criterion verifies their bicausality. For finite products of Polish spaces write $x_{0:t}=(x_0,\ldots,x_t)$ and $y_{0:t}=(y_0,\ldots,y_t)$.

\begin{proposition}[{\cite[Proposition~5.1]{BackhoffBeiglbockLinZalashko2017}}]\label{prop:sequential-bicausal}
A coupling $\pi$ of $\mathbb{P}$ and $\mathbb{Q}$ is bicausal if and only if it factorizes as
\begin{equation}\label{eq:sequential-factorization}
\pi(\mathrm{d}x,\mathrm{d}y)=\kappa_0(\mathrm{d}x_0,\mathrm{d}y_0)\prod_{t=1}^N\kappa_t(\mathrm{d}x_t,\mathrm{d}y_t\mid x_{0:t-1},y_{0:t-1}),
\end{equation}
where $\kappa_0$ is a coupling of the laws of $X_0$ and $Y_0$ and, for every $t$ and $\pi$-almost every $(x_{0:t-1},y_{0:t-1})$, the kernel $\kappa_t(\cdot\mid x_{0:t-1},y_{0:t-1})$ is a coupling of $\mathbb{P}(\mathrm{d}x_t\mid x_{0:t-1})$ and $\mathbb{Q}(\mathrm{d}y_t\mid y_{0:t-1})$.
\end{proposition}

We only sketch the proof and refer to \cite[Proposition~5.1]{BackhoffBeiglbockLinZalashko2017} for the details.

\begin{proof}[Sketch of proof]
If $\pi$ is bicausal, causality from $X$ to $Y$ makes $X_t$ and $Y_{0:t-1}$ conditionally independent given $X_{0:t-1}$. Hence the conditional law of $X_t$ given $(X_{0:t-1},Y_{0:t-1})$ is $\mathbb{P}(\mathrm{d}x_t\mid X_{0:t-1})$. The same holds for $Y_t$ by symmetry. Thus, \eqref{eq:sequential-factorization} holds with $\kappa_0$ the law of $(X_0,Y_0)$ and $\kappa_t$ the conditional law of $(X_t,Y_t)$ given $(X_{0:t-1},Y_{0:t-1})$ under $\pi$.

\medskip
Conversely, assume \eqref{eq:sequential-factorization} and fix $t\le N-1$. By the characterization of causality recalled above, it suffices to show that the conditional law of $X_{t+1:N}$ given $(X_{0:t},Y_{0:t})$ is $\mathbb{P}(\mathrm{d}x_{t+1:N}\mid X_{0:t})$. This conditional law is obtained from $\prod_{s=t+1}^{N}\kappa_s$ by integrating out $y_N,y_{N-1},\ldots,y_{t+1}$ in this order. At the step of $y_s$ the integrand no longer depends on $y_s$, and the first marginal of $\kappa_s$ is $\mathbb{P}(\mathrm{d}x_s\mid x_{0:s-1})$, which does not depend on $y_{0:s-1}$. We are left with
\[
\prod_{s=t+1}^{N}\mathbb{P}(\mathrm{d}x_s\mid x_{0:s-1})=\mathbb{P}(\mathrm{d}x_{t+1:N}\mid x_{0:t}),
\]
which does not depend on $y_{0:t}$. Finally, causality from $Y$ to $X$ follows by exchanging the roles of the two coordinates.
\end{proof}

The key property of bicausal couplings for our purposes is that they preserve the martingale structure of the two price processes. This is the content of the following lemma.

\begin{lemma}[{\cite[Lemma~3.1]{BackhoffBartlBeiglbockEder2020Finance}}]\label{lem:martingale-under-bicausal}
Let $\mathbb{P},\mathbb{Q}\in\mathcal{M}_N(p)$ and $\pi\in\operatorname{Cpl}_{\mathrm{bc}}(\mathbb{P},\mathbb{Q})$. Then $X$ and $Y$ are $\pi$-martingales with respect to the product filtration $\mathcal{H}_t=\sigma(X_0,\ldots,X_t,Y_0,\ldots,Y_t)$. Consequently $X-Y$ is a bounded $(\mathcal{H}_t)$-martingale.
\end{lemma}

The lemma is a special case of the fact that bicausal couplings preserve the martingale property through immersion of filtrations \cite{BremaudYor1978,AcciaioBackhoffZalashko2020}. We include the short proof for the reader's convenience.

\begin{proof}
Fix $t<N$. By a monotone class argument it suffices to show $\mathbb{E}_\pi[(X_{t+1}-X_t)Z]=0$ for $Z=f(X_{0:t})g(Y_{0:t})$ with $f,g$ bounded Borel. Disintegrate $\pi(\mathrm{d}x,\mathrm{d}y)=\mathbb{P}(\mathrm{d}x)\pi_x(\mathrm{d}y)$. Causality from $X$ to $Y$ implies that $\psi_t(x)=\int g(y_{0:t})\,\pi_x(\mathrm{d}y)$ admits a version depending only on $x_{0:t}$. Hence
\[
\mathbb{E}_\pi[(X_{t+1}-X_t)Z]=\int(x_{t+1}-x_t)f(x_{0:t})\psi_t(x_{0:t})\,\mathbb{P}(\mathrm{d}x)=0
\]
by the martingale property of $X$ under $\mathbb{P}$. The statement for $Y$ follows by symmetry, using causality from $Y$ to $X$.
\end{proof}

\subsection{Adapted and ordinary Wasserstein distances}\label{subsec:distances}

A coupling compares two models scenario by scenario, and the cost has to say what it means for two price paths to be close. Since we want to consider the hedging problem,
we measure closeness by the squared differences of the price increments, see Remark~\ref{rem:level-convention} for this choice and its relation to the cost on the path values. The resulting control of hedging gains is recalled in Proposition~\ref{prop:hedge-stability}. For $\mathbb{P},\mathbb{Q}\in\mathcal{M}_N(p)$ we set
\begin{equation}\label{eq:aw2-def}
\begin{aligned}
\mathcal{AW}_2^2(\mathbb{P},\mathbb{Q})&=\inf_{\pi\in\operatorname{Cpl}_{\mathrm{bc}}(\mathbb{P},\mathbb{Q})}\mathbb{E}_\pi\Bigl[\sum_{t=1}^N|\Delta X_t-\Delta Y_t|^2\Bigr],\\
\mathcal{W}_2^2(\mathbb{P},\mathbb{Q})&=\inf_{\pi\in\operatorname{Cpl}(\mathbb{P},\mathbb{Q})}\mathbb{E}_\pi\Bigl[\sum_{t=1}^N|\Delta X_t-\Delta Y_t|^2\Bigr],
\end{aligned}
\end{equation}
where $\Delta X_t=X_t-X_{t-1}$ and $\Delta Y_t=Y_t-Y_{t-1}$. The two distances use the same cost and differ only in the admissible couplings. Since every bicausal coupling is a coupling,
\begin{equation}\label{eq:W-below-AW}
\mathcal{W}_2(\mathbb{P},\mathbb{Q})\le\mathcal{AW}_2(\mathbb{P},\mathbb{Q}),
\end{equation}
and any gap between the two is produced by the prohibition to anticipate.

\begin{remark}[Choice of the cost]\label{rem:level-convention}
The canonical definition of the adapted Wasserstein distance in \cite{BartlBeiglbockPammer2024} uses the cost on the path values, $\sum_t|X_t-Y_t|^2$. Following \cite[Section~3.3]{BackhoffBartlBeiglbockEder2020Finance} and \cite[Remark~6.5]{BartlBeiglbockPammer2024}, we use instead, for both the adapted and the ordinary distance, a cost built on the Doob decomposition of the processes: the quadratic variation of the difference of the martingale parts and the total variation of the difference of the predictable parts. For martingales, the predictable parts vanish, and with quadratic cost in discrete time the expected quadratic variation of $X-Y$ is $\mathbb{E}\sum_t|\Delta X_t-\Delta Y_t|^2$, which is \eqref{eq:aw2-def}. This choice is the more natural one in mathematical finance, for the reason explained in \cite[Section~3.3]{BackhoffBartlBeiglbockEder2020Finance} and \cite[Remark~6.5]{BartlBeiglbockPammer2024}. The nested distance with the cost on the path values between discretizations of two Brownian models with different volatilities diverges as the number of periods grows, while the increment-based distance converges to its continuous-time value \cite[Section~3.3]{BackhoffBartlBeiglbockEder2020Finance}. Moreover, the quantities of interest there are stochastic integrals, such as the gains $\sum_tH_t\Delta X_t$ of a trading strategy, and these are controlled by the quadratic variation through the It\^o isometry, uniformly in the number of periods. In contrast, closeness of the price paths does not control them in the continuous-time limit. In discrete time the two costs are equivalent: since $X_0=Y_0$, Lemma~\ref{lem:increment-path} and the inequality $|\Delta x_t-\Delta y_t|^2\le2(|x_t-y_t|^2+|x_{t-1}-y_{t-1}|^2)$ give equivalence up to the factors $N^2$ and $4$, so the two costs define the same topology. The exact formulas change by explicit terms: in the model with one review, the cost on the path values gives $\mathbb{E}|Q-p|^2+\mathbb{E}|Q-p|$ for the squared adapted distance and $\mathbb{E}|Q-p|^2$ for the ordinary one (Theorem~\ref{thm:exact-w2-one-rating}). The square-root scaling discussed below holds under either convention.
\end{remark}

The following four elementary facts are used throughout.

\begin{lemma}[Discrete martingale isometry]\label{lem:martingale-isometry}
If $M$ is a square-integrable martingale with deterministic $M_0$, then $\mathbb{E}|M_n-M_0|^2=\sum_{t=1}^n\mathbb{E}|\Delta M_t|^2$.
\end{lemma}

\begin{proof}
For $s<t$, by the tower property of conditional expectation we have $\mathbb{E}[\Delta M_s\Delta M_t]=\mathbb{E}[\Delta M_s\mathbb{E}[\Delta M_t\mid\mathcal{F}_{t-1}]]=0$.
\end{proof}

\begin{lemma}[Increment-to-path estimate]\label{lem:increment-path}
If $x,y\in\mathbb{R}^{N+1}$ satisfy $x_0=y_0$, then $\sum_{t=1}^N|x_t-y_t|^2\le N^2\sum_{t=1}^N|\Delta x_t-\Delta y_t|^2$.
\end{lemma}

\begin{proof}
Note that $x_t-y_t=\sum_{s\le t}(\Delta x_s-\Delta y_s)$, so $|x_t-y_t|^2\le N\sum_{s=1}^N|\Delta x_s-\Delta y_s|^2$ by Cauchy-Schwarz. We conclude by summing over $t$.
\end{proof}

\begin{lemma}[Bernoulli mismatch]\label{lem:bernoulli-mismatch}
If $U\sim\mathrm{Bernoulli}(a)$ and $V\sim\mathrm{Bernoulli}(b)$, then $\inf\mathbb{P}[U\neq V]=|a-b|$ over all couplings, attained by the monotone coupling, which uses a single uniform variable $W$ and sets $U=\mathbf{1}_{\{W<a\}}$, $V=\mathbf{1}_{\{W<b\}}$.
\end{lemma}

\begin{proof}
Note that $a-b=\mathbb{P}[U=1,V=0]-\mathbb{P}[U=0,V=1]$, so $|a-b|\le\mathbb{P}[U\neq V]$, with equality for the monotone coupling.
\end{proof}

The next result is the structural identity behind all exact formulas for defaultable bond prices.

\begin{proposition}[Terminal representation]\label{prop:terminal-representation}
For $\mathbb{P},\mathbb{Q}\in\mathcal{M}_N(p)$,
\begin{equation}\label{eq:terminal-representation}
\mathcal{AW}_2^2(\mathbb{P},\mathbb{Q})
=\inf_{\pi\in\operatorname{Cpl}_{\mathrm{bc}}(\mathbb{P},\mathbb{Q})}\pi[X_N\neq Y_N]
=\inf_{\pi\in\operatorname{Cpl}_{\mathrm{bc}}(\mathbb{P},\mathbb{Q})}\mathbb{E}_\pi|X_{N-1}-Y_{N-1}|.
\end{equation}
Moreover, for every $\pi\in\operatorname{Cpl}_{\mathrm{bc}}(\mathbb{P},\mathbb{Q})$ the process $(|X_t-Y_t|)_{t\le N}$ is an $(\mathcal{H}_t)$-submartingale, so that $\mathcal{AW}_2^2(\mathbb{P},\mathbb{Q})\ge\inf_{\pi\in\operatorname{Cpl}_{\mathrm{bc}}(\mathbb{P},\mathbb{Q})}\mathbb{E}_\pi|X_t-Y_t|$ for every $t$.
\end{proposition}

Thus the squared adapted distance between the laws of two defaultable bond price processes is the minimal bicausal probability that the two default indicators disagree, and it equals the minimal bicausal $L^1$-distance of the last pre-maturity survival probabilities.

\begin{proof}
Fix $\pi\in\operatorname{Cpl}_{\mathrm{bc}}(\mathbb{P},\mathbb{Q})$. By Lemma~\ref{lem:martingale-under-bicausal}, $M=X-Y$ is a bounded $(\mathcal{H}_t)$-martingale with $M_0=0$, so its increments are orthogonal in $L^2(\pi)$ by Lemma~\ref{lem:martingale-isometry}. Hence $\mathbb{E}_\pi\sum_{t=1}^N|\Delta M_t|^2=\mathbb{E}_\pi|M_N|^2=\pi[X_N\neq Y_N]$, because $X_N,Y_N\in\{0,1\}$. This is the first identity, and $|M|$ is a submartingale by Jensen's inequality.

\medskip
For the second identity, Lemma~\ref{lem:martingale-under-bicausal} and the fact that $X_N,Y_N$ are binary show that, conditionally on $\mathcal{H}_{N-1}$, $X_N$ and $Y_N$ are Bernoulli with parameters $X_{N-1}$ and $Y_{N-1}$. Lemma~\ref{lem:bernoulli-mismatch} gives $\pi[X_N\neq Y_N\mid\mathcal{H}_{N-1}]\ge|X_{N-1}-Y_{N-1}|$. Conversely, let $\pi'$ agree with $\pi$ on $(X_{0:N-1},Y_{0:N-1})$ and couple $X_N,Y_N$ conditionally on $\mathcal{H}_{N-1}$ by the monotone Bernoulli coupling. The conditional law of $X_N$ given $(X_{0:N-1},Y_{0:N-1})$ under $\pi'$ is $\mathrm{Bernoulli}(X_{N-1})$, which is the $\mathbb{P}$-kernel and does not depend on $Y_{0:N-1}$. The same holds for $Y_N$. Hence $\pi'\in\operatorname{Cpl}_{\mathrm{bc}}(\mathbb{P},\mathbb{Q})$ and $\pi'[X_N\neq Y_N]=\mathbb{E}_\pi|X_{N-1}-Y_{N-1}|$.
\end{proof}

The first identity in \eqref{eq:terminal-representation} only uses that the two laws are martingales, and it is the reduction employed in \cite[Proposition~3.6]{BackhoffBartlBeiglbockEder2020Finance} for geometric Brownian motions. The specific feature of defaultable bond prices is that the terminal values are default indicators, so that the terminal cost is a probability of disagreement and can be reduced once more to the prices at the last pre-maturity date. This second reduction is what makes the information structures of the next section explicitly solvable.

\begin{remark}[Unconstrained couplings]\label{rem:unconstrained-terminal}
Without the bicausality constraint, the infimum of $\pi[X_N\neq Y_N]$ is zero by Lemma~\ref{lem:terminal-law}. All positive values of $\mathcal{AW}_2$ below are therefore produced by the information constraint and not by the terminal law. The continuous-time analogue of \eqref{eq:terminal-representation} for Cox bond prices is Theorem~\ref{thm:aw-cox}, which identifies $\mathcal{AW}_2^2$ with the closed-form default mismatch $\Psi$ of Section~\ref{sec:cox-hedging}.
\end{remark}

\section{Information revealed by ratings}\label{sec:ratings}

This section computes the distances for the information structures that appear in rating-based credit models: a single review, a review whose date is changed, and a rating chain observed over many periods. 
Finally, we show that the distances measure the size of the hedge triggered by a rating review, and we recall how they control prices and hedging errors.

\subsection{A single rating review}\label{sec:one-rating}

We specialize to three dates $0,1,2$, with maturity $2$ and common initial price $p\in(0,1)$.

\label{subsec:models}
\begin{definition}[Coarse credit model]\label{def:coarse}
The coarse model $\mathbb{P}^{0}\in\mathcal{M}_2(p)$ is the law of $X_0=X_1=p$, $X_2=D$, where $D\sim\mathrm{Bernoulli}(p)$.
\end{definition}

The coarse model contains no intermediate credit signal. Let $\mu$ be a probability measure on $[0,1]$ with
\begin{equation}\label{eq:mu-mean}
\int_0^1q\,\mu(\,\mathrm{d} q)=p.
\end{equation}

\begin{definition}[One-rating model]\label{def:rating-model}
The one-rating model $\mathbb{P}^{\mu}\in\mathcal{M}_2(p)$ is the law of $Y_0=p$, $Y_1=Q$, $Y_2=D^{\mu}$, where $Q\sim\mu$ and
\begin{equation}\label{eq:conditional-default}
\mathbb{P}[D^{\mu}=1\mid Q]=Q.
\end{equation}
\end{definition}

The variable $Q$ is the posterior survival probability implied by the rating. Condition \eqref{eq:mu-mean} ensures that the rating model has the same price today as the coarse model. The rating model can be realized explicitly on the unit square. Let $\Omega=[0,1]^2$ be equipped with Lebesgue measure $\mathbb{P}$, let $U_1,U_2$ be the coordinates, and let $F_\mu^{-1}$ denote a generalized inverse of the distribution function of $\mu$. Setting $Q=F_\mu^{-1}(U_1)$ and $D^\mu=\mathbf{1}_{\{U_2\le Q\}}$, we obtain $Q\sim\mu$ and $\mathbb{P}[D^\mu=1\mid Q]=Q$, so that the law $\mathbb{P}^\mu$ of $(p,Q,D^\mu)$ under $\mathbb{P}$ realizes Definition~\ref{def:rating-model} for any probability measure $\mu$ satisfying \eqref{eq:mu-mean}.

\begin{lemma}[Well-posedness]\label{lem:rating-wellposed}
The process of Definition~\ref{def:rating-model} is a martingale, $\mathbb{P}^{\mu}\in\mathcal{M}_2(p)$, and its terminal law is $\mathrm{Bernoulli}(p)$.
\end{lemma}

\begin{proof}
Note that by \eqref{eq:mu-mean} $\mathbb{E}[Y_1\mid Y_0]=\mathbb{E} Q=p$. Moreover, by \eqref{eq:conditional-default}
\[
\mathbb{E}[Y_2\mid Y_0,Y_1]=\mathbb{E}[D^\mu\mid Q]=Q=Y_1.
\]
The terminal law follows from Lemma~\ref{lem:terminal-law}.
\end{proof}

\begin{remark}[Every three-date model is a one-rating model]\label{rem:M2-exhausted}
By Lemma~\ref{lem:rating-wellposed}, every one-rating model belongs to $\mathcal{M}_2(p)$, and the converse also holds. Indeed, let $\mathbb{P}\in\mathcal{M}_2(p)$ and let $\mu$ be the law of $X_1$ under $\mathbb{P}$. The martingale property gives $\mathbb{E}_{\mathbb{P}}[X_1]=p$, so that $\mu$ satisfies \eqref{eq:mu-mean}, and, since $X_2$ takes values in $\{0,1\}$, it also gives $\mathbb{P}[X_2=1\mid X_0,X_1]=X_1$. Hence $\mathbb{P}=\mathbb{P}^\mu$. In particular, the formulas of this section describe the adapted distance between the coarse model and every three-date law of a defaultable bond price process with initial price $p$.
\end{remark}

\begin{remark}[Relative entropy between the two models]\label{rem:entropy-singular}
Relative entropy is the most common way of comparing two probabilistic models, and it underlies the pooling of model views in \cite{Meucci2008}, so it is natural to ask whether it could replace the adapted distance here. Between the terminal laws of the coarse and the rating model, which are both Bernoulli with parameter $p$, it vanishes. However, between their path laws it is infinite as soon as $\mu\neq\delta_p$, since $\mathbb{P}^\mu[X_1\neq p]>0=\mathbb{P}^0[X_1\neq p]$ and hence $\mathbb{P}^\mu$ is not absolutely continuous with respect to $\mathbb{P}^0$. This holds however small the revision is, whereas the adapted distance tends to zero with it. The same obstruction motivates the transport criteria of \cite{Vecer2026}.
\end{remark}

\begin{theorem}[Exact information-revelation formula]\label{thm:exact-aw-one-rating}
Let $p\in(0,1)$ and let $\mu$ satisfy \eqref{eq:mu-mean}. Then
\begin{equation}\label{eq:exact-aw-one-rating}
\mathcal{AW}_2^2\bigl(\mathbb{P}^0,\mathbb{P}^{\mu}\bigr)=\int_0^1|q-p|\,\mu(\,\mathrm{d} q),
\qquad\text{i.e.}\qquad
\mathcal{AW}_2\bigl(\mathbb{P}^0,\mathbb{P}^{\mu}\bigr)=\bigl(\mathbb{E}|Q-p|\bigr)^{1/2}.
\end{equation}
\end{theorem}

\begin{proof}
Both laws belong to $\mathcal{M}_2(p)$, by Definition~\ref{def:coarse} and Lemma~\ref{lem:rating-wellposed}, and the product coupling is bicausal, so that $\operatorname{Cpl}_{\mathrm{bc}}(\mathbb{P}^0,\mathbb{P}^\mu)\neq\emptyset$. Under every coupling $\pi$ we have $X_1=p$ $\pi$-almost surely and $Y_1\sim\mu$, hence $\mathbb{E}_\pi|X_1-Y_1|=\mathbb{E}|Q-p|$ does not depend on $\pi$. Proposition~\ref{prop:terminal-representation} with $N=2$ gives the claim.
\end{proof}

An optimal coupling samples $Q\sim\mu$ and $X_2\sim\mathrm{Bernoulli}(p)$ independently and then couples $Y_2\sim\mathrm{Bernoulli}(Q)$ monotonically with $X_2$. The independence of $Q$ and $X_2$ is not a choice: causality from $X$ to $Y$ at time $1$ forces it, because $(X_0,X_1)$ is deterministic. Given $Q=q$, the two defaults then disagree with probability $|q-p|$, and each disagreement costs a jump of size one.

\begin{corollary}[Information is invisible to static calibration]\label{cor:static-invisible}
The following are equivalent.
\begin{enumerate}
\item\label{it:aw-zero} $\mathcal{AW}_2(\mathbb{P}^0,\mathbb{P}^{\mu})=0$.
\item\label{it:dirac} $\mu=\delta_p$.
\item\label{it:deterministic} The intermediate rating posterior is deterministic.
\end{enumerate}
If $\mu\neq\delta_p$, the two models have the same initial bond price and the same terminal Bernoulli law, but a strictly positive adapted distance.
\end{corollary}

\begin{proof}
By Theorem~\ref{thm:exact-aw-one-rating}, condition~\ref{it:aw-zero} holds if and only if $q=p$ for $\mu$-almost every $q$, that is, if and only if condition~\ref{it:dirac} holds. Conditions~\ref{it:dirac} and~\ref{it:deterministic} are equivalent because $\mu$ has mean $p$, so a deterministic posterior must equal $p$. The last statement follows from Lemma~\ref{lem:terminal-law} and Lemma~\ref{lem:rating-wellposed}.
\end{proof}

\begin{theorem}[Ordinary Wasserstein distance]\label{thm:exact-w2-one-rating}
Under the assumptions of Theorem~\ref{thm:exact-aw-one-rating},
\begin{equation}\label{eq:exact-w2-one-rating}
\mathbb{E}|Q-p|^2\le\mathcal{W}_2^2\bigl(\mathbb{P}^0,\mathbb{P}^{\mu}\bigr)\le2\,\mathbb{E}|Q-p|^2,
\end{equation}
and the upper bound is an equality if $|Q-p|\le\frac12$ almost surely. With the cost on the path values of Remark~\ref{rem:level-convention}, the ordinary and adapted squared distances are exactly $\mathbb{E}|Q-p|^2$ and $\mathbb{E}|Q-p|^2+\mathbb{E}|Q-p|$.
\end{theorem}

\begin{proof}
For any coupling, the increments satisfy $\Delta X_1-\Delta Y_1=p-Q$ and $\Delta X_2-\Delta Y_2=(X_2-p)-(Y_2-Q)$, so the cost in \eqref{eq:aw2-def} is
\[
|Q-p|^2+|X_2-Y_2+Q-p|^2 .
\]

\emph{Lower bound.} The first term does not depend on the coupling, and the second is nonnegative, so $\mathcal{W}_2^2(\mathbb{P}^0,\mathbb{P}^\mu)\ge\mathbb{E}|Q-p|^2$.

\medskip
\emph{Upper bound.} Sample $Y\sim\mathbb{P}^\mu$ and set $X_0=X_1=p$ and $X_2=Y_2$. Since $Y_2\sim\mathrm{Bernoulli}(p)$ by Lemma~\ref{lem:rating-wellposed}, the law of $X$ is $\mathbb{P}^0$. For this coupling the second term equals $|Q-p|^2$, so $\mathcal{W}_2^2(\mathbb{P}^0,\mathbb{P}^\mu)\le2\,\mathbb{E}|Q-p|^2$.

\medskip
\emph{Equality.} Assume $|Q-p|\le\frac12$ almost surely, and consider any coupling. On the event $\{X_2=Y_2\}$ the second term equals $|Q-p|^2$. On the event $\{X_2\neq Y_2\}$ we have $X_2-Y_2=\pm1$, and the second term is at least $(1-|Q-p|)^2\ge|Q-p|^2$. Hence the total cost is at least $2|Q-p|^2$ for every coupling, and the upper bound is attained.

\medskip
\emph{Cost on the path values.} With the cost $\sum_{t=1}^2|X_t-Y_t|^2$ of Remark~\ref{rem:level-convention}, the two terms become $|X_1-Y_1|^2=|Q-p|^2$ and $|X_2-Y_2|^2=\mathbf{1}_{\{X_2\neq Y_2\}}$. The first term again does not depend on the coupling. The second vanishes for the coupling of the upper bound, which gives the ordinary squared distance $\mathbb{E}|Q-p|^2$. Over bicausal couplings its minimal expectation is $\mathbb{E}|Q-p|$, by Proposition~\ref{prop:terminal-representation} and Theorem~\ref{thm:exact-aw-one-rating}, which gives the adapted squared distance $\mathbb{E}|Q-p|^2+\mathbb{E}|Q-p|$.
\end{proof}

The anticipating coupling in the proof sets the two default indicators equal, which it can do only because it correlates the intermediate posterior with the future default of the coarse model. Bicausal couplings cannot use this information.

\begin{corollary}[Symmetric two-rating model]\label{cor:two-rating}
Let $0<\delta<\min\{p,1-p\}$ and $\mu_{\delta}=\tfrac12\delta_{p-\delta}+\tfrac12\delta_{p+\delta}$. Then
\begin{equation}\label{eq:two-rating-distances}
\mathcal{AW}_2\bigl(\mathbb{P}^0,\mathbb{P}^{\mu_\delta}\bigr)=\sqrt{\delta},
\qquad
\mathcal{W}_2\bigl(\mathbb{P}^0,\mathbb{P}^{\mu_\delta}\bigr)=\sqrt2\,\delta,
\qquad
\frac{\mathcal{AW}_2(\mathbb{P}^0,\mathbb{P}^{\mu_\delta})}{\mathcal{W}_2(\mathbb{P}^0,\mathbb{P}^{\mu_\delta})}=(2\delta)^{-1/2}.
\end{equation}
\end{corollary}

\begin{proof}
Since $\delta<\min\{p,1-p\}$,  $p\pm\delta$ belongs to $(0,1)$, so $\mu_\delta$ is a probability measure on $[0,1]$, and its mean is $1/2 (p-\delta)+1/2 (p+\delta)=p$. Hence $\mu_\delta$ satisfies \eqref{eq:mu-mean}. Under $\mu_\delta$ we have $|Q-p|=\delta$ almost surely, so that $\mathbb{E}|Q-p|=\delta$ and $\mathbb{E}|Q-p|^2=\delta^2$. Theorem~\ref{thm:exact-aw-one-rating} then gives $\mathcal{AW}_2^2(\mathbb{P}^0,\mathbb{P}^{\mu_\delta})=\delta$. Moreover, $\delta<\min\{p,1-p\}\le 1/2$, so the condition $|Q-p|\le 1/2$ of Theorem~\ref{thm:exact-w2-one-rating} holds, and the upper bound there is an equality, $\mathcal{W}_2^2(\mathbb{P}^0,\mathbb{P}^{\mu_\delta})=2\delta^2$. Taking square roots and the quotient gives \eqref{eq:two-rating-distances}.
\end{proof}

Close to the coarse model, the two distances therefore live on different scales: a revision of size $\delta$ is of first order for the ordinary distance and of square-root order for the adapted one. The squared adapted distance also has a direct pricing interpretation. The rating-date straddle with payoff $|X_1-X_0|$, struck at the forward survival probability, is worthless in the coarse model and is worth $\mathbb{E}|Q-p|=\mathcal{AW}_2^2(\mathbb{P}^0,\mathbb{P}^\mu)$ in the rating model. Its value is the expected size of the revision of the survival probability caused by the review, and is therefore a natural measure of how informative the review is.

\medskip
In practice $\mu$ is unknown and has to be estimated from observed posterior survival probabilities. Their empirical law need not have mean $p$, so we rescale it to obtain an estimated rating model that is still calibrated to the quoted price.

\begin{theorem}[Calibration-preserving estimation of the rating discrepancy]
\label{thm:posterior-estimation}
Let $Q_1,\ldots,Q_n$ be independent observations with common law
$\mu$ on $[0,1]$ and mean $p\in(0,1)$. Denote their empirical mean by
$\overline Q_n$ and their empirical law by $\mu_n$.
For $m\in[0,1]$ define the map
\[
T_{m,p}(q)=
\begin{cases}
\frac p m q,&m\ge p,\\
1-\dfrac{1-p}{1-m}(1-q),&m<p.
\end{cases}
\]
Write $\mu_n^{\mathrm{cal}}=(T_{\overline Q_n,p})_\#\mu_n$.
Then $\mu_n^{\mathrm{cal}}$ is supported on $[0,1]$ and has mean exactly
$p$ for every sample. Moreover,
\begin{equation}\label{eq:posterior-statistical}
\mathbb E\left|
\mathcal{AW}_2^2(\mathbb P^0,\mathbb P^{\mu_n^{\mathrm{cal}}})
-\mathcal{AW}_2^2(\mathbb P^0,\mathbb P^\mu)
\right|\le \frac1{\sqrt n}.
\end{equation}
The expectation on the left is with respect to the sample, and the
transport distances are computed between the corresponding pricing laws.
\end{theorem}
\begin{proof}
If $m\ge p>0$, the map $T_{m,p}$ sends $[0,1]$ into $[0,p/m]$, which is contained in $[0,1]$. If $m<p<1$, it sends $[0,1]$ into $[(p-m)/(1-m),1]$, again a subset of $[0,1]$. Its expectation under any law of mean $m$ equals $p$ by direct substitution in the corresponding branch. Thus the transformed empirical measure is exactly calibrated.

\medskip
By Theorem~\ref{thm:exact-aw-one-rating}, the absolute difference on the left of \eqref{eq:posterior-statistical} equals 
\[
\left|\int|q-p|\,\mu_n^{\mathrm{cal}}(\mathrm{d}q)-\int|q-p|\,\mu(\mathrm{d}q)\right|. 
\]
Since $q\mapsto|q-p|$ is $1$-Lipschitz, integrating $\bigl||x-p|-|y-p|\bigr|\le|x-y|$ against any coupling of $\mu_n^{\mathrm{cal}}$ and $\mu$ shows that this difference is bounded by the Wasserstein distance of order one on the unit interval,
\begin{equation}\label{eq:W1-def}
W_1(\mu_n^{\mathrm{cal}},\mu)=\inf_{\gamma\in\operatorname{Cpl}(\mu_n^{\mathrm{cal}},\mu)}\int|x-y|\,\gamma(\mathrm{d}x,\mathrm{d}y).
\end{equation}
%
The triangle inequality yields
\[
W_1(\mu_n^{\mathrm{cal}},\mu)\le W_1(\mu_n^{\mathrm{cal}},\mu_n)+W_1(\mu_n,\mu).
\]

\medskip
For the first term, couple $q$ with $T_{m,p}(q)$ under the empirical measure. If $m\ge p$, then $T_{m,p}(q)\le q$ and
\[
\int|T_{m,p}(q)-q|\,\mu_n(\mathrm{d}q)=\frac{m-p}{m}\int q\,\mu_n(\mathrm{d}q)=m-p.
\]
If $m<p$, the transform is greater than $q$ and the same computation gives $p-m$. Hence $W_1(\mu_n^{\mathrm{cal}},\mu_n)\le|\overline Q_n-p|\le W_1(\mu_n,\mu)$, where the last inequality follows by taking $f(q)=q$, since $\mu$ has mean $p$. Therefore $W_1(\mu_n^{\mathrm{cal}},\mu)\le2W_1(\mu_n,\mu)$.

\medskip
Finally, let $F_n$ and $F$ be the empirical and population distribution functions. Recall that on the real line, by Dall'Aglio's formula \cite{DallAglio1956}, 
\[
W_1(\mu_n,\mu)=\int_0^1|F_n(x)-F(x)|\,\mathrm{d}x.
\] 
For fixed $x$, the random variable $nF_n(x)$ has binomial distribution with success probability $F(x)$, so
\[
\mathbb{E}|F_n(x)-F(x)|\le\bigl[\mathbb{E}|F_n(x)-F(x)|^2\bigr]^{1/2}=\sqrt{\frac{F(x)(1-F(x))}{n}}\le\frac1{2\sqrt n}.
\]
Tonelli's theorem yields $\mathbb{E}\,W_1(\mu_n,\mu)\le1/(2\sqrt n)$, and combining the preceding estimates proves \eqref{eq:posterior-statistical}.
\end{proof}
%

The result controls the estimation error of the squared adapted discrepancy without estimating a bicausal optimizer. Extending it to estimated migration matrices requires control of the conditional transition laws at successive dates, and cannot be obtained by treating all rating observations as independent.

\subsection{The timing of information}\label{sec:timing}

We now keep the rating distribution fixed and alter only the date at which the signal becomes available. This isolates the timing component of credit information.

\medskip
Fix four dates $0,1,2,3$ and a probability measure $\mu$ on $[0,1]$ with mean $p$.

\begin{definition}[Early and late revelation models]\label{def:early-late}
Let $(\Omega,\mathcal{F},\mathbb{P})$ be a probability space supporting random variables $Q,\widetilde Q$ with law $\mu$ and $D,\widetilde D\in\{0,1\}$ such that $\mathbb{P}[D=1\mid Q]=Q$ and $\mathbb{P}[\widetilde D=1\mid\widetilde Q]=\widetilde Q$. We define:
\begin{itemize}
\item the \emph{early revelation model} $\mathbb{P}^{\mathrm E}\in\mathcal{M}_3(p)$ as the law of $(X_0,X_1,X_2,X_3)=(p,\,Q,\,Q,\,D)$,
\item the \emph{late revelation model} $\mathbb{P}^{\mathrm L}\in\mathcal{M}_3(p)$ as the law of $(Y_0,Y_1,Y_2,Y_3)=(p,\,p,\,\widetilde Q,\,\widetilde D)$.
\end{itemize}
\end{definition}

Both processes are martingales. The early model reveals the posterior one period before the late model, and from time $2$ onward the two have the same mechanism. Terminal survival calibration cannot distinguish the date at which the posterior was learned.

\begin{lemma}[Independence forced by bicausality]\label{lem:forced-independence}
If $\pi\in\operatorname{Cpl}_{\mathrm{bc}}(\mathbb{P}^{\mathrm E},\mathbb{P}^{\mathrm L})$, then $X_1$ is independent of the whole path $Y$ under $\pi$. In particular $(X_1,Y_2)\sim\mu\otimes\mu$.
\end{lemma}

\begin{proof}
Causality from $Y$ to $X$ at time $1$ implies that the conditional law of $X_1$ given the complete $Y$ path depends only on $(Y_0,Y_1)$, which is deterministic. Hence this conditional law equals $\mu$.
\end{proof}

\begin{theorem}[Exact information-date formula]\label{thm:early-late-aw}
Let $Q$ and $\widetilde Q$ be independent with common law $\mu$. Then
\begin{equation}\label{eq:early-late-aw}
\mathcal{AW}_2^2\bigl(\mathbb{P}^{\mathrm E},\mathbb{P}^{\mathrm L}\bigr)=\mathbb{E}|Q-\widetilde Q|.
\end{equation}
\end{theorem}

\begin{proof}
Both laws belong to $\mathcal{M}_3(p)$, so Proposition~\ref{prop:terminal-representation} with $N=3$ gives
\[
\mathcal{AW}_2^2\bigl(\mathbb{P}^{\mathrm E},\mathbb{P}^{\mathrm L}\bigr)=\inf_{\pi\in\operatorname{Cpl}_{\mathrm{bc}}(\mathbb{P}^{\mathrm E},\mathbb{P}^{\mathrm L})}\mathbb{E}_\pi|X_2-Y_2|.
\]
Let $\pi$ be any bicausal coupling. In the early model $X_2=X_1$, and by Lemma~\ref{lem:forced-independence} $X_1$ is independent of $Y_2$ under $\pi$, with both variables distributed according to $\mu$. Hence $\mathbb{E}_\pi|X_2-Y_2|=\mathbb{E}|Q-\widetilde Q|$ for every bicausal $\pi$. Since the product coupling is bicausal, the infimum is taken over a nonempty set, and \eqref{eq:early-late-aw} follows.
\end{proof}

By Theorem~\ref{thm:exact-aw-one-rating}, $\mathcal{AW}_2^2(\mathbb{P}^0,\mathbb{P}^\mu)=\mathbb{E}|Q-p|$, while $\mathbb{E}|Q-p|\le\mathbb{E}|Q-\widetilde Q|\le2\,\mathbb{E}|Q-p|$ by Jensen's inequality and the triangle inequality. Delaying information by one period thus costs, in squared distance, between one and two times as much as suppressing it altogether.

\medskip
A related benchmark is the model that reveals nothing at all before maturity.

\begin{definition}[Completely coarse law]\label{def:coarse-N}
For $N\ge2$, let $\mathbb{P}^{0,N}\in\mathcal{M}_N(p)$ be the law of $X_t=p$ for $0\le t<N$ and $X_N=D\sim\mathrm{Bernoulli}(p)$.
\end{definition}

\begin{theorem}[Distance from the completely coarse benchmark]\label{thm:multidate-coarse}
For every $\mathbb{Q}\in\mathcal{M}_N(p)$, with canonical process $Y$,
\begin{equation}\label{eq:multidate-coarse}
\mathcal{AW}_2^2\bigl(\mathbb{P}^{0,N},\mathbb{Q}\bigr)=\mathbb{E}_{\mathbb{Q}}|Y_{N-1}-p|.
\end{equation}
\end{theorem}

\begin{proof}
Both laws belong to $\mathcal{M}_N(p)$, so Proposition~\ref{prop:terminal-representation} gives
\[
\mathcal{AW}_2^2\bigl(\mathbb{P}^{0,N},\mathbb{Q}\bigr)=\inf_{\pi\in\operatorname{Cpl}_{\mathrm{bc}}(\mathbb{P}^{0,N},\mathbb{Q})}\mathbb{E}_\pi|X_{N-1}-Y_{N-1}|.
\]
Under every coupling $\pi$ we have $X_{N-1}=p$ $\pi$-almost surely, because the completely coarse model reveals nothing before maturity, while $Y_{N-1}$ has the same law as under $\mathbb{Q}$. Hence $\mathbb{E}_\pi|X_{N-1}-Y_{N-1}|=\mathbb{E}_{\mathbb{Q}}|Y_{N-1}-p|$ does not depend on $\pi$. Since the product coupling is bicausal, the infimum is taken over a nonempty set, and \eqref{eq:multidate-coarse} follows.
\end{proof}

Against the model that reveals nothing before maturity, the adapted distance therefore depends only on the law of the last pre-maturity price, and not on how this price was reached. Between two informative models, the timing of revelation also matters, as Theorem~\ref{thm:early-late-aw} shows.

\subsection{Multi-period rating migration}\label{sec:migration}

The formulas of Sections~\ref{sec:one-rating} and~\ref{sec:timing} involve at most one informative date. We now move to the more general case of defaultable bond prices driven by finite-state rating chains over many periods, where both models may reveal information at every date. The terminal representation reduces the adapted distance to a backward recursion over pairs of current ratings, and under the standard monotonicity of rating migration the recursion is solved by driving both chains with the same uniform innovations.

\medskip
We consider bond price processes that take finitely many values at each date and evolve as Markov chains. Let $E_0=\{p\}$, let $E_1,\ldots,E_{N-1}$ be finite subsets of $[0,1]$, and let $E_N=\{0,1\}$. For $t<N$, a transition kernel $\kappa_t$ from $E_t$ to $E_{t+1}$ assigns to every $x\in E_t$ a probability measure $\kappa_t(x,\cdot)$ on $E_{t+1}$. We say that $\mathbb{P}\in\mathcal{M}_N(p)$ is \emph{Markov with kernels} $\kappa=(\kappa_t)_{t<N}$ if $X_t\in E_t$ $\mathbb{P}$-almost surely for every $t\le N$ and
\[
\mathbb{P}[X_{t+1}=y\mid X_0,\ldots,X_t]=\kappa_t(X_t,y),\qquad y\in E_{t+1},\ t<N.
\]
We impose the martingale condition $\sum_{y\in E_{t+1}}y\,\kappa_t(x,y)=x$ for every $x\in E_t$. At the last step it determines the kernel completely: a law on $\{0,1\}$ with mean $x$ is the Bernoulli law with parameter $x$, so that $\kappa_{N-1}(x,\cdot)=\mathrm{Bernoulli}(x)$.

\medskip
Rating models provide a natural class of such laws. Let $\{0,1,\ldots,m\}$ be the set of ratings, ordered by credit quality, where $0$ denotes the absorbing default state. Let $R$ be a time-inhomogeneous Markov chain on this set, started at $r_0$, with risk-neutral transition matrices $(P_t)_{t<N}$. The probability of surviving until $N$ given rating $i$ at date $t$ is
\[
q_t(i)=\mathbb{P}[R_N\neq0\mid R_t=i],
\]
and the price process of the defaultable zero-coupon bond maturing at $N$ is
\[
X_t=q_t(R_t)\quad(t<N),\qquad X_N=\mathbf{1}_{\{R_N\neq0\}},
\]
with initial price $X_0=q_0(r_0)$. We assume that, for $1\le t\le N-2$, $q_t$ is injective on the ratings with $0<q_t(i)<1$. Thus an interior price identifies the current rating. If the price equals $0$ or $1$, the martingale property makes all subsequent prices equal to that value. No injectivity is needed at $t=N-1$, since the terminal conditional law is Bernoulli with parameter $X_{N-1}$. Consequently $X$ is Markov in the above sense, with $E_t=q_t(\{0,\ldots,m\})$ for $0<t<N$, kernels
\[
\kappa_t\bigl(q_t(i),z\bigr)
=\sum_{j:\,q_{t+1}(j)=z}P_t(i,j),\qquad z\in E_{t+1},\quad t<N-1,
\]
and last kernel $\kappa_{N-1}(q_{N-1}(i),\cdot)=\mathrm{Bernoulli}(q_{N-1}(i))$. At $t=0$ the formula uses only $i=r_0$. At a price $0$ or $1$, the aggregated transition law is the corresponding point mass, so it does not depend on which rating represents that price. Two rating models, possibly with different rating sets and transition matrices, are called \emph{calibrated} if they have the same initial price.

\medskip
For Markov models, the infimum in Proposition~\ref{prop:terminal-representation} can be computed by a backward recursion over pairs of current states.

\begin{theorem}[Dynamic programming principle]\label{thm:dpp}
Let $\mathbb{P},\mathbb{Q}\in\mathcal{M}_N(p)$ be Markov with kernels $\kappa,\lambda$ on state spaces $(E_t),(F_t)$. Define $V_{N-1}(x,y)=|x-y|$ and, for $t\le N-2$,
\begin{equation}\label{eq:dpp}
V_t(x,y)=\min_{\gamma\in\operatorname{Cpl}(\kappa_t(x,\cdot),\lambda_t(y,\cdot))}\ \sum_{u\in E_{t+1},\,v\in F_{t+1}}\gamma(u,v)\,V_{t+1}(u,v),
\qquad(x,y)\in E_t\times F_t .
\end{equation}
Then $\mathcal{AW}_2^2(\mathbb{P},\mathbb{Q})=V_0(p,p)$. The Markov coupling that uses an optimal $\gamma_t(x,y)$ in \eqref{eq:dpp} at each $t\le N-2$ and the monotone Bernoulli coupling at $t=N-1$ is optimal, and $V_t(x,y)\ge|x-y|$ for all $t$.
\end{theorem}

Theorem~\ref{thm:dpp} is the dynamic programming principle for bicausal transport of \cite[Proposition~5.2]{BackhoffBeiglbockLinZalashko2017}, specialised in two ways. First, by Proposition~\ref{prop:terminal-representation}, the cost depends only on the values at date $N-1$, so that the recursion starts there. Second, for Markov laws, the value functions of \cite{BackhoffBeiglbockLinZalashko2017}, which in general depend on the whole past, depend only on the current pair of states.

\begin{proof}
By Proposition~\ref{prop:terminal-representation}, $\mathcal{AW}_2^2(\mathbb{P},\mathbb{Q})$ is the value of the bicausal transport problem between the laws of $(X_0,\ldots,X_{N-1})$ and $(Y_0,\ldots,Y_{N-1})$ with cost $|x_{N-1}-y_{N-1}|$. Indeed, the restriction of a bicausal coupling to the first $N$ dates is bicausal. Conversely, every bicausal coupling of the restricted laws extends to a bicausal coupling of $\mathbb{P}$ and $\mathbb{Q}$ by the monotone Bernoulli coupling at the last step, as in the proof of Proposition~\ref{prop:terminal-representation}. By \cite[Proposition~5.2]{BackhoffBeiglbockLinZalashko2017}, this value is obtained by a backward recursion whose value functions depend on the histories $(x_{0:t},y_{0:t})$, and whose step at date $t$ is a transport problem between the conditional laws of $X_{t+1}$ and $Y_{t+1}$ given these histories. For Markov laws these conditional laws are $\kappa_t(x_t,\cdot)$ and $\lambda_t(y_t,\cdot)$, and the terminal cost depends only on $(x_{N-1},y_{N-1})$. Backward induction therefore shows that each value function depends only on $(x_t,y_t)$ and satisfies \eqref{eq:dpp}, and the minima are attained because the state spaces are finite. Composing optimal kernels gives a coupling of the sequential form of Proposition~\ref{prop:sequential-bicausal}, hence a bicausal coupling with cost $V_0(p,p)$. Finally, applying the submartingale property of Proposition~\ref{prop:terminal-representation} to the problem started at $(x,y)$ at date $t$ gives $V_t(x,y)\ge|x-y|$.
\end{proof}

Each step of \eqref{eq:dpp} is a small transportation problem, so that the adapted distance between two rating models with $m$ and $m'$ states is computed exactly by solving $O(N\,m\,m')$ linear programs of size $m\times m'$. Moreover, the rating labels enter only through the bond prices. Relabelling the states, or splitting a state into substates with the same conditional survival probabilities and transitions, therefore leaves $V_0$ unchanged.

\medskip
A kernel $\kappa_t$ is \emph{stochastically monotone} if $x\le x'$ implies $\kappa_t(x,\cdot)\preceq_{\mathrm{st}}\kappa_t(x',\cdot)$. For rating chains, the natural condition is that each $P_t(i,\cdot)$ is stochastically nondecreasing in $i$ for the credit-quality order, which means that a better current rating makes it more likely to end up at or above any given rating. Such monotone migration matrices are standard in the rating literature \cite{JarrowLandoTurnbull1997,Lando2004,KalkbrenerPackham2026}. They preserve nondecreasing functions \cite{KeilsonKester1977}, and in particular each $q_t$ is nondecreasing.

\begin{lemma}[From ratings to prices]\label{lem:rating-monotone}
If the matrices $P_t$ are stochastically monotone and each $q_t$, $1\le t\le N-2$, is injective on the ratings with prices in $(0,1)$, then the price kernels $\kappa_t$ are stochastically monotone.
\end{lemma}

\begin{proof}
At $t=0$ there is only one current price, and at $t=N-1$ the kernels are Bernoulli and hence stochastically monotone. For $1\le t\le N-2$, prices $0$ and $1$ have the absorbing transition laws $\delta_0$ and $\delta_1$, so comparisons involving either endpoint are immediate. Let $0<x\le x'<1$ be two prices at date $t$. By monotonicity of $q_t$ and the injectivity assumption, $x=q_t(i)$ and $x'=q_t(i')$ for ratings $i\le i'$. The law $\kappa_t(q_t(i),\cdot)$ of the next price is the image of $P_t(i,\cdot)$ under the map $j\mapsto q_{t+1}(j)$ if $t<N-1$, and under $j\mapsto\mathbf{1}_{\{j\neq0\}}$ if $t=N-1$. Both maps are nondecreasing. Since $P_t(i,\cdot)\preceq_{\mathrm{st}}P_t(i',\cdot)$ by assumption, and the image of two stochastically ordered laws under a nondecreasing map is again stochastically ordered, we obtain $\kappa_t(x,\cdot)\preceq_{\mathrm{st}}\kappa_t(x',\cdot)$.
\end{proof}

For a stochastically monotone kernel let $G_{t,x}$ denote the quantile function of $\kappa_t(x,\cdot)$, and $H_{t,y}$ that of $\lambda_t(y,\cdot)$. Let $U_1,\ldots,U_N$ be independent uniform variables. The \emph{synchronous coupling}\footnote{Sometimes referred as the Knothe-Rosenblatt coupling.} is the law of
\begin{equation}\label{eq:synchronous}
X_0=Y_0=p,\qquad X_{t+1}=G_{t,X_t}(U_{t+1}),\qquad Y_{t+1}=H_{t,Y_t}(U_{t+1}).
\end{equation}
In rating terms, both chains are driven by the same uniform innovation through their inverse cumulative transition probabilities, taken in the credit-quality order.

\begin{theorem}[Synchronous coupling]\label{thm:synchronous}
Let $\mathbb{P},\mathbb{Q}\in\mathcal{M}_N(p)$ be Markov with stochastically monotone kernels. Then the synchronous coupling $\pi^{\mathrm s}$ is bicausal and optimal, and
\begin{equation}\label{eq:synchronous-value}
\mathcal{AW}_2^2(\mathbb{P},\mathbb{Q})=\pi^{\mathrm s}[X_N\neq Y_N]=\mathbb{E}_{\pi^{\mathrm s}}|X_{N-1}-Y_{N-1}| .
\end{equation}
In particular, by Lemma~\ref{lem:rating-monotone}, for two calibrated rating models with stochastically monotone migration matrices, $\mathcal{AW}_2^2$ is the probability that the two synchronously driven chains disagree on default at maturity.
\end{theorem}

A function $C$ on a product of two finite subsets of $\mathbb{R}$ is \emph{submodular} if $C(x,y)+C(x',y')\le C(x,y')+C(x',y)$ whenever $x\le x'$ and $y\le y'$. For submodular costs, the comonotone coupling is optimal between any two probability measures on finite subsets of $\mathbb{R}$ \cite{Lorentz1953,CambanisSimonsStout1976,BurkardKlinzRudolf1996}.

\begin{proof}
The synchronous coupling has the sequential form of Proposition~\ref{prop:sequential-bicausal}, since $G_{t,x}(U)$ has law $\kappa_t(x,\cdot)$, and it is therefore bicausal. By Theorem~\ref{thm:dpp}, it suffices to show that the comonotone coupling attains the minimum in \eqref{eq:dpp} at every date, and by the result just recalled it suffices to show that every $V_t$ is submodular. We argue by backward induction. The function $V_{N-1}(x,y)=|x-y|$ is submodular. If $V_{t+1}$ is submodular, the comonotone coupling is optimal in \eqref{eq:dpp}, so that
\[
V_t(x,y)=\int_0^1V_{t+1}\bigl(G_{t,x}(u),H_{t,y}(u)\bigr)\,\mathrm{d}u .
\]
By stochastic monotonicity, the maps $x\mapsto G_{t,x}(u)$ and $y\mapsto H_{t,y}(u)$ are nondecreasing for every $u$. Composing a submodular function with nondecreasing maps in each variable preserves submodularity, and so does integration in $u$, hence $V_t$ is submodular. At $t=N-1$ we have $G_{N-1,x}(u)=\mathbf{1}_{\{u>1-x\}}$, so the synchronous step is the monotone Bernoulli coupling, and Theorem~\ref{thm:dpp} concludes.
\end{proof}

Theorem~\ref{thm:synchronous} is an instance of the optimality of the synchronous, or Knothe-Rosenblatt, coupling between stochastically monotone Markov processes, which is due to R\"uschendorf \cite{Ruschendorf1985} and is proved for bicausal transport in \cite{BackhoffBeiglbockLinZalashko2017}, applied to the terminal cost of Proposition~\ref{prop:terminal-representation}. Without monotonicity the synchronous coupling need not be optimal \cite[Example~3.5]{BackhoffBartlBeiglbockEder2020Finance}, and the dynamic programme \eqref{eq:dpp} has to be solved by linear programming, as we do in Section~\ref{sec:examples}.

\subsection{Hedging interpretation and stability}\label{sec:hedging}

To give the distances a hedging interpretation, we consider again the three-date models of Section~\ref{sec:one-rating}. The discounted defaultable zero-coupon bond is traded together with the numeraire. A self-financing strategy consists of an initial capital $m\in\mathbb{R}$, a holding $H_1\in\mathbb{R}$ in the bond between dates $0$ and $1$, and a holding $H_2=h(S_0,S_1)$ between dates $1$ and $2$, where $h$ is a Borel function. Its terminal wealth is
\begin{equation}\label{eq:wealth-3date}
V_2^{m,H}=m+H_1(S_1-S_0)+H_2(S_2-S_1).
\end{equation}

As a test claim, we consider the \emph{momentum payoff}
\begin{equation}\label{eq:cross-claim}
\Gamma(S)=(S_1-S_0)(S_2-S_1),
\end{equation}
the product of the price revision caused by the review and the subsequent revision at maturity. It is the profit and loss of a desk that, after the review, holds the bond in proportion to the price revision, buying after an upgrade and selling after a downgrade. Equivalently, $\Gamma(S)$ is one half of the cross term in $(S_2-S_0)^2=(S_1-S_0)^2+2\Gamma(S)+(S_2-S_1)^2$, that is, the realized autocovariance of the two price increments. It is the simplest claim whose replication requires a position that depends on the outcome of the review.

\begin{theorem}[Exact dynamic replication]\label{thm:dynamic-replication}
Under every law in $\mathcal{M}_2(p)$, the momentum payoff $\Gamma$ is replicated pathwise by the strategy
\begin{equation}\label{eq:canonical-hedge}
m=0,\qquad H_1=0,\qquad H_2=S_1-S_0.
\end{equation}
Its holding after the review vanishes under $\mathbb{P}^0$ and equals $Q-p$ under $\mathbb{P}^{\mu}$. Moreover, every strategy that replicates $\Gamma$ under $\mathbb{P}^\mu$ holds $q-p$ units of the bond after a review with outcome $q$, for $\mu$-almost every $q\in(0,1)$. If $\mu$ charges at least two points of $(0,1)$ and $\mu(\{0,1\})=0$, then \eqref{eq:canonical-hedge} is the unique replicating strategy, up to almost-sure equality.
\end{theorem}

\begin{proof}
\emph{Replication.} Inserting \eqref{eq:canonical-hedge} into \eqref{eq:wealth-3date} gives $V_2=(S_1-S_0)(S_2-S_1)=\Gamma(S)$ on every path, whatever the law of $S$. Under $\mathbb{P}^0$, the price does not move at the review, $S_1=S_0=p$, so the holding $H_2$ vanishes. Under $\mathbb{P}^\mu$, the review moves the price to $S_1=Q$, so $H_2=Q-p$.

\medskip
\emph{Uniqueness.} Let a strategy replicate $\Gamma$ under $\mathbb{P}^\mu$, with initial capital $m$, constant holding $K_1$ before the review, and holding $k(q)$ after a review with outcome $S_1=q$. Fix an outcome $q\in(0,1)$. Given $S_1=q$, the bond survives, $S_2=1$, with probability $q$ and defaults, $S_2=0$, with probability $1-q$. Both probabilities are positive, so replication must hold in both cases:
\begin{equation}\label{eq:replication-at-q}
m+K_1(q-p)+k(q)(d-q)=(q-p)(d-q),\qquad d\in\{0,1\}.
\end{equation}
Both sides are affine in $d$. Comparing the coefficients of $d$ gives $k(q)=q-p$, so the holding after the review is determined by its outcome. Comparing the constant terms then gives $m+K_1(q-p)=0$. This holds for $\mu$-almost every $q\in(0,1)$. If $\mu$ charges two distinct points of $(0,1)$, the affine function $q\mapsto m+K_1(q-p)$ vanishes at two points, hence everywhere, and $m=K_1=0$. If moreover $\mu(\{0,1\})=0$, the already established identity $k(q)=q-p$ holds $\mu$-almost surely. At $q\in\{0,1\}$, by contrast, $S_2-S_1=0$, so the holding after the review is immaterial.
\end{proof}

\begin{corollary}[Transport cost equals hedge size]\label{cor:hedge-turnover}
Under the rating model,
\begin{equation}\label{eq:turnover-aw}
\mathbb{E}_{\mathbb{P}^{\mu}}|H_2|=\mathcal{AW}_2^2\bigl(\mathbb{P}^0,\mathbb{P}^{\mu}\bigr),
\qquad
\bigl(\mathbb{E}_{\mathbb{P}^{\mu}}|H_2|^2\bigr)^{1/2}=\bigl(\mathbb{E}|Q-p|^2\bigr)^{1/2},
\end{equation}
and the second quantity equals $\mathcal{W}_2(\mathbb{P}^0,\mathbb{P}^\mu)/\sqrt2$ if $|Q-p|\le 1/2$. Since $H_1=0$, $|H_2|$ is also the trading turnover of the hedge at the rating date.
\end{corollary}

\begin{proof}
Combine $H_2=Q-p$ with Theorems~\ref{thm:exact-aw-one-rating} and~\ref{thm:exact-w2-one-rating}.
\end{proof}

The adapted distance is tied to the absolute size of the position created by the information event, the ordinary distance to its quadratic size.

\medskip
The remaining question is how the distance controls prices and hedges in general. Here we rely on \cite{BackhoffBartlBeiglbockEder2020Finance} and only record the two bounds we use later, with explicit constants for defaultable bond prices.

\medskip
A functional $G:\Omega_N\to\mathbb{R}$ is \emph{$L_G$-Lipschitz in increment norm} if
\begin{equation}\label{eq:increment-lip-claim}
|G(x)-G(y)|\le L_G\Bigl(\sum_{t=1}^N|\Delta x_t-\Delta y_t|^2\Bigr)^{1/2}
\end{equation}
for all paths with the same initial value.

\begin{remark}[Prices]\label{rem:price-stability}
If $G$ satisfies \eqref{eq:increment-lip-claim}, then $|\mathbb{E}_{\mathbb{P}}G(X)-\mathbb{E}_{\mathbb{Q}}G(Y)|\le\mathbb{E}_\pi|G(X)-G(Y)|\le L_G(\mathbb{E}_\pi\sum_t|\Delta X_t-\Delta Y_t|^2)^{1/2}$ for every coupling $\pi$, so that
\[
|\mathbb{E}_{\mathbb{P}}G(X)-\mathbb{E}_{\mathbb{Q}}G(Y)|\le L_G\,\mathcal{W}_2(\mathbb{P},\mathbb{Q})\le L_G\,\mathcal{AW}_2(\mathbb{P},\mathbb{Q}).
\]
The price of a fixed path-dependent claim is thus controlled already by the ordinary distance. Adaptedness becomes essential only for quantities that involve trading strategies, such as hedging errors, superhedging prices and utility maximization \cite{BackhoffBartlBeiglbockEder2020Finance}.
\end{remark}

\label{subsec:hedge-stability}For $t\ge1$ let $h_t:[0,1]^t\to\mathbb{R}$ be Borel and $H_t(x)=h_t(x_0,\ldots,x_{t-1})$. Assume that there are constants $K,L_H\ge0$ with
\begin{equation}\label{eq:strategy-bounds}
|H_t(x)|\le K,
\qquad
|H_t(x)-H_t(y)|\le L_H\Bigl(\sum_{s=0}^{t-1}|x_s-y_s|^2\Bigr)^{1/2}.
\end{equation}
The gains process is $(H\cdot X)_N=\sum_{t=1}^NH_t(X)\Delta X_t$.

\medskip
For a claim $G$ and initial capital $m$ let $\mathfrak R_{\mathbb{P}}(m,H;G)=\|G(X)-m-(H\cdot X)_N\|_{L^2(\mathbb{P})}$ be the quadratic hedging error under $\mathbb{P}$.

\begin{proposition}[Stability of a common hedge, after {\cite{BackhoffBartlBeiglbockEder2020Finance}}]\label{prop:hedge-stability}
Let $\mathbb{P},\mathbb{Q}\in\mathcal{M}_N(p)$, let $H$ satisfy \eqref{eq:strategy-bounds}, let $G$ satisfy \eqref{eq:increment-lip-claim}, and set $c_N=\sqrt{N(N-1)/2}$. Then
\begin{equation}\label{eq:risk-difference}
\bigl|\mathfrak R_{\mathbb{P}}(m,H;G)-\mathfrak R_{\mathbb{Q}}(m,H;G)\bigr|\le\bigl(L_G+K+c_NL_H\bigr)\mathcal{AW}_2(\mathbb{P},\mathbb{Q}).
\end{equation}
\end{proposition}

The proof follows the argument of \cite[Lemma~3.9]{BackhoffBartlBeiglbockEder2020Finance}, which underlies the strong hedging inequality of \cite[Theorem~1.5]{BackhoffBartlBeiglbockEder2020Finance}. The difference of the gains is split into a stochastic integral against $X-Y$ and a stochastic integral of $H(X)-H(Y)$ against $Y$, and both are controlled through the quadratic variation of $X-Y$. We repeat it because Section~\ref{sec:examples} uses the explicit constant, which in our setting follows from the bound $|\Delta Y_t|\le1$ on the increments of defaultable bond prices.

\begin{proof}
Fix $\pi\in\operatorname{Cpl}_{\mathrm{bc}}(\mathbb{P},\mathbb{Q})$ and set $M=X-Y$. By Lemma~\ref{lem:martingale-under-bicausal}, $X$, $Y$, $M$ are bounded $(\mathcal{H}_t)$-martingales. Write
\[
(H\cdot X)_N-(H\cdot Y)_N=\underbrace{\sum_{t=1}^NH_t(X)\Delta M_t}_{=:A}+\underbrace{\sum_{t=1}^N[H_t(X)-H_t(Y)]\Delta Y_t}_{=:B}.
\]
The integrands are $(\mathcal{H}_t)$-predictable, so orthogonality of martingale increments gives
\begin{gather*}
\mathbb{E}_\pi A^2=\sum_t\mathbb{E}_\pi\bigl[H_t(X)^2|\Delta M_t|^2\bigr]\le K^2\,\mathbb{E}_\pi\sum_t|\Delta M_t|^2,\\
\mathbb{E}_\pi B^2=\sum_t\mathbb{E}_\pi\bigl[|H_t(X)-H_t(Y)|^2|\Delta Y_t|^2\bigr].
\end{gather*}
Since $|\Delta Y_t|\le1$, \eqref{eq:strategy-bounds} and the martingale isometry $\mathbb{E}_\pi|M_s|^2=\mathbb{E}_\pi\sum_{r\le s}|\Delta M_r|^2$ give
\[
\mathbb{E}_\pi B^2\le L_H^2\sum_{t=1}^N\sum_{s=1}^{t-1}\mathbb{E}_\pi|M_s|^2\le\frac{N(N-1)}2\,L_H^2\,\mathbb{E}_\pi\sum_{r=1}^N|\Delta M_r|^2 .
\] Adding the pathwise bound $\|G(X)-G(Y)\|_{L^2(\pi)}\le L_G(\mathbb{E}_\pi\sum_t|\Delta M_t|^2)^{1/2}$, the triangle inequality in $L^2(\pi)$ bounds the $L^2(\pi)$-norm of the difference of the two residuals by $(L_G+K+c_NL_H)(\mathbb{E}_\pi\sum_t|\Delta M_t|^2)^{1/2}$. Since the residuals have $L^2(\pi)$-norms $\mathfrak R_{\mathbb{P}}$ and $\mathfrak R_{\mathbb{Q}}$, the reverse triangle inequality and the infimum over $\pi$ give \eqref{eq:risk-difference}.
\end{proof}

In Section~\ref{subsec:empirical-ratings} we measure on agency data how far these generic bounds are from the realised price differences and hedging errors.

\section{Intensity models and model-transfer hedging}\label{sec:cox-hedging}

We now pass to an intensity representation. We first compute the adapted distance between the price processes of two Cox bonds with deterministic hazards, and we then study a different hedging question. A strategy is selected because it replicates under a calibrated reference model. The same strategy is then marked and implemented under another calibrated model, interpreted as the market law. The resulting cash residual is a model-transfer hedging error. As before, adapted transport costs are used as discrepancies between filtered pricing models and not as intrinsic amounts of information.

\subsection{The adapted distance between Cox bond prices}\label{subsec:cox-mismatch}

Let $a,b\in L^1([0,T];[0,\infty))$ be deterministic hazard paths and define
\begin{equation}\label{eq:cox-cumulative-hazards}
\mathsf A_t=\int_0^t a_s\,\mathrm{d} s,
\qquad
\mathsf B_t=\int_0^t b_s\,\mathrm{d} s,
\qquad
\mathsf M_t=\int_0^t\max\{a_s,b_s\}\,\mathrm{d} s.
\end{equation}
Let $\mathcal E\sim\mathrm{Exp}(1)$, let $\sigma=\inf\{s\in[0,T]:\mathsf A_s\ge\mathcal E\}$, with $\inf\emptyset=+\infty$, and let $H^a_s=\mathbf{1}_{\{\sigma\le s\}}$. With zero interest rates and zero recovery, the price at $s$ of the defaultable zero-coupon bond maturing at $T$, given the default history up to $s$, is
\begin{equation}\label{eq:cox-price}
X^a_s=\mathbf{1}_{\{\sigma>s\}}\,g_a(s),
\qquad
g_a(s)=e^{-(\mathsf A_T-\mathsf A_s)},
\qquad s\in[0,T],
\end{equation}
see Lemma~\ref{lem:cox-price} below. We call $X^a$ the \emph{Cox bond price} with hazard $a$. Let $\mathbb D_T$ be the space of c\`adl\`ag paths on $[0,T]$ with canonical process $X$ and canonical filtration $\mathcal{F}^X_s=\sigma(X_r:r\le s)$, and let $\mathbb{P}^a$ be the law of $X^a$ on $\mathbb D_T$. On $\mathbb D_T\times\mathbb D_T$ we write $(X,Y)$ for the canonical process and $\mathcal G_s=\sigma(X_r,Y_r:r\le s)$. Causal and bicausal couplings are defined as in Definition~\ref{def:bicausal}, with $s\in[0,T]$ in place of $t\in\{0,\ldots,N\}$ and conditioning on the whole path, and $\operatorname{Cpl}_{\mathrm{bc}}$ denotes again the set of bicausal couplings. Equivalently, $\pi$ is causal from $X$ to $Y$ if and only if $\mathbb F^X$ is immersed in $\mathbb G$, that is, every $\mathbb F^X$-martingale is a $\mathbb G$-martingale \cite{BremaudYor1978,AcciaioBackhoffZalashko2020}.

\medskip
As cost we take, as in Remark~\ref{rem:level-convention}, the quadratic variation of the difference of the two martingales. Cox bond prices have paths of finite variation, so this quadratic variation is the sum of the squared jumps of $X-Y$. Since the two prices start from $e^{-\mathsf A_T}$ and $e^{-\mathsf B_T}$, which differ in general, we also include the initial discrepancy and set
\begin{equation}\label{eq:aw-cox-def}
\mathcal{AW}_2^2(\mathbb{P}^a,\mathbb{P}^b)=\inf_{\pi\in\operatorname{Cpl}_{\mathrm{bc}}(\mathbb{P}^a,\mathbb{P}^b)}\mathbb{E}_\pi\Bigl[|X_0-Y_0|^2+\sum_{0<s\le T}|\Delta X_s-\Delta Y_s|^2\Bigr],
\end{equation}
where $\Delta X_s=X_s-X_{s-}$. When $\mathsf A_T=\mathsf B_T$ the two bonds have the same price today, the first term vanishes, and \eqref{eq:aw-cox-def} is the continuous-time counterpart of \eqref{eq:aw2-def}.

\begin{lemma}[Cox bond prices]\label{lem:cox-price}
Let $a\in L^1([0,T];[0,\infty))$.
\begin{enumerate}
\item\label{it:cox-path} $\mathbb{P}^a$-almost surely, $X_s\in[0,1]$ for all $s$ and $X_T\in\{0,1\}$, with $\mathbb{P}^a[X_T=1]=e^{-\mathsf A_T}$. Moreover, $X=X^c+X^d$, where $X^c$ is continuous and nondecreasing with total variation at most $1$, and $X^d$ is a step function with at most one jump, of size at most $1$.
\item\label{it:cox-filtration} $\mathbb{P}^a$-almost surely, $X_s=(1-\chi_s)\,g_a(s)$ for all $s$, where $\chi_s=\mathbf{1}_{\{X_s=0\}}$. Hence $\mathbb F^X$ coincides with the natural filtration of $\chi$ up to $\mathbb{P}^a$-null sets, and $\chi$ has the law of $H^a$.
\item\label{it:cox-mart} $X$ is a $\mathbb{P}^a$-martingale with respect to $\mathbb F^X$, and $\chi$ has $\mathbb F^X$-compensator $\int_0^s(1-\chi_{r-})\,a_r\,\mathrm{d} r$.
\end{enumerate}
\end{lemma}

\begin{proof}
Since $g_a$ is continuous and nondecreasing with $g_a\ge e^{-\mathsf A_T}>0$, we have $X^a_s=0$ if and only if $\sigma\le s$. This gives~\ref{it:cox-filtration}, as well as $X^a_T=\mathbf{1}_{\{\sigma>T\}}$ and $\mathbb{P}[\sigma>T]=\mathbb{P}[\mathcal E>\mathsf A_T]=e^{-\mathsf A_T}$. The decomposition in~\ref{it:cox-path} is $X^c_s=g_a(s\wedge\sigma)$ and $X^d_s=-\mathbf{1}_{\{\sigma\le s\}}\,g_a(\sigma)$, and the total variation of $X^c$ is at most $1-e^{-\mathsf A_T}\le1$.

\medskip
For~\ref{it:cox-mart}, fix $s<u$. By~\ref{it:cox-filtration}, the trace of $\mathcal{F}^X_s$ on $\{\sigma>s\}$ is trivial up to null sets, because $\chi$ vanishes on $[0,s]$ there. On $\{\sigma\le s\}$ we have $X_u=X_s=0$. On $\{\sigma>s\}$,
\[
\mathbb{E}[X_u\mid\mathcal{F}^X_s]=\mathbb{P}[\sigma>u\mid\sigma>s]\,g_a(u)=e^{-(\mathsf A_u-\mathsf A_s)}\,e^{-(\mathsf A_T-\mathsf A_u)}=X_s .
\]
The compensator of the indicator of a random time with hazard rate $a$ in its own filtration is standard \cite[Ch.~III]{Bremaud1981}.
\end{proof}

The lower bound in the main result of this section rests on the following estimate, which only uses that two default times keep their hazard rates in a common filtration.

\begin{lemma}[Default mismatch in a common filtration]\label{lem:joint-survival}
Let $\sigma,\tau$ be stopping times of a filtration $\mathbb H$ on $(\Omega,\mathcal H,\pi)$ with $\pi[\sigma>0,\tau>0]=1$ such that $H^a_t=\mathbf{1}_{\{\sigma\le t\}}$ and $H^b_t=\mathbf{1}_{\{\tau\le t\}}$ have $\mathbb H$-compensators
\begin{equation}\label{eq:cox-marginal-compensators}
\int_0^t\mathbf{1}_{\{\sigma\ge s\}}a_s\,\mathrm{d} s,
\qquad
\int_0^t\mathbf{1}_{\{\tau\ge s\}}b_s\,\mathrm{d} s.
\end{equation}
Then $\pi[\sigma>t]=e^{-\mathsf A_t}$, $\pi[\tau>t]=e^{-\mathsf B_t}$ and $\pi[\sigma>t,\tau>t]\le e^{-\mathsf M_t}$. Consequently, for all $t\in[0,T]$,
\begin{equation}\label{eq:Psi-exact}
\pi\bigl[\mathbf{1}_{\{\sigma>t\}}\neq\mathbf{1}_{\{\tau>t\}}\bigr]\ge\Psi_t(a,b),
\qquad
\Psi_t(a,b)=e^{-\mathsf A_t}+e^{-\mathsf B_t}-2e^{-\mathsf M_t}.
\end{equation}
\end{lemma}

\begin{proof}
Set $S^a_t=\mathbf{1}_{\{\sigma>t\}}$ and $S^b_t=\mathbf{1}_{\{\tau>t\}}$.

\medskip
\emph{Marginals.} By \eqref{eq:cox-marginal-compensators}, $u_a(t)=\mathbb{E}_\pi[S^a_t]$ satisfies $u_a(t)=1-\int_0^ta_su_a(s)\,\mathrm{d} s$, hence $u_a(t)=e^{-\mathsf A_t}$ by uniqueness for this linear integral equation. Similarly, $\mathbb{E}_\pi[S^b_t]=e^{-\mathsf B_t}$.

\medskip
\emph{Common jumps.} Let $J_t=\sum_{0<s\le t}\Delta H^a_s\Delta H^b_s$ record a simultaneous default, and let $\mathcal K$ be its $\mathbb H$-compensator. For every nonnegative predictable $U$,
$\mathbb{E}_\pi\int U\,\mathrm{d} J\le\mathbb{E}_\pi\int U\,\mathrm{d} H^a=\mathbb{E}_\pi\int U_sS^a_{s-}a_s\,\mathrm{d} s$, and analogously with $b$. Hence $\mathrm{d}\mathcal K\le S^a_{s-}a_s\,\mathrm{d} s$ and $\mathrm{d}\mathcal K\le S^b_{s-}b_s\,\mathrm{d} s$ as random measures \cite[Ch.~I.3]{JacodShiryaev2003}. Since $J=\int S^a_{-}S^b_{-}\,\mathrm{d} J$, also $\mathcal K=\int S^a_{-}S^b_{-}\,\mathrm{d}\mathcal K$. By the Radon-Nikodym theorem on the predictable $\sigma$-field there is a predictable $c\ge0$ with
\begin{equation}\label{eq:common-jump-compensator}
\mathcal K_t=\int_0^tS^a_{s-}S^b_{s-}c_s\,\mathrm{d} s,
\qquad
0\le c_s\le\min\{a_s,b_s\}
\end{equation}
$\mathrm{d} s\otimes\pi$-a.e.\ on $\{S^a_{s-}S^b_{s-}=1\}$.

\medskip
\emph{Joint survival.} Integration by parts for the finite-variation processes $S^a,S^b$ gives
$S^a_tS^b_t=1-\int_0^tS^b_{s-}\,\mathrm{d} H^a_s-\int_0^tS^a_{s-}\,\mathrm{d} H^b_s+J_t$. Taking expectations, $q(t)=\mathbb{E}_\pi[S^a_tS^b_t]$ is absolutely continuous and
\begin{equation}\label{eq:q-integral-equation}
q(t)=1-\int_0^t(a_s+b_s)q(s)\,\mathrm{d} s+\int_0^t\mathbb{E}_\pi[S^a_{s-}S^b_{s-}c_s]\,\mathrm{d} s.
\end{equation}
By \eqref{eq:common-jump-compensator}, $\mathbb{E}_\pi[S^a_{s-}S^b_{s-}c_s]\le\min\{a_s,b_s\}q(s)$, whence $q'(s)\le-\max\{a_s,b_s\}q(s)$ a.e.\ and, by Gr\"onwall's inequality,
\begin{equation}\label{eq:joint-survival-upper}
q(t)\le e^{-\mathsf M_t}.
\end{equation}
Since $\pi[S^a_t\neq S^b_t]=\mathbb{E}_\pi S^a_t+\mathbb{E}_\pi S^b_t-2q(t)$, \eqref{eq:Psi-exact} follows.
\end{proof}

We call $\Psi_t(a,b)$ the \emph{adapted default mismatch}. The next result identifies it with the squared adapted distance between Cox bond prices.

\begin{theorem}[Adapted distance between Cox bond prices]\label{thm:aw-cox}
For all $a,b\in L^1([0,T];[0,\infty))$,
\begin{equation}\label{eq:aw-cox}
\mathcal{AW}_2^2(\mathbb{P}^a,\mathbb{P}^b)
=\inf_{\pi\in\operatorname{Cpl}_{\mathrm{bc}}(\mathbb{P}^a,\mathbb{P}^b)}\pi[X_T\neq Y_T]
=\Psi_T(a,b)
=e^{-\mathsf A_T}+e^{-\mathsf B_T}-2e^{-\mathsf M_T}.
\end{equation}
The infimum is attained by the Cox bond prices \eqref{eq:cox-price} built from the common-shock default times
\begin{equation}\label{eq:common-shock}
\sigma=\rho_c\wedge\rho_a,
\qquad
\tau=\rho_c\wedge\rho_b,
\end{equation}
where $\rho_c,\rho_a,\rho_b$ are the first jump times of independent inhomogeneous Poisson processes $N^c,N^a,N^b$ with intensities $\min\{a,b\}$, $(a-b)_+$ and $(b-a)_+$. The common shock has the largest admissible simultaneous default intensity
\begin{equation}\label{eq:maximal-common-shock}
c_t=\min\{a_t,b_t\},
\end{equation}
and it does not depend on $T$, so that it is optimal for the bonds of all maturities in $(0,T]$ simultaneously.
\end{theorem}

\begin{proof}
\emph{Terminal representation.} Fix $\pi\in\operatorname{Cpl}_{\mathrm{bc}}(\mathbb{P}^a,\mathbb{P}^b)$ and set $M=X-Y$. The argument of Lemma~\ref{lem:martingale-under-bicausal} shows that causality from $X$ to $Y$ turns every bounded $\mathbb F^X$-martingale, and in particular $X$, into a $\mathbb G$-martingale, and the same holds for $Y$. Hence $M$ is a $\mathbb G$-martingale with $|M|\le1$ and deterministic $M_0=e^{-\mathsf A_T}-e^{-\mathsf B_T}$. Let $t^n_i=iT/n$. Lemma~\ref{lem:martingale-isometry}, applied to the discrete martingale $(M_{t^n_i})_{i\le n}$, gives
\[
\mathbb{E}_\pi[M_T^2]=M_0^2+\mathbb{E}_\pi[S_n],
\qquad
S_n=\sum_{i<n}\bigl|M_{t^n_{i+1}}-M_{t^n_i}\bigr|^2 .
\]
By Lemma~\ref{lem:cox-price}\ref{it:cox-path}, $\pi$-almost surely $M=M^c+M^d$, where $M^c$ is continuous with total variation at most $2$, and $M^d$ is a step function with at most two jumps, each of size at most $1$. Write $\alpha_i$ and $\beta_i$ for the increments of $M^c$ and $M^d$ over $(t^n_i,t^n_{i+1}]$, and set $\delta_n=\max_i|\alpha_i|$, which tends to $0$ by uniform continuity of $M^c$. For $n$ large enough, every interval of the partition contains at most one jump time of $M^d$, so that $\sum_i\beta_i^2=\sum_{0<s\le T}|\Delta M_s|^2$ and
\[
\Bigl|S_n-\sum_{0<s\le T}|\Delta M_s|^2\Bigr|\le2\delta_n\sum_i|\beta_i|+\delta_n\sum_i|\alpha_i|\le6\,\delta_n .
\]
Moreover, $S_n\le8$, since the increments of $M$ are bounded by $2$ and its total variation by $4$. By dominated convergence, $\mathbb{E}_\pi[M_T^2]=M_0^2+\mathbb{E}_\pi\sum_{0<s\le T}|\Delta M_s|^2$, which is the cost in \eqref{eq:aw-cox-def}. Since $X_T,Y_T\in\{0,1\}$, we have $M_T^2=\mathbf{1}_{\{X_T\neq Y_T\}}$, and the first identity in \eqref{eq:aw-cox} follows.

\medskip
\emph{Lower bound.} Let $\sigma$ and $\tau$ be the jump times of $\chi^X=\mathbf{1}_{\{X=0\}}$ and $\chi^Y=\mathbf{1}_{\{Y=0\}}$. By Lemma~\ref{lem:cox-price}\ref{it:cox-mart}, $\chi^X-\int_0^\cdot(1-\chi^X_{r-})a_r\,\mathrm{d} r$ is a bounded $\mathbb F^X$-martingale, hence a $\mathbb G$-martingale by the first step. Its second term is continuous and $\mathbb G$-adapted, so it is the $\mathbb G$-compensator of $\chi^X$, and the same holds for $\chi^Y$ with $b$ in place of $a$. Since $X_T=\mathbf{1}_{\{\sigma>T\}}$ and $Y_T=\mathbf{1}_{\{\tau>T\}}$, Lemma~\ref{lem:joint-survival} gives $\pi[X_T\neq Y_T]\ge\Psi_T(a,b)$.

\medskip
\emph{Attainment.} In the filtration $\mathbb H$ generated by $N^c,N^a,N^b$ the intensity of $\sigma$ before default is $\min\{a,b\}+(a-b)_+=a$, and that of $\tau$ is $b$, so $H^a$ and $H^b$ have the $\mathbb H$-compensators \eqref{eq:cox-marginal-compensators}. Both survive to $t$ exactly when no process has jumped, so $\pi[\sigma>t,\tau>t]=\exp(-\int_0^t[\min\{a,b\}+|a-b|]\,\mathrm{d} s)=e^{-\mathsf M_t}$, and $\pi[\mathbf{1}_{\{\sigma>t\}}\neq\mathbf{1}_{\{\tau>t\}}]=\Psi_t(a,b)$ by the marginal identities of Lemma~\ref{lem:joint-survival}. It remains to show that the law of the two Cox bond prices built from $\sigma$ and $\tau$ is bicausal. By Lemma~\ref{lem:cox-price}\ref{it:cox-filtration}, it suffices to show that the natural filtrations $\mathbb F^a$ and $\mathbb F^b$ of $H^a$ and $H^b$ are immersed in their joint filtration $\mathbb G^{ab}$. Write $L^a_t=\int_0^t\mathbf{1}_{\{\sigma\ge s\}}a_s\,\mathrm{d} s$. The process $H^a-L^a$ is an $\mathbb H$-martingale adapted to $\mathbb G^{ab}\subseteq\mathbb H$, hence a $\mathbb G^{ab}$-martingale. Every bounded $\mathbb F^a$-martingale $n$ can be written as $n_t=n_0+\int_0^t\varphi_s\,\mathrm{d}(H^a-L^a)_s$, with an $\mathbb F^a$-predictable integrand $\varphi$ bounded by $2\sup|n|$ \cite[Ch.~III]{Bremaud1981}. This integrand is also $\mathbb G^{ab}$-predictable, so $n$ is a $\mathbb G^{ab}$-martingale. Hence $\mathbb F^a$ is immersed in $\mathbb G^{ab}$. By the same argument, so is $\mathbb F^b$, and the coupling is bicausal. Finally, the construction does not depend on $T$.
\end{proof}

\begin{remark}[Relation with the discrete-time results]\label{rem:aw-cox-discrete}
The first identity in \eqref{eq:aw-cox} is the continuous-time analogue of the first identity in \eqref{eq:terminal-representation} and is proved in the same way. The second reduction in \eqref{eq:terminal-representation} conditions on the last date before maturity. There is no such date in continuous time, and its role is taken by the Gr\"onwall argument of Lemma~\ref{lem:joint-survival}, which produces the closed form. The optimal coupling is the Marshall-Olkin common-shock construction \cite{MarshallOlkin1967} with the largest admissible common component. In the language of \cite{BurdzyKendall2000,ConnorJacka2008} it is the optimal co-adapted coupling of two killing times, and Theorem~\ref{thm:aw-cox} shows that this classical coupling solves the adapted transport problem between Cox bond prices. The initial term in \eqref{eq:aw-cox-def} is $|e^{-\mathsf A_T}-e^{-\mathsf B_T}|^2$, so that without it the infimum is $\Psi_T(a,b)-|e^{-\mathsf A_T}-e^{-\mathsf B_T}|^2$.
\end{remark}

For a finite nonnegative Borel measure $\theta$ on $[0,T]$ define
\begin{equation}\label{eq:Phi-theta-definition}
\Phi_\theta(a,b)=\int_{[0,T]}\Psi_t(a,b)\,\theta(\mathrm{d} t),
\end{equation}
and, for $t\in(0,T]$, let $\mathbb{P}^a_t$ and $\mathbb{P}^b_t$ be the laws of the Cox bond prices maturing at $t$. Since $\Psi_0=0$, the point $t=0$ does not contribute to \eqref{eq:Phi-theta-definition}.

\begin{corollary}[Several maturities]\label{cor:weighted-mismatch-representation}
For every finite nonnegative Borel measure $\theta$ on $[0,T]$,
\begin{equation}\label{eq:weighted-mismatch-optimal}
\Phi_\theta(a,b)=\int_{(0,T]}\mathcal{AW}_2^2(\mathbb{P}^a_t,\mathbb{P}^b_t)\,\theta(\mathrm{d} t)=\inf_{\pi}\mathbb{E}_\pi\Bigl[\int_{[0,T]}\bigl|\mathbf{1}_{\{\sigma>t\}}-\mathbf{1}_{\{\tau>t\}}\bigr|\,\theta(\mathrm{d} t)\Bigr],
\end{equation}
where the infimum is over bicausal couplings $\pi$ of the laws of $H^a$ and $H^b$, and $\sigma,\tau$ denote the two default times. The common-shock default times \eqref{eq:common-shock} attain the infimum.
\end{corollary}

\begin{proof}
The first identity is Theorem~\ref{thm:aw-cox} applied at each maturity $t$. For the second, Lemma~\ref{lem:cox-price} \ref{it:cox-filtration} shows that a bicausal coupling of the laws of $H^a$ and $H^b$ induces, for every $t$, a bicausal coupling of $\mathbb{P}^a_t$ and $\mathbb{P}^b_t$, under which the two bond prices disagree at $t$ exactly when the two survival indicators do. Tonelli's theorem and Theorem~\ref{thm:aw-cox} therefore bound the expectation from below by $\Phi_\theta(a,b)$, and the common-shock default times attain $\Psi_t(a,b)$ for every $t$ simultaneously.
\end{proof}

The unconstrained mismatch of the Bernoulli variables at time $t$ is the Fr\'echet bound
\begin{equation}\label{eq:frechet-mismatch}
\phi_t(a,b)=\bigl|e^{-\mathsf A_t}-e^{-\mathsf B_t}\bigr|.
\end{equation}
By Lemma~\ref{lem:bernoulli-mismatch}, it is the minimal probability that the two survival indicators disagree at $t$ over all couplings. The difference between \eqref{eq:Psi-exact} and \eqref{eq:frechet-mismatch} is therefore the cost of respecting the flow of information.

\begin{proposition}[Exact adaptation slack]\label{prop:adaptation-slack}
Set $\kappa_t(a,b)=\mathsf M_t-\max\{\mathsf A_t,\mathsf B_t\}$. Then
\begin{equation}\label{eq:kappa-positive-parts}
\kappa_t(a,b)=\min\Bigl\{\int_0^t(a_s-b_s)_+\,\mathrm{d} s,\int_0^t(b_s-a_s)_+\,\mathrm{d} s\Bigr\}
\end{equation}
and
\begin{equation}\label{eq:Psi-decomposition}
\Psi_t(a,b)=\phi_t(a,b)+2e^{-\max\{\mathsf A_t,\mathsf B_t\}}\bigl(1-e^{-\kappa_t(a,b)}\bigr).
\end{equation}
In particular $\phi_t\le\Psi_t$, with equality if and only if $a\le b$ a.e.\ on $[0,t]$ or $b\le a$ a.e.\ on $[0,t]$.
\end{proposition}

\begin{proof}
Write
\[
P_t=\int_0^t(a_s-b_s)_+\,\mathrm{d} s,
\qquad
N_t=\int_0^t(b_s-a_s)_+\,\mathrm{d} s
\]
for the cumulated amounts by which each hazard exceeds the other. Since $\max\{a,b\}=b+(a-b)_+=a+(b-a)_+$, we have
\[
\mathsf M_t=\mathsf B_t+P_t=\mathsf A_t+N_t,
\qquad
\mathsf A_t-\mathsf B_t=P_t-N_t .
\]
If $\mathsf A_t\ge\mathsf B_t$, then $P_t\ge N_t$ and $\kappa_t=\mathsf M_t-\mathsf A_t=N_t=\min\{P_t,N_t\}$, and symmetrically $\kappa_t=P_t=\min\{P_t,N_t\}$ if $\mathsf B_t\ge\mathsf A_t$. This proves \eqref{eq:kappa-positive-parts}.

\medskip
To obtain \eqref{eq:Psi-decomposition}, assume again $\mathsf A_t\ge\mathsf B_t$, the other case being symmetric. Then $\phi_t=e^{-\mathsf B_t}-e^{-\mathsf A_t}$, and \eqref{eq:Psi-exact} gives
\[
\Psi_t-\phi_t=2e^{-\mathsf A_t}-2e^{-\mathsf M_t}=2e^{-\mathsf A_t}\bigl(1-e^{-\kappa_t}\bigr),
\]
since $\mathsf M_t-\mathsf A_t=\kappa_t$ and $\mathsf A_t=\max\{\mathsf A_t,\mathsf B_t\}$.

\medskip
Finally, the second term in \eqref{eq:Psi-decomposition} is nonnegative, so $\phi_t\le\Psi_t$, and since its exponential factor is positive, it vanishes if and only if $\kappa_t=0$. This happens exactly when $P_t=0$ or $N_t=0$, that is, when $a\le b$ a.e.\ on $[0,t]$ or $b\le a$ a.e.\ on $[0,t]$.
\end{proof}

\medskip
Calibration to the prices at the ends of a bucket fixes only the cumulative hazards $\mathsf A_T$ and $\mathsf B_T$, whereas the adaptation slack of Proposition~\ref{prop:adaptation-slack} depends on how the hazard is distributed within the bucket. The next result determines the full range of the adapted default mismatch that is compatible with the calibrated values.

\begin{theorem}[Sharp range of the adapted default mismatch under calibration]\label{thm:hazard-envelope}
Fix a bucket of length $T>0$, an intensity cap $L>0$, and numbers
$h_a,h_b\in[0,LT]$. Let $\mathcal H_L(h_a,h_b)$ denote the class of
pairs of measurable functions $a,b:[0,T]\to[0,L]$ whose respective
integrals are $h_a$ and $h_b$. Then
\begin{align}
\inf_{(a,b)\in\mathcal H_L(h_a,h_b)}\Psi_T(a,b)
 &=\left|e^{-h_a}-e^{-h_b}\right|,\label{eq:hazard-envelope-min}\\
\sup_{(a,b)\in\mathcal H_L(h_a,h_b)}\Psi_T(a,b)
 &=e^{-h_a}+e^{-h_b}-2e^{-\min\{h_a+h_b,LT\}}.\label{eq:hazard-envelope-max}
\end{align}
Both extrema are attained.
If the hazards are unbounded, the same result holds with the last exponent
replaced by $h_a+h_b$.
\end{theorem}
\begin{proof}
\emph{Upper and lower bounds.} Let $(a,b)\in\mathcal H_L(h_a,h_b)$, so that $\mathsf A_T=h_a$ and $\mathsf B_T=h_b$. The pointwise inequalities
\[
\max\{a_s,b_s\}\ge a_s,\qquad\max\{a_s,b_s\}\ge b_s,\qquad\max\{a_s,b_s\}\le a_s+b_s,\qquad\max\{a_s,b_s\}\le L
\]
give, after integration over $[0,T]$,
\begin{equation}\label{eq:M-range}
\max\{h_a,h_b\}\le\mathsf M_T\le\min\{h_a+h_b,LT\}.
\end{equation}
By \eqref{eq:Psi-exact}, $\Psi_T(a,b)=e^{-h_a}+e^{-h_b}-2e^{-\mathsf M_T}$ is strictly increasing in $\mathsf M_T$. Hence \eqref{eq:M-range} yields the inequalities $\ge$ in \eqref{eq:hazard-envelope-min} and $\le$ in \eqref{eq:hazard-envelope-max}, since $e^{-h_a}+e^{-h_b}-2e^{-\max\{h_a,h_b\}}=|e^{-h_a}-e^{-h_b}|$.

\medskip
\emph{Attainment of the lower bound.} Take the constant intensities $a_s=h_a/T$ and $b_s=h_b/T$. They belong to $\mathcal H_L(h_a,h_b)$ and are ordered on $[0,T]$, so that $\mathsf M_T=\max\{h_a,h_b\}$ and equality holds in \eqref{eq:hazard-envelope-min}.

\medskip
\emph{Attainment of the upper bound.} Set
\[
E_a=[0,h_a/L],\qquad E_b=[T-h_b/L,T],\qquad a=L\mathbf{1}_{E_a},\qquad b=L\mathbf{1}_{E_b}.
\]
Since $h_a,h_b\le LT$, both sets are subintervals of $[0,T]$, and the intensities take values in $[0,L]$ with integrals $h_a$ and $h_b$. The two intervals lie at opposite ends of $[0,T]$, so their union has Lebesgue measure $\min\{T,(h_a+h_b)/L\}$. Since $\max\{a,b\}=L$ on $E_a\cup E_b$ and vanishes outside,
\[
\mathsf M_T=L\,|E_a\cup E_b|=\min\{h_a+h_b,LT\},
\]
and equality holds in \eqref{eq:hazard-envelope-max}.

\medskip
Finally, without the cap the last inequality in \eqref{eq:M-range} becomes $\mathsf M_T\le h_a+h_b$. Two nonnegative intensities with disjoint supports and integrals $h_a$ and $h_b$ attain it.
\end{proof}

\begin{corollary}[Identical calibration, maximal intra-bucket discrepancy]
\label{cor:invisible-calibrated-mismatch}
If $h_a=h_b=h\in[0,LT]$, the two models agree on the survival probability
at the end of the bucket, but their
adapted default mismatch $\Psi_T(a,b)$
can attain any value in the interval
\[
\left[0,\;2\bigl(e^{-h}-e^{-\min\{2h,LT\}}\bigr)\right].
\]
In particular, when $LT\ge2h$, the largest such mismatch equals
$2e^{-h}(1-e^{-h})$ despite identical terminal survival probabilities.
The lower endpoint is attained by identical flat hazards and the upper
endpoint by the bang-bang construction of
Theorem~\ref{thm:hazard-envelope}.
\end{corollary}
\begin{proof}
For equal integrated hazards, the
Fr\'echet mismatch $\phi_T(a,b)=|e^{-h_a}-e^{-h_b}|$ of \eqref{eq:frechet-mismatch}
vanishes identically. Both endpoints are
attained by Theorem~\ref{thm:hazard-envelope}. The class of admissible
intensities is convex, and interpolation between a lower and an upper
extremizer yields a continuous map of the interpolation parameter into
$\Psi_T$: indeed the cumulative hazards remain constant, while the
integral of the pointwise maximum depends continuously on the two
intensities in $L^1$. The intermediate-value theorem supplies every
value between the endpoints. The asserted numerical expression follows
by inserting $\min\{2h,LT\}=2h$ whenever $LT\ge2h$.
\end{proof}


Theorem~\ref{thm:hazard-envelope} separates two sources of mismatch. The lower value \eqref{eq:hazard-envelope-min} is the Fr\'echet mismatch, which is due only to different calibrated survival probabilities. The excess above it is due to the allocation of hazard within the bucket. The theorem also shows that a bound on $\Psi_T$ in terms of the two endpoint prices alone cannot be better than \eqref{eq:hazard-envelope-max} unless the admissible hazard schedules are restricted. In the equal-calibration case the endpoint roll residual is zero, whereas the adapted default mismatch may be strictly positive, as already seen in Example~\ref{ex:pure-slack}. By Theorem~\ref{thm:aw-cox}, two Cox bonds with the same price today and hazards bounded by $L$ with $LT\ge2h$ can therefore be at squared adapted distance up to $2e^{-h}(1-e^{-h})$.

\begin{theorem}[Sharp $L^1$ stability of the adapted default mismatch]
\label{thm:psi-L1-stability}
Let $a,b,\widetilde a,\widetilde b$ be nonnegative integrable intensities
on $[0,T]$. For every $t\le T$,
\begin{equation}\label{eq:psi-L1-stability}
\left|\Psi_t(a,b)-\Psi_t(\widetilde a,\widetilde b)\right|
\le\int_0^t\left(|a_s-\widetilde a_s|+
|b_s-\widetilde b_s|\right)\,\mathrm ds.
\end{equation}
The coefficient one cannot be reduced uniformly. For a finite
nonnegative measure $\theta$ on $[0,T]$,
the weighted mismatch $\Phi_\theta$ of \eqref{eq:Phi-theta-definition}
satisfies
\[
\left|\Phi_\theta(a,b)-\Phi_\theta(\widetilde a,\widetilde b)\right|
\le \int_{[0,T]}\!\int_0^t
\left(|a_s-\widetilde a_s|+|b_s-\widetilde b_s|\right)
\mathrm ds\,\theta(\mathrm dt).
\]
\end{theorem}
\begin{proof}
\emph{Interpolation.} Fix $t\le T$, put $u=\widetilde a-a$ and $v=\widetilde b-b$, and for $r\in[0,1]$ let $a^r=a+ru$ and $b^r=b+rv$. These are convex combinations of nonnegative integrable intensities, so they are again nonnegative and integrable. Write
\[
\mathsf A^r=\int_0^ta^r_s\,\mathrm{d} s,\qquad\mathsf B^r=\int_0^tb^r_s\,\mathrm{d} s,\qquad\mathsf M^r=\int_0^t\max\{a^r_s,b^r_s\}\,\mathrm{d} s.
\]
The first two are affine in $r$. Since $|\max\{x,y\}-\max\{x',y'\}|\le|x-x'|+|y-y'|$, the third is Lipschitz in $r$ with constant $\int_0^t(|u_s|+|v_s|)\,\mathrm{d} s$. In particular all three are absolutely continuous, and so is $r\mapsto\Psi_t(a^r,b^r)=e^{-\mathsf A^r}+e^{-\mathsf B^r}-2e^{-\mathsf M^r}$.

\medskip
\emph{Derivative of the maximum.} For each $s$, the map $r\mapsto\max\{a^r_s,b^r_s\}$ is the maximum of two affine functions, so it is differentiable except at most at one value of $r$, and its difference quotients are bounded by $|u_s|+|v_s|$. By Fubini's theorem, for almost every $r$ it is differentiable at $r$ for almost every $s$, and dominated convergence gives
\[
\frac{\mathrm{d}\mathsf M^r}{\mathrm{d} r}=\int_0^t\bigl(\alpha^r_su_s+(1-\alpha^r_s)v_s\bigr)\,\mathrm{d} s,
\qquad\alpha^r_s=\mathbf{1}_{\{a^r_s>b^r_s\}}.
\]
Indeed, at a point of differentiability the derivative equals $u_s$ if $a^r_s>b^r_s$ and $v_s$ if $a^r_s<b^r_s$. If $a^r_s=b^r_s$, differentiability forces $u_s=v_s$, so the formula holds there too.

\medskip
\emph{Lipschitz bound.} For almost every $r$ the chain rule gives
\begin{align*}
\frac{\mathrm{d}}{\mathrm{d} r}\Psi_t(a^r,b^r)
={}&\int_0^t\bigl[-e^{-\mathsf A^r}+2\alpha^r_se^{-\mathsf M^r}\bigr]u_s\,\mathrm{d} s\\
&+\int_0^t\bigl[-e^{-\mathsf B^r}+2(1-\alpha^r_s)e^{-\mathsf M^r}\bigr]v_s\,\mathrm{d} s.
\end{align*}
Since $\mathsf M^r\ge\mathsf A^r$, the first coefficient satisfies $-1\le-e^{-\mathsf A^r}\le-e^{-\mathsf A^r}+2\alpha^r_se^{-\mathsf M^r}\le-e^{-\mathsf A^r}+2e^{-\mathsf M^r}\le e^{-\mathsf M^r}\le1$. In the same way, since $\mathsf M^r\ge\mathsf B^r$, the second coefficient lies in $[-1,1]$. Hence
\[
\Bigl|\frac{\mathrm{d}}{\mathrm{d} r}\Psi_t(a^r,b^r)\Bigr|\le\int_0^t\bigl(|u_s|+|v_s|\bigr)\,\mathrm{d} s
\]
for almost every $r$, and integrating over $r\in[0,1]$ gives \eqref{eq:psi-L1-stability}.

\medskip
\emph{Sharpness.} Take $a=b=\widetilde b=0$ and $\widetilde a=\varepsilon/T$ with $\varepsilon>0$. Then $\Psi_T(a,b)=0$ and $\Psi_T(\widetilde a,\widetilde b)=1-e^{-\varepsilon}$, while the right-hand side of \eqref{eq:psi-L1-stability} at $t=T$ equals $\varepsilon$. The ratio $(1-e^{-\varepsilon})/\varepsilon$ tends to one as $\varepsilon\downarrow0$, so the constant one cannot be reduced.

\medskip
Finally, $t\mapsto\Psi_t$ is continuous for both pairs of intensities. Integrating \eqref{eq:psi-L1-stability} against $\theta$ and using $|\int f\,\mathrm{d}\theta|\le\int|f|\,\mathrm{d}\theta$ gives the weighted inequality.
\end{proof}

\subsection{Random intensities: the default-layer reduction}\label{subsec:default-layer}

We now allow both intensities to be driven by factors whose laws may differ. For $i\in\{A,B\}$, a Cox model consists of a c\`adl\`ag factor process $\xi^i$ with values in a Polish space and law $L^i$, a nonnegative intensity $a^i$ that is predictable for the natural filtration of $\xi^i$ with $\int_0^Ta^i_s\,\mathrm{d} s<\infty$, and the default time $\inf\{t:\int_0^ta^i_s\,\mathrm{d} s\ge\mathcal E^i\}$ with $\mathcal E^i\sim\mathrm{Exp}(1)$ independent of $\xi^i$. Write $a=a^A$, $b=a^B$.

\begin{definition}[Doubly stochastic coupling]\label{def:ds-coupling}
Let $\gamma\in\operatorname{Cpl}(L^A,L^B)$. A \emph{doubly stochastic coupling over $\gamma$} is a probability space carrying $(\xi^A,\xi^B)\sim\gamma$ and random times $\sigma,\tau>0$ almost surely such that, with $\mathcal G_T=\sigma(\xi^A_s,\xi^B_s:s\le T)$ and $\overline{\mathcal H}_t=\mathcal G_T\vee\sigma(H^a_s,H^b_s:s\le t)$, the $\overline{\mathbb H}$-compensators of $H^a,H^b$ are $\int_0^t\mathbf{1}_{\{\sigma\ge s\}}a_s\,\mathrm{d} s$ and $\int_0^t\mathbf{1}_{\{\tau\ge s\}}b_s\,\mathrm{d} s$. The class is denoted by $\mathfrak D(\gamma)$.
\end{definition}

Conditionally on both factor paths, each default time then has its model hazard, so $(\xi^A,\sigma)$ and $(\xi^B,\tau)$ have the laws of the two Cox models: this is the doubly stochastic property of each marginal \cite{Lando1998,ElliottJeanblancYor2000}, imposed jointly.

\begin{theorem}[Default-layer reduction]\label{thm:default-layer}
Fix $T$ and $\gamma\in\operatorname{Cpl}(L^A,L^B)$.
\begin{enumerate}
\item\label{it:dl-lower} For every coupling in $\mathfrak D(\gamma)$, $\pi[\mathbf{1}_{\{\sigma>T\}}\neq\mathbf{1}_{\{\tau>T\}}\mid\mathcal G_T]\ge\Psi_T(a,b)$ almost surely.
\item\label{it:dl-attain} The conditional common-shock construction given in the proof belongs to $\mathfrak D(\gamma)$ and attains equality, simultaneously for all $T$.
\item\label{it:dl-bicausal} If $\gamma$ is bicausal, the construction in~\ref{it:dl-attain} is a bicausal coupling of the laws of $(\xi^A,H^a)$ and $(\xi^B,H^b)$.
\end{enumerate}
Consequently
\begin{equation}\label{eq:default-layer}
\inf_{\gamma\in\operatorname{Cpl}_{\mathrm{bc}}(L^A,L^B)}\ \inf_{\mathfrak D(\gamma)}\pi\bigl[\mathbf{1}_{\{\sigma>T\}}\neq\mathbf{1}_{\{\tau>T\}}\bigr]
=\inf_{\gamma\in\operatorname{Cpl}_{\mathrm{bc}}(L^A,L^B)}\mathbb{E}_\gamma\bigl[\Psi_T(a,b)\bigr],
\end{equation}
an adapted transport problem between the factor laws with the path-dependent, non-additive cost $\Psi_T$.
\end{theorem}

\begin{proof}
\ref{it:dl-lower}. Since $a,b$ are $\overline{\mathcal H}_0$-measurable, for $G\in\mathcal G_T$ and a bounded $\overline{\mathbb H}$-predictable $U$ the process $\mathbf{1}_GU$ is again $\overline{\mathbb H}$-predictable, so the compensator identities hold after multiplication by $\mathbf{1}_G$. The filtrations are countably generated, so a regular conditional probability $\pi^\omega=\pi(\cdot\mid\mathcal G_T)(\omega)$ exists. For $\pi$-a.e.\ $\omega$, the indicators $H^a,H^b$ have under $\pi^\omega$ the $\overline{\mathbb H}$-compensators \eqref{eq:cox-marginal-compensators} with the deterministic intensities $a(\omega),b(\omega)$. Lemma~\ref{lem:joint-survival} applied under $\pi^\omega$ gives the claim.

\medskip
\ref{it:dl-attain}. Let $(\xi^A,\xi^B)\sim\gamma$ and let $\widetilde N^c,\widetilde N^a,\widetilde N^b$ be independent unit Poisson processes, independent of the factors. Set $N^c_t=\widetilde N^c(\int_0^t\min\{a,b\})$, $N^a_t=\widetilde N^a(\int_0^t(a-b)_+)$, $N^b_t=\widetilde N^b(\int_0^t(b-a)_+)$, and let $\sigma$ and $\tau$ be the first jumps of $N^c+N^a$ and of $N^c+N^b$, respectively. Conditionally on $\mathcal G_T$, these are independent Poisson processes with deterministic intensities, so the compensators in $\overline{\mathbb H}$ are as required, and the conditional joint survival probability is $e^{-\mathsf M_T}$ as in Theorem~\ref{thm:aw-cox}.

\medskip
\ref{it:dl-bicausal}. Let $X=(\xi^A,H^a)$ and $Y=(\xi^B,H^b)$. We show that the conditional law of $X_{(t,T]}$ given $(X_{[0,t]},Y_{[0,t]})$ depends only on $X_{[0,t]}$. The symmetric statement is proved in the same way, and together the two give bicausality. The past $(X_{[0,t]},Y_{[0,t]})$ is a function of $\xi_{[0,t]}=(\xi^A,\xi^B)_{[0,t]}$ and of the Poisson processes restricted to the random, $\xi_{[0,t]}$-measurable intervals $[0,\int_0^t\cdot]$. First, $\widetilde N$ is independent of the factors, so the conditional law of $\xi^A_{(t,T]}$ given $\xi_{[0,t]}$ and the Poisson past equals its law given $\xi_{[0,t]}$, which by causality of $\gamma$ equals its law given $\xi^A_{[0,t]}$. Second, on $\{\sigma>t\}$ and conditionally on all factors and the Poisson past, $\sigma$ is the first point after $t$ of $N^c+N^a$, which by independent increments is a Poisson process with intensity $a$ on $(t,T]$, independent of the past. On $\{\sigma\le t\}$ we simply have $H^a_{(t,T]}\equiv1$. Hence the conditional law of $H^a_{(t,T]}$ given everything else depends only on $H^a_t$ and $a_{(t,T]}$, that is, on $X_{[0,t]}$ and $\xi^A_{(t,T]}$. Combining the two steps and using the tower property yields the claim.

\medskip
We now turn to \eqref{eq:default-layer}. Integrating~\ref{it:dl-lower} gives the inequality $\ge$ for every bicausal $\gamma$, and~\ref{it:dl-attain} and~\ref{it:dl-bicausal} show that it is attained by a coupling that is bicausal for the full processes.
\end{proof}

\begin{remark}[Couplings outside the doubly stochastic class]\label{rem:feedback}
Theorem~\ref{thm:default-layer} minimizes the default mismatch over doubly stochastic couplings, in which the two factor paths are coupled first, through $\gamma$, and the default times are then constructed conditionally on both factor paths. A bicausal coupling of the full processes $(\xi^A,H^a)$ and $(\xi^B,H^b)$ need not have this form. Bicausality preserves each model's conditional future law in the joint filtration, but the coupling of future factor innovations may depend on the observed default history. For example, suppose that both factor innovations at a deterministic review are symmetric Bernoulli variables independent of their respective pasts. One may couple them comonotonically or antimonotonically according to whether model~A has already defaulted. Each marginal conditional law is unchanged, whereas observing both innovations reveals that past default event. Such a construction can be bicausal while failing the doubly stochastic condition. Consequently, the right-hand side of \eqref{eq:default-layer} is an upper bound for the minimal mismatch over all bicausal couplings of the full processes, and we do not know whether the two coincide. We restrict to doubly stochastic couplings because they preserve the Cox structure of both models.
\end{remark}

\subsection{A calibrated rating-review bucket}\label{subsec:rating-review-bucket}

Fix $0<t_1<T$. Interest rates and recovery are zero. All factor variables live on $(\Omega^F,\mathcal{F}^F,\mathbb F^F,\mathbb{Q})$, and the unit exponential variables used to construct Cox default times are independent of $\mathcal{F}^F_T$. Let $\alpha\ge0$ be an $\mathbb F^F$-predictable common intensity on $[0,t_1]$ with $\int_0^{t_1}\alpha_s\,\mathrm ds<\infty$ almost surely and set
\[
\Xi=\int_0^{t_1}\alpha_s\,\mathrm{d} s .
\]
At $t_1$ a rating review reveals a $\sigma$-field $\mathcal R\subseteq\mathcal{F}^F_{t_1}$. We compare two models that share the intensity $\alpha$ before the review and differ afterwards. In the \emph{market model} the post-review hazard schedule $(\lambda_s)_{t_1<s\le T}$ is $\mathcal R$-measurable with $\int_{t_1}^T\lambda_s\,\mathrm{d} s<\infty$ almost surely, and the cumulative hazard is
\begin{equation}\label{eq:Lambda-u-definition}
A_u=\Xi+\Lambda_u,
\qquad
\Lambda_u=\int_{t_1}^u\lambda_s\,\mathrm{d} s,
\qquad t_1\le u\le T.
\end{equation}
Define the survival-weighted measure and the calibrated forward survival factor
\begin{equation}\label{eq:survival-weighted-measure}
\frac{\mathrm{d}\widehat{\mathbb{Q}}}{\mathrm{d}\mathbb{Q}}=\frac{e^{-\Xi}}{\mathbb{E}_{\mathbb{Q}}[e^{-\Xi}]},
\qquad
F=\widehat{\mathbb{E}}\bigl[e^{-\Lambda_T}\bigr]\in(0,1],
\qquad
\overline\Lambda=-\log F.
\end{equation}
The \emph{bootstrap model} replaces the post-review hazard by the flat forward hazard $\beta=\overline\Lambda/(T-t_1)$, so that its cumulative hazard is
\begin{equation}\label{eq:flat-bootstrap-intensity}
B_u=\Xi+(u-t_1)\beta,
\qquad t_1\le u\le T.
\end{equation}
Throughout, the superscripts $A$ and $B$ refer to the market model \eqref{eq:Lambda-u-definition} and to the bootstrap model \eqref{eq:flat-bootstrap-intensity}. In each model, the default time is constructed by comparing the cumulative hazard with an independent unit exponential variable, and the factor filtration is enlarged by the default indicator. Standard Cox conditioning \cite{Lando1998,BieleckiRutkowski2002} gives
\begin{equation}\label{eq:timezero-zcb-prices}
P^A_0(u)=\mathbb{E}_{\mathbb{Q}}[e^{-A_u}],
\qquad
P^B_0(u)=\mathbb{E}_{\mathbb{Q}}[e^{-B_u}],
\end{equation}
and, on survival to $t_1$, the pre-default prices at $t_1$ of the bond maturing at $u$ are
\begin{equation}\label{eq:review-date-predefault-prices}
\widetilde P^A_{t_1}(u)=e^{-\Lambda_u},
\qquad
\widetilde P^B_{t_1}(u)=e^{-(u-t_1)\beta}.
\end{equation}

\begin{remark}[Interpretation of $\widehat{\mathbb{Q}}$]\label{rem:hatQ}
For every $E\in\mathcal{F}^F_T$, $\mathbb{Q}[E\mid\tau^A>t_1]=\mathbb{E}_{\mathbb{Q}}[e^{-\Xi}\mathbf{1}_E]/\mathbb{E}_{\mathbb{Q}}[e^{-\Xi}]$ $=\widehat{\mathbb{Q}}[E]$. Thus $\widehat{\mathbb{Q}}$ is the law of the factors conditional on survival to the review date, for an observer who does not see the factors.
\end{remark}

\begin{proposition}[Calibration pins do not determine the bucket law]\label{prop:calibration-pins}
The two models satisfy
\begin{equation}\label{eq:pin-calibration}
P^A_0(t_1)=P^B_0(t_1)=\mathbb{E}_{\mathbb{Q}}[e^{-\Xi}],
\qquad
P^A_0(T)=P^B_0(T)=\mathbb{E}_{\mathbb{Q}}[e^{-\Xi}]\,F .
\end{equation}
If in addition $\lambda_s=L$ on $(t_1,T]$ for a nonnegative $\mathcal R$-measurable $L$, then for $u=t_1+\vartheta(T-t_1)$ with $\vartheta\in(0,1)$,
\begin{equation}\label{eq:intermediate-survival-comparison}
\frac{P^A_0(u)}{P^A_0(t_1)}=\widehat{\mathbb{E}}\bigl[e^{-\vartheta L(T-t_1)}\bigr]\le F^\vartheta=\frac{P^B_0(u)}{P^B_0(t_1)},
\end{equation}
with strict inequality unless $L$ is $\widehat{\mathbb{Q}}$-almost surely constant.
\end{proposition}

\begin{proof}
The two cumulative hazards coincide up to $t_1$, which gives the first identity. At maturity, the definition of $\widehat{\mathbb{Q}}$ gives
\[
P^A_0(T)=\mathbb{E}_{\mathbb{Q}}[e^{-\Xi}]\,\widehat{\mathbb{E}}[e^{-\Lambda_T}]=\mathbb{E}_{\mathbb{Q}}[e^{-\Xi}]\,e^{-\overline\Lambda}=P^B_0(T).
\]

\medskip
Under the additional assumption, set $Z=e^{-L(T-t_1)}\in[0,1]$, so that $F=\widehat{\mathbb{E}} Z$ and $P^A_0(u)/P^A_0(t_1)$ $=\widehat{\mathbb{E}}[Z^\vartheta]$. Since $z\mapsto z^\vartheta$ is strictly concave, Jensen's inequality gives $\widehat{\mathbb{E}}[Z^\vartheta]\le(\widehat{\mathbb{E}} Z)^\vartheta=F^\vartheta$, with equality only if $Z$, and hence $L$, is $\widehat{\mathbb{Q}}$-almost surely constant.
\end{proof}

Equality of the quoted prices at the pins therefore fixes every payoff that depends only on survival at the pins. Continuous premium legs, protection paid at default, and deterministic-exposure valuation adjustments may still differ. Proposition~\ref{prop:filtered-projection} below constructs the deterministic-hazard model that removes this discrepancy.

\subsection{The bootstrap roll and its exact error}\label{subsec:bootstrap-roll}

At time zero, sell one defaultable bond maturing at $T$ and buy
\begin{equation}\label{eq:roll-notional}
N_0=\frac{P_0(T)}{P_0(t_1)}=F
\end{equation}
units of the defaultable bond maturing at $t_1$. The initial value is zero. If default occurs before $t_1$, both positions are extinguished. On survival to $t_1$, the short position is closed at the market price, leaving the cash residual
\begin{equation}\label{eq:roll-residual}
\varepsilon_{t_1}=F-e^{-\Lambda_T}.
\end{equation}
A positive residual is cash released by the hedge, a negative residual requires a cash injection. We measure financing flows by the absolute residual. Calibration implies zero expected signed residual, so the expected positive injection alone is one half of the expected absolute residual.

In the bootstrap model, the residual is identically zero, so the strategy hedges the short position in the bootstrap model and is only a partial hedge in the market model.

For each factor realization, define the intensity paths
\begin{equation}\label{eq:random-intensity-paths}
a_s=\alpha_s\mathbf{1}_{\{s\le t_1\}}+\lambda_s\mathbf{1}_{\{s>t_1\}},
\qquad
b_s=\alpha_s\mathbf{1}_{\{s\le t_1\}}+\beta\mathbf{1}_{\{s>t_1\}},
\end{equation}
of the two models, and evaluate $\Psi_t(a,b)$, $\phi_t(a,b)$ and $\kappa_t(a,b)$ pathwise through \eqref{eq:Psi-exact}, \eqref{eq:frechet-mismatch} and Proposition~\ref{prop:adaptation-slack}. By Theorem~\ref{thm:aw-cox}, $\Psi_t(a,b)$ is then, for each factor realization, the squared adapted distance between the prices of the two Cox bonds maturing at $t$.

\begin{remark}[Meaning of $\mathbb{E}\Psi$ for random intensities]\label{rem:random-Psi}
The market and bootstrap models are driven by the same factor path. By Theorem~\ref{thm:default-layer} applied with the identical factor coupling, $\mathbb{E}_{\mathbb{Q}}[\Psi_T(a,b)]$ is the minimal default mismatch among doubly stochastic couplings of the two models over that factor coupling. It is therefore an upper bound for the right-hand side of \eqref{eq:default-layer}.
\end{remark}

\begin{theorem}[Cox mismatch bound for a model-transfer hedge]\label{thm:cox-hedge-error}
The expected absolute cash residual in the market model satisfies
\begin{align}
\mathbb{E}_{\mathbb{Q}}\bigl[\mathbf{1}_{\{\tau^A>t_1\}}|\varepsilon_{t_1}|\bigr]
&=\mathbb{E}_{\mathbb{Q}}\bigl|e^{-A_T}-e^{-B_T}\bigr|
\label{eq:roll-error-frechet}\\
&=\mathbb{E}_{\mathbb{Q}}[\Psi_T(a,b)]-2\,\mathbb{E}_{\mathbb{Q}}\Bigl[e^{-\max\{A_T,B_T\}}\bigl(1-e^{-\kappa_T(a,b)}\bigr)\Bigr]
\label{eq:roll-error-decomposition}\\
&\le\mathbb{E}_{\mathbb{Q}}[\Psi_T(a,b)].
\label{eq:roll-error-Psi-bound}
\end{align}
Equality holds in \eqref{eq:roll-error-Psi-bound} if and only if, almost surely, $\lambda_s\le\beta$ for a.e.\ $s\in[t_1,T]$ or $\lambda_s\ge\beta$ for a.e.\ $s\in[t_1,T]$.
\end{theorem}

\begin{proof}
The residual is $\mathcal R$-measurable, and conditionally on the factors, the probability of surviving to $t_1$ in the market model is $e^{-\Xi}$. Hence
\[
\mathbb{E}_{\mathbb{Q}}\bigl[\mathbf{1}_{\{\tau^A>t_1\}}|\varepsilon_{t_1}|\bigr]=\mathbb{E}_{\mathbb{Q}}\bigl[e^{-\Xi}|e^{-\overline\Lambda}-e^{-\Lambda_T}|\bigr]=\mathbb{E}_{\mathbb{Q}}\bigl|e^{-B_T}-e^{-A_T}\bigr|,
\]
which is \eqref{eq:roll-error-frechet}.

\medskip
The right-hand side is the expected Fr\'echet mismatch $\mathbb{E}_{\mathbb{Q}}[\phi_T(a,b)]$, so Proposition~\ref{prop:adaptation-slack}, applied pathwise, gives \eqref{eq:roll-error-decomposition} and \eqref{eq:roll-error-Psi-bound}.

\medskip
Finally, the subtracted term vanishes if and only if $\kappa_T(a,b)=0$ almost surely. Since the two intensity paths coincide before $t_1$, this is the stated ordering condition on $[t_1,T]$.
\end{proof}

\begin{corollary}[Tightness for rating-dependent flat hazards]\label{cor:flat-rating-tightness}
If the review selects a nonnegative random rate $L$ with $\lambda_s=L$ on $(t_1,T]$, then $\mathbb{E}_{\mathbb{Q}}[\mathbf{1}_{\{\tau^A>t_1\}}|\varepsilon_{t_1}|]=\mathbb{E}_{\mathbb{Q}}[\Psi_T(a,b)]$.
\end{corollary}

\begin{proof}
For each factor realization, the difference $L-\beta$ is constant in time, so the post-review intensities are ordered, and Theorem~\ref{thm:cox-hedge-error} gives equality.
\end{proof}

\begin{example}[Pure adaptation slack]\label{ex:pure-slack}
Let the post-review horizon have length one, let $\beta>0$ be the bootstrap rate, and let the market hazard be deterministic, $\lambda_s=\beta+c$ on the first half of the bucket and $\beta-c$ on the second half, with $0<c<\beta$. The two cumulative hazards agree at maturity, so the review-date prices of the $T$-bond coincide, and the roll residual and the Fr\'echet mismatch at $T$ both vanish. The intensities cross at the midpoint, $\kappa_T=c/2$, and Proposition~\ref{prop:adaptation-slack} gives
\begin{equation}\label{eq:pure-shape-Psi}
\Psi_T=2e^{-\Xi-\beta}\bigl(1-e^{-c/2}\bigr)>0,
\end{equation}
approximately $0.03656$ for $\beta=0.08$, $c=0.04$ and $\Xi=0$. A single rebalance at the bucket boundary sees no error, while the adapted default mismatch is positive, so the whole of \eqref{eq:pure-shape-Psi} is adaptation slack. By Theorem~\ref{thm:aw-cox}, for $\Xi=0$ it is the squared adapted distance between the prices of two bonds that have the same price today and the same terminal survival law. A claim depending only on terminal survival cannot detect it, while the bond maturing at mid-bucket can.
\end{example}

The calibration ratio $F$ is a mean under $\widehat{\mathbb{Q}}$ and is not, in general, the best notional under an absolute-error criterion.

\begin{proposition}[Forward sizing and loss-optimal sizing]\label{prop:forward-versus-median}
Let $Z=e^{-\Lambda_T}$ and, for $x\in\mathbb{R}$, $L_j(x)=\mathbb{E}_{\mathbb{Q}}[e^{-\Xi}|x-Z|^j]$, $j\in\{1,2\}$. The minimizers of $L_1$ are the $\widehat{\mathbb{Q}}$-medians of $Z$, and the unique minimizer of $L_2$ is $F=\widehat{\mathbb{E}} Z$. Hence the bootstrap notional is optimal for the weighted quadratic loss, and it is optimal for the weighted absolute loss if and only if $F$ is a $\widehat{\mathbb{Q}}$-median of $Z$.
\end{proposition}

\begin{proof}
By the definition of $\widehat{\mathbb{Q}}$, $L_j(x)=\mathbb{E}_{\mathbb{Q}}[e^{-\Xi}]\,\widehat{\mathbb{E}}|x-Z|^j$, so it suffices to minimize $\widehat{\mathbb{E}}|x-Z|^j$. For $j=2$, this is the variance identity, which gives the unique minimizer $\widehat{\mathbb{E}} Z=F$. For $j=1$, the function $x\mapsto\widehat{\mathbb{E}}|x-Z|$ is convex with right and left derivatives $2\widehat{\mathbb{Q}}[Z\le x]-1$ and $2\widehat{\mathbb{Q}}[Z<x]-1$, so its minimizers are exactly the medians.
\end{proof}

\begin{proposition}[The roll loss as a forward credit straddle]\label{prop:forward-credit-straddle}
The survival-contingent payoff $\mathbf{1}_{\{\tau^A>t_1\}}|\widetilde P^A_{t_1}(T)-F|$ at $t_1$ has time-zero value in the market model equal to the expected absolute roll residual of Theorem~\ref{thm:cox-hedge-error}, and value zero in the bootstrap model. Moreover, with $Z=e^{-\Lambda_T}$,
\begin{equation}\label{eq:straddle-equal-legs}
\mathbb{E}_{\mathbb{Q}}\bigl[e^{-\Xi}(Z-F)_+\bigr]=\mathbb{E}_{\mathbb{Q}}\bigl[e^{-\Xi}(F-Z)_+\bigr]=\tfrac12\mathbb{E}_{\mathbb{Q}}\bigl[e^{-\Xi}|Z-F|\bigr].
\end{equation}
\end{proposition}

\begin{proof}
The payoff equals $\mathbf{1}_{\{\tau^A>t_1\}}|\varepsilon_{t_1}|$, so its value in the market model is the left-hand side of \eqref{eq:roll-error-frechet}. In the bootstrap model, the forward price is the constant $F$, and the payoff vanishes. Finally, calibration gives $\mathbb{E}_{\mathbb{Q}}[e^{-\Xi}(Z-F)]=\mathbb{E}_{\mathbb{Q}}[e^{-\Xi}](\widehat{\mathbb{E}} Z-F)=0$, so the two legs in \eqref{eq:straddle-equal-legs} have equal value.
\end{proof}

\subsection{The survival-filtered projection}\label{subsec:filtered-projection}

The flat bootstrap is calibrated only at the pins. The next result identifies the deterministic post-review hazard that reproduces the entire survival curve of the market model, which is the posterior-mean hazard given survival. That the intensity of a default time in a smaller filtration is the projection of its intensity onto that filtration is standard in credit risk with partial information \cite{ElliottJeanblancYor2000,BieleckiRutkowski2002,DuffieLando2001,FreyRoeslerLu2019}. Here it is the Markov projection of the mixture of rating scenarios, in analogy with the posterior-mean drift of the arithmetic pool of diffusion experts in \cite[Prop.~3.3]{Vecer2026}.

\begin{proposition}[Survival-filtered projection]\label{prop:filtered-projection}
For $u\in[t_1,T]$ let $\mathsf S(u)=\widehat{\mathbb{E}}[e^{-\Lambda_u}]$ and
\begin{equation}\label{eq:filtered-hazard}
h_u=\frac{\widehat{\mathbb{E}}\bigl[\lambda_ue^{-\Lambda_u}\bigr]}{\mathsf S(u)} .
\end{equation}
Then the following hold.
\begin{enumerate}
\item\label{it:fp-curve} $h\in L^1([t_1,T])$ and $\mathsf S(u)=\exp(-\int_{t_1}^uh_s\,\mathrm{d} s)$. In particular $\int_{t_1}^Th_s\,\mathrm{d} s=\overline\Lambda$.
\item\label{it:fp-calibrated} The model $B^\sharp$ with intensity $b^\sharp_s=\alpha_s\mathbf{1}_{\{s\le t_1\}}+h_s\mathbf{1}_{\{s>t_1\}}$ satisfies $P^{B^\sharp}_0(u)=P^A_0(u)$ for every $u\in[0,T]$, and $h$ is the hazard rate of $\tau^A$ on $(t_1,T]$ under $\mathbb{Q}[\,\cdot\mid\tau^A>t_1]$.
\item\label{it:fp-unbiased} The bootstrap roll has the same residual under $B^\sharp$ as under the bootstrap model, so \eqref{eq:roll-error-frechet} is unchanged. Moreover, for every linear claim of the form \eqref{eq:linear-credit-value} the expected signed model-transfer residual vanishes, $\mathbb{E}_{\mathbb{Q}}[\mathbf{1}_{\{\tau^A>t_1\}}(V^{B^\sharp}_{t_1}-V^A_{t_1})]=0$.
\item\label{it:fp-crossing} Assume $\lambda_s=L$ on $(t_1,T]$ with $L$ bounded and not $\widehat{\mathbb{Q}}$-almost surely constant. Then $h$ is continuous and strictly decreasing, with $h_{t_1}=\widehat{\mathbb{E}} L$, and $\kappa_T(a,b^\sharp)>0$ exactly on the event $\{h_T<L<h_{t_1}\}$. In particular, $\mathbb{E}_{\mathbb{Q}}[\Psi_T(a,b^\sharp)]>\mathbb{E}_{\mathbb{Q}}[\mathbf{1}_{\{\tau^A>t_1\}}|\varepsilon_{t_1}|]$ if and only if $\widehat{\mathbb{Q}}[h_T<L<\widehat{\mathbb{E}} L]>0$. If $L$ takes only two values, this event is empty.
\end{enumerate}
\end{proposition}

\begin{proof}
\emph{Proof of~\ref{it:fp-curve}.} Since $\int_{t_1}^T\lambda_ue^{-\Lambda_u}\,\mathrm{d} u=1-e^{-\Lambda_T}\le1$, Tonelli's theorem gives $\mathsf S(u)=1-\int_{t_1}^u\widehat{\mathbb{E}}[\lambda_se^{-\Lambda_s}]\,\mathrm{d} s$. Hence $\mathsf S$ is absolutely continuous with $\mathsf S'=-h\mathsf S$ almost everywhere, and $\mathsf S\ge F>0$ on $[t_1,T]$. It follows that $h\in L^1$ and $\mathsf S=\exp(-\int h)$.

\medskip
\emph{Proof of~\ref{it:fp-calibrated}.} For $u\le t_1$ both prices equal $\mathbb{E}_{\mathbb{Q}}[e^{-\int_0^u\alpha}]$. For $u>t_1$, part~\ref{it:fp-curve} gives $P^{B^\sharp}_0(u)=\mathbb{E}_{\mathbb{Q}}[e^{-\Xi}]\mathsf S(u)=\mathbb{E}_{\mathbb{Q}}[e^{-\Xi-\Lambda_u}]=P^A_0(u)$. By Remark~\ref{rem:hatQ}, $\mathbb{Q}[\tau^A>u\mid\tau^A>t_1]=\mathsf S(u)$, whose hazard rate is $h$.

\medskip
\emph{Proof of~\ref{it:fp-unbiased}.} The first claim follows from $\int_{t_1}^Th_s\,\mathrm{d} s=\overline\Lambda$, which shows that $B^\sharp$ and the bootstrap model assign the same survival factor to the bucket. For the second, the computation in the proof of Theorem~\ref{thm:general-instrument-Psi-bound} expresses the signed residual as
\[
\int\mathbb{E}_{\mathbb{Q}}\bigl[e^{-B^\sharp_u}-e^{-A_u}\bigr]\,\nu(\mathrm{d} u)=\int\bigl(P^{B^\sharp}_0(u)-P^A_0(u)\bigr)\,\nu(\mathrm{d} u),
\]
which vanishes by part~\ref{it:fp-calibrated}.

\medskip
\emph{Proof of~\ref{it:fp-crossing}.} Let $\widehat{\mathbb{Q}}_u$ have density proportional to $e^{-L(u-t_1)}$ with respect to $\widehat{\mathbb{Q}}$. These measures are equivalent to $\widehat{\mathbb{Q}}$, and $h_u=\widehat{\mathbb{E}}_u[L]$ with $h'_u=-\operatorname{Var}_{\widehat{\mathbb{Q}}_u}(L)<0$. For a given value of $L$, the function $s\mapsto L-h_s$ is therefore continuous and strictly increasing on $[t_1,T]$, and it has constant sign if and only if $L\ge h_{t_1}$ or $L\le h_T$. By Proposition~\ref{prop:adaptation-slack}, $\kappa_T>0$ if and only if this function changes sign, that is, if and only if $h_T<L<h_{t_1}$. The strict inequality then follows from \eqref{eq:roll-error-decomposition} with $b^\sharp$ in place of $b$. Finally, if $L\in\{\ell_1,\ell_2\}$ with $\ell_1<\ell_2$, then $h_s\in(\ell_1,\ell_2)$ for all $s$, so no value of $L$ lies strictly between $h_T$ and $h_{t_1}$.
\end{proof}

Proposition~\ref{prop:filtered-projection} exhibits a genuine trade-off. For flat rating-dependent market hazards, the flat bootstrap misprices intermediate survival but keeps the hazard paths ordered, so that the adapted bound is tight. The filtered projection is unbiased for every linear claim, but intermediate ratings cross its hazard and generate strictly positive adaptation slack. A numerical example is given in Example~\ref{ex:three-ratings}.

\begin{example}[Three ratings]\label{ex:three-ratings}
Take a five-year bucket, normalize $e^{-\Xi}=1$, and let the review select one of three flat hazards $L\in\{0.02,\,0.06,\,0.20\}$ with probabilities $\{0.5,\,0.3,\,0.2\}$. Then $F\approx0.748240$ and the flat bootstrap rate is $\beta\approx0.058006$. The survival-filtered hazard of Proposition~\ref{prop:filtered-projection} decreases from $h_{t_1}=\widehat{\mathbb{E}} L=0.068$ to $h_T\approx0.049581$ and crosses the middle rating $L=0.06$ about $1.87$ years into the bucket. The expected absolute roll residual is $\approx0.156597$ under both reference models. Under the flat bootstrap, all three hazards are ordered relative to $\beta$ and $\mathbb{E}[\Psi_T(a,b)]\approx0.156597$ is exact, but the model misprices intermediate survival, for instance $0.865009$ against the market value $0.855133$ at mid-bucket. Under the filtered projection, all bucket survival probabilities are exact, but $\mathbb{E}[\Psi_T(a,b^\sharp)]\approx0.159812$. The adaptation slack $\approx0.003215$ comes entirely from the middle rating, whose pathwise mismatch rises from $0.007422$ to $0.018138$.
\end{example}

\subsection{Linear credit values and claim-weighted mismatch}\label{subsec:linear-credit-mismatch}

Let $\nu$ be a finite signed Borel measure on $[t_1,T]$ and $c\in\mathbb{R}$. Consider pre-default values
\begin{equation}\label{eq:linear-credit-value}
V^i_{t_1}=c+\int_{[t_1,T]}\widetilde P^i_{t_1}(u)\,\nu(\mathrm{d} u),
\qquad i\in\{A,B\}.
\end{equation}
This class contains zero-recovery coupon bonds with deterministic discounted cash flows and, after integration by parts in the maturity variable, premium and protection legs with deterministic recovery \cite{Schonbucher2003,OKane2008}. Define the model-transfer residual $\varepsilon^\nu_{t_1}=V^B_{t_1}-V^A_{t_1}$, let $|\nu|$ be the total-variation measure and $s_\nu=\mathrm{d}\nu/\mathrm{d}|\nu|$ its polar density, and, for each factor realization, set
\begin{equation}\label{eq:d-u-definition}
d_u=e^{-B_u}-e^{-A_u},
\qquad
\Delta_\nu=\int_{[t_1,T]}|d_u|\,|\nu|(\mathrm{d} u)-\Bigl|\int_{[t_1,T]}d_u\,\nu(\mathrm{d} u)\Bigr|\ \ge0 .
\end{equation}

\begin{theorem}[Claim-weighted adapted mismatch bound]\label{thm:general-instrument-Psi-bound}
The expected absolute model-transfer residual satisfies
\begin{align}
\mathbb{E}_{\mathbb{Q}}\bigl[\mathbf{1}_{\{\tau^A>t_1\}}|\varepsilon^\nu_{t_1}|\bigr]
&=\mathbb{E}_{\mathbb{Q}}\Bigl|\int_{[t_1,T]}d_u\,\nu(\mathrm{d} u)\Bigr|
\label{eq:general-error-identity}\\
&\le\int_{[t_1,T]}\mathbb{E}_{\mathbb{Q}}\bigl|e^{-A_u}-e^{-B_u}\bigr|\,|\nu|(\mathrm{d} u)
\label{eq:general-error-frechet-bound}\\
&\le\mathbb{E}_{\mathbb{Q}}\bigl[\Phi_{|\nu|}(a,b)\bigr],
\label{eq:general-error-Psi-bound}
\end{align}
and the total slack admits the exact decomposition
\begin{equation}\label{eq:general-slack-decomposition}
\mathbb{E}_{\mathbb{Q}}\bigl[\Phi_{|\nu|}(a,b)\bigr]-\mathbb{E}_{\mathbb{Q}}\bigl[\mathbf{1}_{\{\tau^A>t_1\}}|\varepsilon^\nu_{t_1}|\bigr]
=\mathbb{E}_{\mathbb{Q}}[\Delta_\nu]+2\int_{[t_1,T]}\mathbb{E}_{\mathbb{Q}}\Bigl[e^{-\max\{A_u,B_u\}}\bigl(1-e^{-\kappa_u(a,b)}\bigr)\Bigr]|\nu|(\mathrm{d} u).
\end{equation}
Equality holds in \eqref{eq:general-error-Psi-bound} if and only if both terms on the right of \eqref{eq:general-slack-decomposition} vanish.
\end{theorem}

\begin{proof}
By \eqref{eq:review-date-predefault-prices}, the residual $\varepsilon^\nu_{t_1}=\int[\widetilde P^B_{t_1}(u)-\widetilde P^A_{t_1}(u)]\,\nu(\mathrm{d} u)$ is $\mathcal R$-measurable. Conditioning on the factors replaces $\mathbf{1}_{\{\tau^A>t_1\}}$ by $e^{-\Xi}$, and moving this positive factor inside the integral, with $e^{-\Xi}\widetilde P^A_{t_1}(u)=e^{-A_u}$ and $e^{-\Xi}\widetilde P^B_{t_1}(u)=e^{-B_u}$, gives \eqref{eq:general-error-identity}.

\medskip
The triangle inequality and Tonelli's theorem then give \eqref{eq:general-error-frechet-bound}, and their slack is exactly $\mathbb{E}_{\mathbb{Q}}[\Delta_\nu]$. Finally, Proposition~\ref{prop:adaptation-slack}, integrated against $|\nu|$, gives \eqref{eq:general-error-Psi-bound} together with the second term in \eqref{eq:general-slack-decomposition}.
\end{proof}

\begin{corollary}[Equality conditions for linear credit values]\label{cor:general-equality-conditions}
$\Delta_\nu=0$ if and only if there is a random sign $\xi\in\{-1,1\}$ with $\xi\,d_u\,s_\nu(u)\ge0$ for $|\nu|$-a.e.\ $u$, almost surely. The adaptation term vanishes if and only if $\kappa_u(a,b)=0$ for $|\nu|\otimes\mathbb{Q}$-a.e.\ $(u,\omega)$. For $\nu=\delta_T$ the first condition is automatic and Theorem~\ref{thm:general-instrument-Psi-bound} reduces to Theorem~\ref{thm:cox-hedge-error}.
\end{corollary}

\begin{proof}
Equality in $|\int f\,\mathrm{d}|\nu||\le\int|f|\,\mathrm{d}|\nu|$ holds if and only if $f=d\,s_\nu$ has constant sign $|\nu|$-almost everywhere, which gives the first statement. The adaptation integrand is nonnegative, and its exponential factor is strictly positive, which gives the second.
\end{proof}

The weight in the adapted default mismatch is thus the total variation of the claim's cash-flow measure. By Corollary~\ref{cor:weighted-mismatch-representation}, the bound \eqref{eq:general-error-Psi-bound} is, for each factor realization, a $|\nu|$-weighted sum of squared adapted distances between the bond prices of the two models. The two sources of conservatism, cancellation across maturities and hazard crossings, are separated exactly.

\begin{remark}[Several reset dates and richer hedging universes]\label{rem:several-reset-dates}
Let $R_1,\ldots,R_m$ be discounted residuals of model-transfer resets at deterministic dates, each bounded as in \eqref{eq:general-error-Psi-bound} by $\mathbb{E}[\Phi^{k}_{|\nu_k|}]$. Then $\mathbb{E}|\sum_kR_k|\le\sum_k\mathbb{E}[\Phi^k_{|\nu_k|}]$ by the triangle inequality. When all resets are controlled by the same pathwise mismatch, the right-hand side equals $\mathbb{E}[\Phi_\theta]$ with $\theta=\sum_k|\nu_k|$, by linearity of $\theta\mapsto\Phi_\theta$. The bound also remains valid after enlarging the hedging universe, since the smallest attainable residual cannot exceed that of the bootstrap roll. It may then become loose, because additional instruments span part of the rating-induced state space.
\end{remark}

\subsection{A full term structure: the rolling bootstrap}\label{subsec:term-structure}

We now replace the single bucket by a quoted term structure. Let $0=t_0<t_1<\cdots<t_K=T$ be the calibration pins. The market model is now a general Cox model with nonnegative $\mathbb F^F$-predictable intensity $a$ and cumulative hazard $A_t=\int_0^ta_s\,\mathrm{d} s$, where $A_T<\infty$ almost surely, so that information may arrive at any time. The bootstrap model has the deterministic piecewise-flat hazard
\begin{equation}\label{eq:piecewise-bootstrap}
b_s=\beta_k\ \text{ on }(t_{k-1},t_k],
\qquad
e^{-\beta_k(t_k-t_{k-1})}=\frac{P^A_0(t_k)}{P^A_0(t_{k-1})},
\end{equation}
and therefore reproduces all quoted prices $P^A_0(t_k)$. On survival, the pre-default price at $t_k$ of the bond maturing at $u$ is $\widetilde P^A_{t_k}(u)=\mathbb{E}_{\mathbb{Q}}[e^{-(A_u-A_{t_k})}\mid\mathcal{F}^F_{t_k}]$ in the market model and $e^{-(B_u-B_{t_k})}$ in the bootstrap model.

\medskip
\emph{The rolling hedge.} Sell one $T$-bond at time zero. At each pin $t_{k-1}$, on survival, hold $N_k=e^{-(B_T-B_{t_k})}$ units of the bond maturing at $t_k$, which makes the position worthless in the bootstrap model. At $t_k$, the maturing bonds pay $N_k$, and the desk buys $N_{k+1}$ bonds maturing at $t_{k+1}$ at the market price. The cash residual at $t_k$, $1\le k\le K-1$, is
\begin{equation}\label{eq:rolling-residual}
\varepsilon_k=N_k-N_{k+1}\widetilde P^A_{t_k}(t_{k+1})
=N_{k+1}\Bigl[e^{-\beta_{k+1}(t_{k+1}-t_k)}-\widetilde P^A_{t_k}(t_{k+1})\Bigr],
\end{equation}
and at $t_K$ the long position settles the short bond. In the bootstrap model every residual vanishes, and for $K=2$ this is the roll of Section~\ref{subsec:bootstrap-roll}.

For $0\le k<K$ let $b^{(k)}$ be the \emph{hybrid intensity} that follows the market model up to $t_k$ and the bootstrap model afterwards, $b^{(k)}_s=a_s\mathbf{1}_{\{s\le t_k\}}+b_s\mathbf{1}_{\{s>t_k\}}$, with cumulative hazard $B^{(k)}$.

\begin{theorem}[Bucket-by-bucket decomposition]\label{thm:term-structure}
Each residual has zero mean, namely $\mathbb{E}_{\mathbb{Q}}[\mathbf{1}_{\{\tau^A>t_k\}}\varepsilon_k]$ $=0$, and
\begin{align}
\mathbb{E}_{\mathbb{Q}}\bigl[\mathbf{1}_{\{\tau^A>t_k\}}|\varepsilon_k|\bigr]
&=N_{k+1}\,\mathbb{E}_{\mathbb{Q}}\Bigl|\mathbb{E}_{\mathbb{Q}}\bigl[e^{-A_{t_{k+1}}}-e^{-B^{(k)}_{t_{k+1}}}\bigm|\mathcal{F}^F_{t_k}\bigr]\Bigr|
\label{eq:ts-identity}\\
&\le N_{k+1}\,\mathbb{E}_{\mathbb{Q}}\bigl|e^{-A_{t_{k+1}}}-e^{-B^{(k)}_{t_{k+1}}}\bigr|
\le N_{k+1}\,\mathbb{E}_{\mathbb{Q}}\bigl[\Psi_{t_{k+1}}(a,b^{(k)})\bigr].
\label{eq:ts-bound}
\end{align}
Consequently, the total expected absolute cash residual satisfies
\begin{equation}\label{eq:ts-total}
\sum_{k=1}^{K-1}\mathbb{E}_{\mathbb{Q}}\bigl[\mathbf{1}_{\{\tau^A>t_k\}}|\varepsilon_k|\bigr]
\le\sum_{k=1}^{K-1}N_{k+1}\,\mathbb{E}_{\mathbb{Q}}\bigl[\Psi_{t_{k+1}}(a,b^{(k)})\bigr].
\end{equation}
The first inequality in \eqref{eq:ts-bound} is an equality if $A_{t_{k+1}}-A_{t_k}$ is $\mathcal{F}^F_{t_k}$-measurable, that is, if the hazard on each bucket is revealed at its left pin. The second is an equality if and only if, almost surely, $a-\beta_{k+1}$ does not change sign on $(t_k,t_{k+1}]$. If both conditions hold for every $k$, which is the case for rating reviews at the pins selecting flat hazards on the following bucket, then \eqref{eq:ts-total} holds with equality.
\end{theorem}

\begin{proof}
\emph{The identity.} Conditionally on $\mathcal{F}^F_T$, the probability of surviving to $t_k$ in the market model is $e^{-A_{t_k}}$, and $\varepsilon_k$ is $\mathcal{F}^F_{t_k}$-measurable. Hence
\[
\mathbb{E}_{\mathbb{Q}}\bigl[\mathbf{1}_{\{\tau^A>t_k\}}|\varepsilon_k|\bigr]=N_{k+1}\,\mathbb{E}_{\mathbb{Q}}\Bigl[e^{-A_{t_k}}\bigl|e^{-\beta_{k+1}(t_{k+1}-t_k)}-\widetilde P^A_{t_k}(t_{k+1})\bigr|\Bigr].
\]
Since $e^{-A_{t_k}}$ is $\mathcal{F}^F_{t_k}$-measurable, it can be moved inside the conditional expectation that defines $\widetilde P^A_{t_k}(t_{k+1})$. Together with $B^{(k)}_{t_{k+1}}=A_{t_k}+\beta_{k+1}(t_{k+1}-t_k)$, this gives \eqref{eq:ts-identity}. The same computation without absolute values, combined with \eqref{eq:piecewise-bootstrap}, gives $N_{k+1}[P^A_0(t_k)e^{-\beta_{k+1}(t_{k+1}-t_k)}-P^A_0(t_{k+1})]=0$, so each residual has zero mean.

\medskip
\emph{The bounds.} The first inequality in \eqref{eq:ts-bound} is the conditional Jensen inequality, with equality when the integrand is $\mathcal{F}^F_{t_k}$-measurable. The second follows from Proposition~\ref{prop:adaptation-slack}, applied pathwise to $(a,b^{(k)})$, which coincide on $[0,t_k]$, together with its equality criterion. Summing over $k$ gives \eqref{eq:ts-total}.
\end{proof}

The decomposition is a telescoping comparison, bucket by bucket, of the market model with the hybrid that switches to the bootstrap model one bucket ahead. Each term separates three effects: information that arrives inside a bucket and is averaged by the conditional expectation in \eqref{eq:ts-identity}, the Fr\'echet mismatch visible at the next pin, and the adaptation slack produced by hazard crossings inside the bucket. None of them is detected by the calibration \eqref{eq:piecewise-bootstrap}.

\subsection{Hedging with a window of liquid bonds}\label{subsec:window}

The rolling bootstrap trades only the bond maturing at the next pin. Suppose now that at $t_k$ the desk can trade every bond maturing at $t_{k+1},\ldots,t_{\min\{k+m,K\}}$, for a \emph{liquidity window} $m\ge1$. We compare two desks acting on this universe: a bootstrap desk, which still relies on the bootstrap model, and a market desk, which knows the market model.

The bootstrap desk holds, between $t_{k-1}$ and $t_k$, the longest available bond, namely $N_{j(k)}$ units of the bond maturing at $t_{j(k)}$ with
\[
j(k)=\min\{k-1+m,K\},\qquad N_j=e^{-(B_T-B_{t_j})}.
\]
At $t_k$, on survival, it sells this position at the market price and buys $N_{j(k+1)}$ units of the bond maturing at $t_{j(k+1)}$. In the bootstrap model the position is worthless at every pin, so the strategy hedges the short $T$-bond position there. In the market model, it leaves the cash residual
\begin{equation}\label{eq:window-residual}
\varepsilon^{(m)}_k=N_{j(k)}\widetilde P^A_{t_k}(t_{j(k)})-N_{j(k+1)}\widetilde P^A_{t_k}(t_{j(k+1)}),\qquad 1\le k\le K-1,
\end{equation}
with the convention $\widetilde P^A_{t_k}(t_k)=1$. For $m=1$ this is \eqref{eq:rolling-residual}, and for $m\ge K$ the $T$-bond is bought at time zero and all residuals vanish.

\begin{theorem}[Window bootstrap]\label{thm:window}
For $u\in\{t_k,\ldots,T\}$ let $D_k(u)=e^{-(B_u-B_{t_k})}-\widetilde P^A_{t_k}(u)$ be the bootstrap mispricing at $t_k$, and let
\[
\Delta_k=N_{j(k+1)}\bigl(e^{-B^{(k)}_{t_{j(k+1)}}}-e^{-A_{t_{j(k+1)}}}\bigr)-N_{j(k)}\bigl(e^{-B^{(k)}_{t_{j(k)}}}-e^{-A_{t_{j(k)}}}\bigr),
\]
where the second term is absent if $j(k)=k$. Then $\varepsilon^{(m)}_k=N_{j(k+1)}D_k(t_{j(k+1)})-N_{j(k)}D_k(t_{j(k)})$, $\mathbb{E}_{\mathbb{Q}}[\mathbf{1}_{\{\tau^A>t_k\}}\varepsilon^{(m)}_k]=0$, and
\begin{equation}\label{eq:window-bound}
\begin{aligned}
\mathbb{E}_{\mathbb{Q}}\bigl[\mathbf{1}_{\{\tau^A>t_k\}}|\varepsilon^{(m)}_k|\bigr]
&=\mathbb{E}_{\mathbb{Q}}\bigl|\mathbb{E}_{\mathbb{Q}}[\Delta_k\mid\mathcal{F}^F_{t_k}]\bigr|
\le\mathbb{E}_{\mathbb{Q}}|\Delta_k|\\
&\le N_{j(k+1)}\mathbb{E}_{\mathbb{Q}}\bigl[\Psi_{t_{j(k+1)}}(a,b^{(k)})\bigr]+N_{j(k)}\mathbb{E}_{\mathbb{Q}}\bigl[\Psi_{t_{j(k)}}(a,b^{(k)})\bigr].
\end{aligned}
\end{equation}
\end{theorem}

\begin{proof}
\emph{Form of the residual.} Since $N_je^{-(B_{t_j}-B_{t_k})}$ $=e^{-(B_T-B_{t_k})}$ for every $j$, we have $N_j\widetilde P^A_{t_k}(t_j)=e^{-(B_T-B_{t_k})}-N_jD_k(t_j)$. The common term cancels in \eqref{eq:window-residual}, which gives the stated expression for $\varepsilon^{(m)}_k$.

\medskip
\emph{The identity and the bounds.} By Cox pricing, we have $e^{-A_{t_k}}\widetilde P^A_{t_k}(u)$ $=\mathbb{E}_{\mathbb{Q}}[e^{-A_u}\mid\mathcal{F}^F_{t_k}]$, while $e^{-A_{t_k}}e^{-(B_u-B_{t_k})}$ $=e^{-B^{(k)}_u}$ is $\mathcal{F}^F_{t_k}$-measurable. Hence $e^{-A_{t_k}}\varepsilon^{(m)}_k=\mathbb{E}_{\mathbb{Q}}[\Delta_k\mid\mathcal{F}^F_{t_k}]$, and conditioning on $\mathcal{F}^F_T$ as in the proof of Theorem~\ref{thm:term-structure} gives the identity in \eqref{eq:window-bound}. The first inequality is the conditional Jensen inequality, and the second follows from the triangle inequality and Proposition~\ref{prop:adaptation-slack}, applied pathwise to $(a,b^{(k)})$.

\medskip
\emph{Zero mean.} The bootstrap model is calibrated at all pins, so $\mathbb{E}_{\mathbb{Q}}[e^{-B^{(k)}_u}]=P^A_0(t_k)e^{-(B_u-B_{t_k})}=P^A_0(u)=\mathbb{E}_{\mathbb{Q}}[e^{-A_u}]$ for every pin $u$. Hence $\mathbb{E}_{\mathbb{Q}}[\Delta_k]=0$.
\end{proof}

The last inequality in \eqref{eq:window-bound} discards the cancellation between the two maturities, which is the term $\Delta_\nu$ of Theorem~\ref{thm:general-instrument-Psi-bound} for the signed cash-flow measure $\nu=N_{j(k+1)}\delta_{t_{j(k+1)}}-N_{j(k)}\delta_{t_{j(k)}}$. For $m\ge2$ this cancellation is substantial, and the pathwise bound $\mathbb{E}_{\mathbb{Q}}|\Delta_k|$ is the informative one (Section~\ref{subsec:empirical-multi}).

The market desk needs a model in which the pre-default prices at the pins are computable. We call the market model \emph{rating driven} if there is a process $(R_{t_k})_{k\le K}$ with values in a finite set $\{1,\ldots,n\}$ of non-default ratings such that, conditionally on $\mathcal{F}^F_{t_k}$, the law of $\bigl((a_s)_{s>t_k},(R_{t_l})_{l>k}\bigr)$ depends only on $R_{t_k}$, in the same way for every $k$. Both rating models of Section~\ref{subsec:empirical-hedge} are rating driven. Put
\[
\widehat K(i,i')=\mathbb{E}_{\mathbb{Q}}\bigl[e^{-(A_{t_{k+1}}-A_{t_k})}\mathbf{1}_{\{R_{t_{k+1}}=i'\}}\bigm|R_{t_k}=i\bigr],
\]
the survival-weighted one-period kernel. The tower property and the Cox pricing formula give, by induction on $j-k$,
\begin{equation}\label{eq:rating-prices}
\widetilde P^A_{t_k}(t_j)=\bigl(\widehat K^{\,j-k}\mathbf{1}\bigr)(R_{t_k}),\qquad 0\le k\le j\le K,
\end{equation}
where $\mathbf{1}$ is the vector of ones. At each pin the market desk holds a portfolio of the window bonds whose market value equals the market value $\widetilde P^A_{t_k}(T)$ of the liability, and it chooses the weights to minimize the conditional second moment of the mark-to-market error at $t_{k+1}$,
\[
e_{k+1}=\sum_j\theta_j\widetilde P^A_{t_{k+1}}(t_j)-\widetilde P^A_{t_{k+1}}(T)\quad\text{on }\{\tau^A>t_{k+1}\}.
\]
Because all positions have zero recovery and the portfolio is value matched, a default inside the bucket causes no error. The corresponding cash surplus is released, or the shortfall funded, when the portfolio is reset.

\begin{proposition}[Spanning by short bonds]\label{prop:spanning}
Let the market model be rating driven, and let $d\le n$ be the dimension of the Krylov space $\operatorname{span}\{\widehat K^{\,l}\mathbf{1}:l\ge0\}\subseteq\mathbb{R}^n$. If $m\ge d$, the market desk hedges the short $T$-bond position exactly: every error $e_{k+1}$ vanishes.
\end{proposition}

\begin{proof}
If $T$ lies in the window, the desk holds the $T$-bond itself. Otherwise, by \eqref{eq:rating-prices}, the window bonds bought at $t_k$ have prices $\widehat K^{\,l}\mathbf{1}$, $l=0,\ldots,m-1$, at $t_{k+1}$, and the liability has price $\widehat K^{\,K-k-1}\mathbf{1}$, as functions of $R_{t_{k+1}}$. By the Cayley-Hamilton theorem the Krylov space is spanned by its first $d$ vectors, so there are weights $\theta_l$ with $\sum_{l<m}\theta_l\widehat K^{\,l}\mathbf{1}=\widehat K^{\,K-k-1}\mathbf{1}$, and this portfolio has zero error in every state at $t_{k+1}$. Finally, applying $\widehat K$ to this identity shows that its value at $t_k$ equals that of the liability, so it is value matched and hence optimal.
\end{proof}

With finitely many ratings, zero-coupon bonds of $d$ consecutive maturities therefore span the long-bond liability at the pins. Completeness of a finite-state Markov chain market additionally requires that traded payoffs span the relevant state space: see Norberg \cite{Norberg2003} for the Markov chain market in continuous time and \cite{BieleckiRutkowski2002,BieleckiJeanblancRutkowski2004} for hedging in credit migration and intensity models. Proposition~\ref{prop:spanning} gives the corresponding discrete-time spanning statement for bond liabilities, and we state it because it identifies exactly which bonds span. The error of the market desk for smaller windows is governed by how well $\widehat K^{\,K-k-1}\mathbf{1}$ is approximated by the first $m$ Krylov vectors. The bootstrap desk cannot exploit this spanning, because its hedge ratios ignore the rating state. Comparing the two specified strategies on the same universe measures the improvement achieved by using the market model, it is not an intrinsic decomposition of the absolute hedging loss.

\subsection{Calibration-preserving ambiguity}\label{subsec:ambiguity-set}

Fix $p\in(0,1)$ and $\eta\ge0$, and define
\begin{equation}\label{eq:mu-eta-class}
\mathfrak U_\eta(p)=\Bigl\{\mu\in\mathcal P([0,1]):\int q\,\mu(\mathrm{d} q)=p,\ \int|q-p|\,\mu(\mathrm{d} q)\le\eta\Bigr\}.
\end{equation}
The mean constraint preserves the time-zero bond price, and the second constraint limits the amount of information released by the review.

\begin{proposition}[Adapted balls are posterior-dispersion balls]\label{prop:ball-equivalence}
For every $\eta\ge0$,
\begin{equation}\label{eq:ball-equivalence}
\bigl\{\mathbb{P}\in\mathcal{M}_2(p):\mathcal{AW}_2(\mathbb{P}^0,\mathbb{P})\le\sqrt\eta\bigr\}=\bigl\{\mathbb{P}^{\mu}:\mu\in\mathfrak U_\eta(p)\bigr\}.
\end{equation}
Moreover,
\begin{equation}\label{eq:max-mad}
\sup_{\mu:\int q\,\mathrm{d}\mu=p}\int|q-p|\,\mu(\mathrm{d} q)=2p(1-p),
\end{equation}
and the supremum is attained by $(1-p)\delta_0+p\delta_1$.
\end{proposition}

\begin{proof}
By Remark~\ref{rem:M2-exhausted}, every $\mathbb{P}\in\mathcal{M}_2(p)$ is of the form $\mathbb{P}^\mu$ for a law $\mu$ with mean $p$, and by Theorem~\ref{thm:exact-aw-one-rating}, $\mathcal{AW}_2^2(\mathbb{P}^0,\mathbb{P}^\mu)=\int|q-p|\,\mu(\mathrm{d} q)$. This gives \eqref{eq:ball-equivalence}.

\medskip
For \eqref{eq:max-mad}, let $Q\in[0,1]$ have mean $p$. Since $\mathbb{E}(Q-p)=0$, the positive and negative parts of $Q-p$ have the same expectation, so $\mathbb{E}(Q-p)_+=\frac12\mathbb{E}|Q-p|$. On $[0,1]$ we have $(q-p)_+\le(1-p)q$, since for $q>p$ the difference is $p(1-q)\ge0$. Taking expectations gives $\frac12\mathbb{E}|Q-p|\le(1-p)p$, and the Bernoulli law $(1-p)\delta_0+p\delta_1$ attains equality.
\end{proof}

Only the range $0\le\eta\le2p(1-p)$ is therefore nontrivial. Larger values of $\eta$ allow every calibrated posterior.

\label{subsec:robust-call}
For a strike $k\in[0,1]$ consider the rating-date call $C_k(Q)=(Q-k)_+$, whose price in the coarse model is $(p-k)_+$. Its worst case over the ball \eqref{eq:ball-equivalence} is an instance of a classical moment problem. For a convex function of a random variable on an interval with given mean and mean absolute deviation, Ben-Tal and Hochman identified the tight bounds on its expectation, the upper one being attained by a three-point law on the endpoints and the mean \cite{BenTalHochman1972}. These bounds are used in robust optimization under mean-dispersion ambiguity in \cite{PostekEtAl2018}. For the rating-date call, they give the following explicit formula. Since we constrain the mean absolute deviation by an inequality rather than an equality, we include the short direct proof.

\begin{theorem}[Robust rating-call bounds]\label{thm:robust-call}
Let $0\le\eta\le2p(1-p)$. Then
\begin{equation}\label{eq:robust-call-formula}
\sup_{\mu\in\mathfrak U_\eta(p)}\int_0^1(q-k)_+\,\mu(\mathrm{d} q)=
\begin{cases}
p-k+\dfrac{\eta k}{2p}, & 0\le k\le p,\\[0.9em]
\dfrac{\eta(1-k)}{2(1-p)}, & p\le k\le1,
\end{cases}
\qquad
\inf_{\mu\in\mathfrak U_\eta(p)}\int_0^1(q-k)_+\,\mu(\mathrm{d} q)=(p-k)_+ .
\end{equation}
An upper extremizer, independent of $k$, is
\begin{equation}\label{eq:extremal-mu}
\mu_\eta^*=\frac{\eta}{2p}\delta_0+\Bigl[1-\frac{\eta}{2p}-\frac{\eta}{2(1-p)}\Bigr]\delta_p+\frac{\eta}{2(1-p)}\delta_1 .
\end{equation}
\end{theorem}

\begin{proof}
\emph{The extremizer is admissible.} The weight of $\delta_p$ in \eqref{eq:extremal-mu} is nonnegative if and only if $\eta\le2p(1-p)$. The mean of $\mu^*_\eta$ is $p$, and its mean absolute deviation is $\frac{\eta}{2p}p+\frac{\eta}{2(1-p)}(1-p)=\eta$, so $\mu_\eta^*\in\mathfrak U_\eta(p)$.

\medskip
\emph{Upper bound.} Let $Q\sim\mu\in\mathfrak U_\eta(p)$. Since $\mathbb{E}(Q-p)=0$,
\begin{equation}\label{eq:positive-negative-deviation}
\mathbb{E}(Q-p)_+=\mathbb{E}(p-Q)_+=\tfrac12\mathbb{E}|Q-p|\le\tfrac\eta2 .
\end{equation}
If $k\ge p$, then $(q-k)_+\le\frac{1-k}{1-p}(q-p)_+$ on $[0,1]$, because for $q>k$ the difference is $\frac{(k-p)(1-q)}{1-p}\ge0$. Together with \eqref{eq:positive-negative-deviation} this gives $\mathbb{E}(Q-k)_+\le\frac{\eta(1-k)}{2(1-p)}$. Under $\mu^*_\eta$ only the atom at $1$ contributes, with value $\frac{\eta}{2(1-p)}(1-k)$, so the bound is attained. If $k\le p$, we write $\mathbb{E}(Q-k)_+=p-k+\mathbb{E}(k-Q)_+$. On $[0,1]$ we have $(k-q)_+\le\frac kp(p-q)_+$, because for $q<k$ the difference is $\frac{q(p-k)}{p}\ge0$. Hence $\mathbb{E}(Q-k)_+\le p-k+\frac{\eta k}{2p}$, and under $\mu^*_\eta$
\[
\Bigl[1-\tfrac{\eta}{2p}-\tfrac{\eta}{2(1-p)}\Bigr](p-k)+\tfrac{\eta}{2(1-p)}(1-k)=p-k+\tfrac{\eta k}{2p},
\]
so the bound is again attained.

\medskip
\emph{Lower bound.} By Jensen's inequality, $\mathbb{E}(Q-k)_+\ge(\mathbb{E} Q-k)_+=(p-k)_+$, and $\delta_p\in\mathfrak U_\eta(p)$ attains equality.
\end{proof}

\begin{corollary}[Robust premium in adapted radius]\label{cor:robust-premium-radius}
Let $\rho\ge0$ with $\rho^2\le2p(1-p)$. Over the adapted ball $\{\mathbb{P}\in\mathcal{M}_2(p):\mathcal{AW}_2(\mathbb{P}^0,\mathbb{P})\le\rho\}$, the robust call premium above the coarse price is $\rho^2k/(2p)$ for $k\le p$ and $\rho^2(1-k)/(2(1-p))$ for $k\ge p$.
\end{corollary}

\begin{proof}
By Proposition~\ref{prop:ball-equivalence} with $\eta=\rho^2$, the adapted ball consists of the models $\mathbb{P}^\mu$ with $\mu\in\mathfrak U_{\rho^2}(p)$. Subtracting the coarse price $(p-k)_+$ from the upper bound in Theorem~\ref{thm:robust-call} gives the claim.
\end{proof}

The quadratic dependence on the adapted radius comes from the identity between $\mathcal{AW}_2^2$ and the mean absolute deviation of the posterior. The robust problem is linear in the dispersion budget, and therefore quadratic in the metric radius. This contrasts with the first-order behaviour of Wasserstein distributionally robust problems in the static case \cite{BlanchetMurthy2019,BartlDrapeauOblojWiesel2021}.

\label{subsec:general-rating-claim}
\begin{proposition}[Sharp mean-absolute-deviation bound]\label{prop:lipschitz-rating}
Let $g:[0,1]\to\mathbb{R}$ be $L$-Lipschitz. For every $\mu\in\mathfrak U_\eta(p)$,
\begin{equation}\label{eq:rating-lipschitz-bound}
\Bigl|\int g(q)\,\mu(\mathrm{d} q)-g(p)\Bigr|\le L\eta,
\end{equation}
and the constant is optimal.
\end{proposition}

\begin{proof}
Integrating the Lipschitz bound $|g(q)-g(p)|\le L|q-p|$ against $\mu$ gives \eqref{eq:rating-lipschitz-bound}. For $g(q)=|q-p|$, which is $1$-Lipschitz, and any $\mu$ with mean absolute deviation equal to $\eta$, both sides equal $\eta$, so the constant cannot be improved.
\end{proof}

\subsection{Rating-dependent hazard rates}\label{sec:hazard}

Suppose that a rating assessment takes place at $t_r$, that the bond matures at $T>t_r$, and let $\Delta=T-t_r$. Conditionally on a rating variable $R$, the post-rating intensity is constant and equal to $\lambda_R\ge0$, so that the conditional survival probability is
\begin{equation}\label{eq:q-hazard}
Q=e^{-\lambda_R\Delta},
\qquad
\mathbb{E} e^{-\lambda_R\Delta}=p,
\end{equation}
where the second identity is the calibration to the coarse model before the rating is observed.

\begin{proposition}[Hazard representation]\label{prop:hazard-representation}
Under \eqref{eq:q-hazard},
$\mathcal{AW}_2^2(\mathbb{P}^0,\mathbb{P}^\mu)=\mathbb{E}|e^{-\lambda_R\Delta}-p|$, and $\mathcal{W}_2^2(\mathbb{P}^0,\mathbb{P}^\mu)=2\,\mathbb{E}|e^{-\lambda_R\Delta}-p|^2$ whenever $|e^{-\lambda_R\Delta}-p|\le\frac12$ almost surely, where $\mu$ is the law of $e^{-\lambda_R\Delta}$. The canonical hedge of Theorem~\ref{thm:dynamic-replication} is $H_2=e^{-\lambda_R\Delta}-p$.
\end{proposition}

\begin{proof}
The variable $Q=e^{-\lambda_R\Delta}$ takes values in $[0,1]$ and has mean $p$, so its law $\mu$ satisfies \eqref{eq:mu-mean}. The claims follow from Theorems~\ref{thm:exact-aw-one-rating}, \ref{thm:exact-w2-one-rating} and~\ref{thm:dynamic-replication} applied to $\mu$.
\end{proof}

\label{subsec:hazard-perturbation}
Let $\lambda_*=-\Delta^{-1}\log p$, so that $e^{-\lambda_*\Delta}=p$, and let $\zeta$ be a random variable with $|\zeta|\le\bar\zeta$ almost surely and
\begin{equation}\label{eq:Z-center}
\mathbb{E}\zeta=0 .
\end{equation}
For $\varepsilon$ near zero define
\begin{equation}\label{eq:lambda-epsilon}
c_\varepsilon=\frac1\Delta\log\mathbb{E} e^{-\Delta\varepsilon\zeta},
\qquad
\lambda_R^{\varepsilon}=\lambda_*+\varepsilon\zeta+c_\varepsilon .
\end{equation}
Then $\mathbb{E} e^{-\lambda_R^{\varepsilon}\Delta}=p$ exactly, and, since $\lambda_*>0$, $\lambda^\varepsilon_R\ge0$ for $|\varepsilon|$ small.

\begin{lemma}[Calibration correction]\label{lem:calibration-correction}
There are $\varepsilon_0>0$ and $C_c<\infty$, depending only on $\bar\zeta$ and $\Delta$, such that for $|\varepsilon|\le\varepsilon_0$,
\begin{equation}\label{eq:c-expansion}
\Bigl|c_\varepsilon-\frac{\Delta\varepsilon^2}{2}\mathbb{E}\zeta^2\Bigr|\le C_c|\varepsilon|^3 .
\end{equation}
\end{lemma}

\begin{proof}
Since $\zeta$ is bounded, its moment generating function $m_\zeta(u)=\mathbb{E} e^{u\zeta}$ is entire. By \eqref{eq:Z-center} its expansion has no linear term, so $m_\zeta(u)=1+\frac{u^2}{2}\mathbb{E}\zeta^2+O(|u|^3)$ for small $|u|$, with constants depending only on $\bar\zeta$. Since $\log(1+v)=v+O(v^2)$ and $m_\zeta(u)-1=O(u^2)$, it follows that $\log m_\zeta(u)=\frac{u^2}{2}\mathbb{E}\zeta^2+O(|u|^3)$. Taking $u=-\Delta\varepsilon$ and dividing by $\Delta$ gives \eqref{eq:c-expansion}.
\end{proof}

Let $Q_\varepsilon=e^{-\lambda_R^{\varepsilon}\Delta}=p\exp[-\Delta\varepsilon\zeta-\Delta c_\varepsilon]$, and let $\mathbb{P}^\varepsilon$ be the corresponding one-rating model.

\begin{theorem}[Small-dispersion asymptotics]\label{thm:hazard-asymptotics}
Assume $|\zeta|\le\bar\zeta$, $\mathbb{E}\zeta=0$, and $\mathbb{E}|\zeta|>0$. As $\varepsilon\to0$,
\begin{align}
\mathcal{AW}_2^2\bigl(\mathbb{P}^0,\mathbb{P}^{\varepsilon}\bigr)&=p\Delta|\varepsilon|\,\mathbb{E}|\zeta|+O(\varepsilon^2),
\label{eq:AW-asymptotic-squared}\\
\mathcal{AW}_2\bigl(\mathbb{P}^0,\mathbb{P}^{\varepsilon}\bigr)&=\sqrt{p\Delta\mathbb{E}|\zeta|}\,|\varepsilon|^{1/2}+O(|\varepsilon|^{3/2}),
\label{eq:AW-asymptotic}\\
\mathcal{W}_2\bigl(\mathbb{P}^0,\mathbb{P}^{\varepsilon}\bigr)&=\sqrt2\,p\Delta\bigl(\mathbb{E}\zeta^2\bigr)^{1/2}|\varepsilon|+O(\varepsilon^2).
\label{eq:W-asymptotic}
\end{align}
\end{theorem}

\begin{proof}
By Lemma~\ref{lem:calibration-correction}, $c_\varepsilon=O(\varepsilon^2)$, and a Taylor expansion of the exponential gives, uniformly in the sample point,
\[
Q_\varepsilon-p=-p\Delta\varepsilon\zeta+R_\varepsilon,
\qquad |R_\varepsilon|\le C_Q\varepsilon^2,
\]
for a constant $C_Q$ depending only on $p$, $\Delta$ and $\bar\zeta$.

\medskip
Taking absolute values and expectations, $\mathbb{E}|Q_\varepsilon-p|=p\Delta|\varepsilon|\,\mathbb{E}|\zeta|+O(\varepsilon^2)$, and Theorem~\ref{thm:exact-aw-one-rating} gives \eqref{eq:AW-asymptotic-squared}. Since $\mathbb{E}|\zeta|>0$, taking square roots and using $\sqrt{1+x}=1+O(x)$ gives \eqref{eq:AW-asymptotic}.

\medskip
Finally, $\mathbb{E}|Q_\varepsilon-p|^2=p^2\Delta^2\varepsilon^2\,\mathbb{E}\zeta^2+O(|\varepsilon|^3)$. For $|\varepsilon|$ small we have $|Q_\varepsilon-p|\le\frac12$, so Theorem~\ref{thm:exact-w2-one-rating} gives $\mathcal{W}_2^2(\mathbb{P}^0,\mathbb{P}^\varepsilon)=2\,\mathbb{E}|Q_\varepsilon-p|^2$, and taking square roots gives \eqref{eq:W-asymptotic}.
\end{proof}

\begin{corollary}[Asymptotic separation]\label{cor:asymptotic-separation}
Under the assumptions of Theorem~\ref{thm:hazard-asymptotics},
\begin{equation}\label{eq:ratio-asymptotic}
\frac{\mathcal{AW}_2(\mathbb{P}^0,\mathbb{P}^{\varepsilon})}{\mathcal{W}_2(\mathbb{P}^0,\mathbb{P}^{\varepsilon})}
=\frac{\sqrt{\mathbb{E}|\zeta|}}{\sqrt{2p\Delta}\,(\mathbb{E}\zeta^2)^{1/2}}\,|\varepsilon|^{-1/2}\bigl[1+O(|\varepsilon|)\bigr].
\end{equation}
\end{corollary}

The square-root singularity of Corollary~\ref{cor:two-rating} is therefore not an artifact of a two-point example. It persists for every nontrivial bounded centered hazard shock under exact calibration.

\section{Empirical and numerical illustrations}\label{sec:examples}

We now quantify the effects described above on agency rating data, using the S\&P Global Ratings 2024 annual corporate default and transition study \cite{SP2025}. Section~\ref{subsec:empirical-ratings} illustrates the discrete-time theory of Section~\ref{sec:ratings}. It computes adapted distances between calibrated rating systems with the dynamic programme and the synchronous coupling, tests the monotonicity hypothesis, and measures how the distance translates into pricing and hedging errors. Sections~\ref{subsec:empirical-hedge} and~\ref{subsec:empirical-multi} illustrate the intensity models of Section~\ref{sec:cox-hedging}, namely the rolling bootstrap hedge and its bound, and hedging with several instruments.

\subsection{Rating systems in S\&P transition data}\label{subsec:empirical-ratings}

A rating system, together with a rule that assigns hazard rates to rating classes, defines a model for the price of the defaultable zero-coupon bond as in Section~\ref{sec:migration}. The price of an $N$-year bond is the probability, given the current rating, of surviving until $N$. Two rating systems for the same issuer that are calibrated to the same bond price define two laws in $\mathcal{M}_N(p)$, and Theorems~\ref{thm:dpp} and~\ref{thm:synchronous} compute the adapted distance between them exactly. This subsection uses this to answer three questions raised by the theory. Is the monotonicity hypothesis of Theorem~\ref{thm:synchronous} realistic for agency data? 
How much do rating systems that agree on today's bond price differ in the information they release?
 And what do these discrepancies mean for prices and hedges, through the stability results of Section~\ref{sec:hedging}?

\medskip
\emph{Data.} We use the global and regional average one-year transition matrices for 1981-2024, the 2024 one-year matrix, the annual default rates by rating category, and the annual upgrade and downgrade rates from \cite{SP2025}. Withdrawn ratings are removed by proportional redistribution, and the seven categories AAA to CCC/C together with default define a Markov chain ordered by credit quality. For a bond of maturity $N$ years and an issuer with initial rating $r$, the defaultable bond price process is $X_t=q_t(R_t)$ as in Section~\ref{sec:migration}. The matrices are historical rather than risk neutral. Since every comparison below is carried out between two systems with the same initial bond price, we calibrate the second system by scaling its default column by a constant and renormalizing the remaining transitions, in the spirit of the risk-premium adjustment of \cite{JarrowLandoTurnbull1997}.

\medskip
\emph{Comparisons.} We consider three comparisons, and in each of them we call the first named model \emph{system~1} and the recalibrated one \emph{system~2}. The \emph{regional} comparison opposes the U.S.\ and European average matrices. The \emph{vintage} comparison opposes the long-run global average and the 2024 matrix. The \emph{cycle} comparison opposes a point-in-time model and the through-the-cycle average. Its expansion and recession matrices are obtained from the average one by replacing the default column with the mean default rates of expansion and recession years, and by scaling upgrades and downgrades by the ratio of the regime mean to the long-run upgrade and downgrade rates. The recession years are 1990, 1991, 2001, 2002, 2008, 2009 and 2020, and the regime is a Markov chain estimated from this sequence, with $\mathbb{P}[\mathrm E\to\mathrm R]\approx0.111$ and $\mathbb{P}[\mathrm R\to\mathrm R]\approx0.429$. The point-in-time state is the pair (rating, regime), both observed. Conditioning migration matrices on the state of the business cycle in this way follows Bangia, Diebold, Kronimus, Schagen and Schuermann \cite{BangiaEtAl2002}, and a model in which migration is driven jointly by ratings and economic states is developed in \cite{KalkbrenerPackham2026}.

\medskip
\emph{Monotonicity.} The largest violation of stochastic monotonicity, measured by $\max_{i<i',k}(\mathbb{P}[R_1\le k\mid i']-\mathbb{P}[R_1\le k\mid i])$, is $0.0003$ for the global and U.S.\ matrices, $0.0011$ for Europe, $0$ for 2024 and $0.0016$ for the regime model. This is caused by small upward-jump frequencies of the top categories into CCC/C. Solving the dynamic programme \eqref{eq:dpp} exactly and comparing with the synchronous coupling, the relative difference never exceeds $5\cdot10^{-5}$ in the regional comparison, $6\cdot10^{-11}$ in the vintage comparison and $6.4\cdot10^{-4}$ in the cycle comparison, over all initial ratings from A to CCC/C and maturities from two to ten years. For practical purposes, Theorem~\ref{thm:synchronous} is therefore exact on agency data.

\begin{table}[t]
\centering
\small
\begin{tabular}{l l ccc ccc}
\toprule
& & \multicolumn{3}{c}{$N=2$} & \multicolumn{3}{c}{$N=10$}\\
\cmidrule(lr){3-5}\cmidrule(lr){6-8}
Comparison & & BBB & BB & B & BBB & BB & B\\
\midrule
U.S.\ vs Europe & & 0.033 & 0.044 & 0.125 & 0.135 & 0.200 & 0.246\\
Average vs 2024 vintage & & 0.027 & 0.068 & 0.128 & 0.157 & 0.291 & 0.306\\
PIT (expansion) vs TTC & & 0.029 & 0.056 & 0.106 & 0.129 & 0.220 & 0.277\\
PIT (recession) vs TTC & & 0.053 & 0.103 & 0.215 & 0.166 & 0.287 & 0.364\\
\midrule
TTC vs no information & & 0.063 & 0.134 & 0.294 & 0.261 & 0.488 & 0.666\\
\bottomrule
\end{tabular}
\caption{Adapted Wasserstein distances $\mathcal{AW}_2$ between calibrated rating systems built from S\&P data, by initial rating and bond maturity. The last row is the distance of the through-the-cycle model from the model that reveals no information before maturity (Theorem~\ref{thm:multidate-coarse}). Values for the other maturities are shown in Figure~\ref{fig:empirical-ratings}.}
\label{tab:empirical-ratings}
\end{table}

\begin{figure}[t]
\centering
\includegraphics[width=0.49\textwidth]{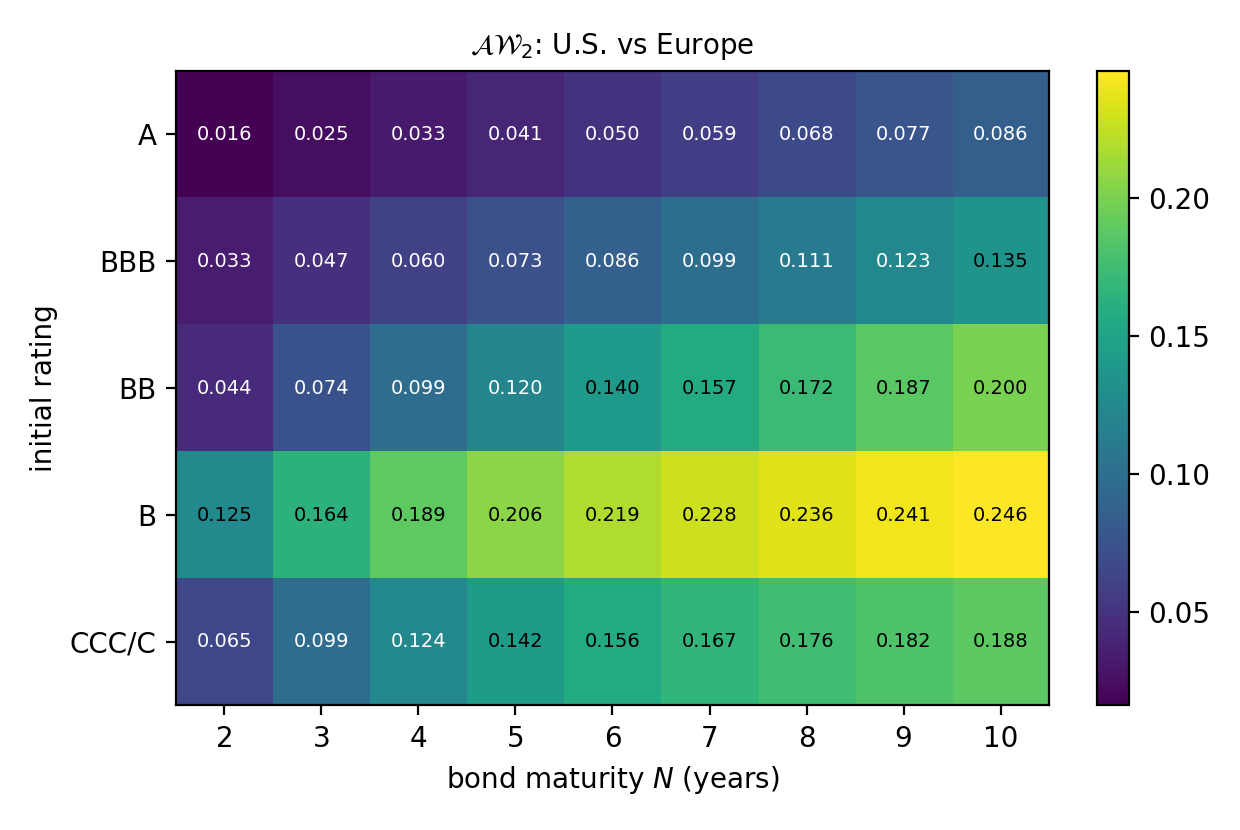}\hfill
\includegraphics[width=0.49\textwidth]{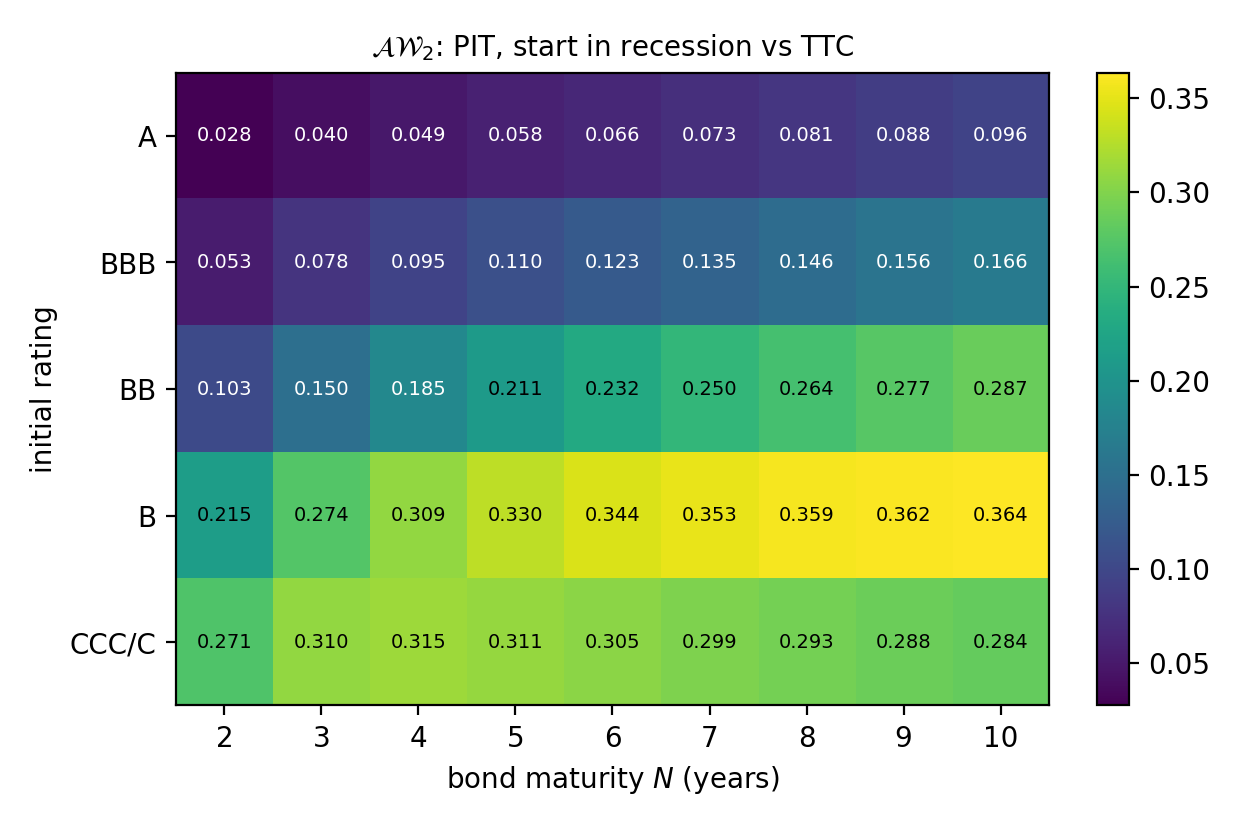}
\caption{$\mathcal{AW}_2$ between calibrated rating systems by initial rating and bond maturity: U.S.\ against European average migration (left), and the point-in-time regime model started in a recession against the through-the-cycle average (right).}
\label{fig:empirical-ratings}
\end{figure}

\medskip
\emph{Distances.} Table~\ref{tab:empirical-ratings} and Figure~\ref{fig:empirical-ratings} report the distances, and three features stand out. First, after calibration to the same bond price, which requires scaling European default probabilities by factors between about $1.4$ and $3$, the U.S.\ and European systems still differ by $\mathcal{AW}_2$ between $0.03$ and $0.25$ for issuers rated BBB to B. The two systems therefore reveal credit information along different paths. Second, at short maturities the point-in-time model started in a recession is roughly twice as far from the through-the-cycle model as when started in an expansion, and the difference fades with maturity as the regime chain mixes. Third, the distance is largest for B issuers and, in the cycle comparison, it is non-monotone in maturity for CCC/C issuers, for whom most information is released by early defaults. Compared with the last row of the table, disagreement between systems accounts for a substantial fraction of the information released by annual reviews. For a B issuer in a recession, it amounts, in squared distance, to about half of it at the two-year maturity and to $30\%$ at the ten-year maturity. For issuers rated A or above, the 2024 vintage has no direct one-year defaults, so calibration can be impossible or require very large default scaling. We therefore report the vintage comparison from BBB downwards.

\medskip
\emph{Comparison with mobility indices.} The adapted distance should be distinguished from the metrics used in the literature to compare migration matrices. Jafry and Schuermann measure the mobility of a matrix $P$ by the average singular value of $P-I$, and compare matrices through the difference of this index \cite{JafrySchuermann2004}. Regional comparisons in agency studies are of the same static kind. Such indices compare one-period transition probabilities, whereas $\mathcal{AW}_2$ compares the filtered laws of the bond prices that the matrices induce over the whole life of the bond, and the two can lead to very different conclusions. On our data, the mobility index is $0.160$ for the U.S.\ and $0.157$ for Europe, a difference of less than $2\%$. At the same time, the calibrated adapted distance between the two systems reaches $0.25$ for B issuers at ten years, a large fraction of their distance from the no-information model. The index is $0.155$ for the long-run average and $0.110$ for the 2024 vintage, and $0.137$ and $0.203$ for the expansion and recession matrices. In these cases, both measures detect a difference, but only the adapted distance says how much of it reaches bond prices and hedges.

\medskip
\emph{Prices and hedges.} We finally turn the distances into statements about prices and hedges. The rating-date straddle $G_t=|X_t-X_0|$ pays the absolute revision of the bond price between today and date $t$. It is $\sqrt t$-Lipschitz in the increment norm, so that $|\mathbb{E}_1G_t-\mathbb{E}_2G_t|\le\sqrt t\,\mathcal{W}_2\le\sqrt t\,\mathcal{AW}_2$ by Remark~\ref{rem:price-stability}, where $\mathbb{E}_1$ and $\mathbb{E}_2$ denote expectations under the two systems. To compare hedges, we compute in system~2 the variance-optimal hedge of $G_t$ with the bond itself \cite{FollmerSchweizer1991}, that is, the initial capital $m=\mathbb{E}_2G_t$ and the hedge ratios
\[
H_s=\frac{\operatorname{Cov}_2(\Delta V_s,\Delta X_s\mid\mathcal{F}_{s-1})}{\operatorname{Var}_2(\Delta X_s\mid\mathcal{F}_{s-1})},
\qquad V_s=\mathbb{E}_2[G_t\mid\mathcal{F}_s].
\]
We set $H_s=0$ whenever the conditional variance in the denominator vanishes.
The desk observes the bond price, so we express $H_s$ as a function of the price at date $s-1$ by linear interpolation on the price grid of system~2, and implement the same rule in system~1. Proposition~\ref{prop:hedge-stability} bounds the change of the $L^2$ hedging error by $(L_G+K+c_NL_H)\,\mathcal{AW}_2$, where $K$ and $L_H$ are the sup norm and the Lipschitz constant of the interpolated rule and $c_N=\sqrt{N(N-1)/2}$. All quantities are computed exactly by matrix recursions on the two rating chains, and Table~\ref{tab:hedge-transfer} reports them for a five-year bond.

\begin{table}[t]
\centering
\footnotesize
\setlength{\tabcolsep}{3.5pt}
\begin{tabular}{l l c cccc ccc}
\toprule
& & & \multicolumn{4}{c}{straddle at $t=1$} & \multicolumn{3}{c}{straddle at $t=4$}\\
\cmidrule(lr){4-7}\cmidrule(lr){8-10}
Comparison & start & $\mathcal{AW}_2$ & $|\Delta\text{price}|$ & $R_2$ & $R_1$ & bound & $|\Delta\text{price}|$ & $R_2$ & $R_1$\\
\midrule
U.S.\ vs Europe & BBB & 0.073 & 0.0004 & 0.0034 & 0.0034 & 0.145 & 0.0006 & 0.0062 & 0.0056\\
 & BB & 0.120 & 0.0015 & 0.0211 & 0.0191 & 0.228 & 0.0032 & 0.0216 & 0.0242\\
 & B & 0.206 & 0.0041 & 0.0611 & 0.0564 & 0.366 & 0.0139 & 0.0554 & 0.0769\\
Average vs 2024 & BBB & 0.075 & 0.0012 & 0.0031 & 0.0029 & 0.148 & 0.0003 & 0.0027 & 0.0217\\
 & BB & 0.173 & 0.0025 & 0.0135 & 0.0164 & 0.337 & 0.0028 & 0.0238 & 0.0323\\
 & B & 0.241 & 0.0003 & 0.0632 & 0.0560 & 0.436 & 0.0167 & 0.0372 & 0.0749\\
PIT (recess.) vs TTC & BBB & 0.110 & 0.0052 & 0.0045 & 0.0101 & 0.217 & 0.0011 & 0.0071 & 0.0088\\
 & BB & 0.211 & 0.0160 & 0.0247 & 0.0388 & 0.397 & 0.0033 & 0.0321 & 0.0358\\
 & B & 0.330 & 0.0479 & 0.0743 & 0.1004 & 0.570 & 0.0010 & 0.0691 & 0.0808\\
\bottomrule
\end{tabular}
\caption{Rating-date straddles on a five-year zero-recovery bond. $\mathcal{AW}_2$ is the distance between the calibrated systems~1 (first named) and~2. The column $|\Delta\text{price}|$ is the difference of the straddle prices, bounded by $\sqrt t\,\mathcal{AW}_2$. $R_2$ is the $L^2$ error of the variance-optimal hedge of system~2 in system~2, and $R_1$ the error of the same rule implemented in system~1. The bound is the bound on $|R_1-R_2|$ of Proposition~\ref{prop:hedge-stability}. At $t=4$ it exceeds $20$ in every case and is not reported.}
\label{tab:hedge-transfer}
\end{table}

\medskip
Three conclusions emerge. First, the bounds are valid and, for the one-period straddle, of the right order. The observed price differences are between $0.1\%$ and $15\%$ of $\mathcal{AW}_2$, and the changes in hedging error are up to about $5\%$ of the hedge bound. These bounds are uniform over all claims and strategies with the given constants, while a particular straddle and hedge are far from the worst case. Second, transferring a hedge between rating systems can matter much more than transferring a price. For the BBB issuer, using the hedge of the 2024 vintage in the long-run average model multiplies the hedging error of the four-year straddle by eight, while its price changes by $0.0003$. Starting in a recession, the through-the-cycle hedge of the one-year straddle has an error $1.35$ to $2.3$ times larger under point-in-time migration. Third, for multi-period strategies the generic bound of Proposition~\ref{prop:hedge-stability} becomes uninformative. A hedge rule defined on a discrete price grid has a large Lipschitz constant wherever two ratings have nearly the same bond price, and $L_H$ ranges from about $60$ to $5\cdot10^4$ in the four-year case. This is precisely the situation in which the exact identities of Section~\ref{sec:cox-hedging}, which do not rely on Lipschitz constants of the strategy, are needed.

\subsection{Monte Carlo study of the rolling bootstrap hedge}\label{subsec:empirical-hedge}

We simulate the rolling hedge of Theorem~\ref{thm:term-structure} for a zero-recovery bond with maturity $K=7$ years and annual pins. The market model is a Cox model driven by the S\&P global average matrix $P$. In rating $i$, the hazard is $h_i=-\log(1-P(i,0))$, where $0$ is the default state, and migrations follow the renormalized non-default transitions. In the \emph{review} version of the market model, the rating for the next year is revealed at each pin, which is the equality case of Theorem~\ref{thm:term-structure}. In the \emph{intra-year} version, a migration takes place at a uniform time inside the year and switches the hazard immediately, so that information arrives within buckets. For comparison, we also simulate a CIR intensity $\mathrm{d}\lambda=\kappa(\theta-\lambda)\,\mathrm{d} t+\sigma\sqrt\lambda\,\mathrm{d} W$ with $\kappa=0.3$, $\theta=0.03$, $\lambda_0=0.02$ and $\sigma\in\{0.04,0.08,0.12\}$, with time step $1/250$. In every case, the bootstrap model is calibrated exactly to the survival curve of the market model at the pins, and all expectations are computed from $2\cdot10^5$ paths. In the tables and numerical summaries below, expectations of absolute pre-default residuals include survival to the reset date: $\mathbb E|\varepsilon_k|$ abbreviates $\mathbb E[\mathbf1_{\{\tau^A>t_k\}}|\varepsilon_k|]$, and similarly for $\varepsilon_k^{(m)}$.

\medskip
As a check on the data, the review model started from BB produces cumulative default probabilities of $0.62\%$, $5.94\%$ and $9.62\%$ at one, five and seven years, against $0.56\%$, $5.75\%$ and $7.94\%$ in the S\&P static-pool data. The Markov chain thus slightly overstates long-horizon defaults, a known feature of time-homogeneous migration models \cite{FrydmanSchuermann2008}.

\begin{table}[t]
\centering
\small
\begin{tabular}{l l cccc}
\toprule
Market model & Start & $\sum_k\mathbb{E}|\varepsilon_k|$ & Fr\'echet & $\sum_kN_{k+1}\mathbb{E}\Psi$ & ratio\\
\midrule
Reviews at pins & BBB & 0.0212 & 0.0212 & 0.0212 & 1.00\\
 & BB & 0.0903 & 0.0903 & 0.0903 & 1.00\\
 & B & 0.1962 & 0.1962 & 0.1962 & 1.00\\
Intra-year migration & BBB & 0.0214 & 0.0237 & 0.0239 & 0.89\\
 & BB & 0.0863 & 0.0959 & 0.0979 & 0.88\\
 & B & 0.1467 & 0.1948 & 0.1993 & 0.74\\
CIR, $\sigma=0.04$ & - & 0.0247 & 0.0281 & 0.0299 & 0.83\\
CIR, $\sigma=0.08$ & - & 0.0481 & 0.0547 & 0.0582 & 0.83\\
CIR, $\sigma=0.12$ & - & 0.0689 & 0.0785 & 0.0835 & 0.83\\
\bottomrule
\end{tabular}
\caption{Rolling bootstrap hedge of a seven-year zero-recovery bond: total expected absolute cash residual, its Fr\'echet upper bound, the adapted bound of Theorem~\ref{thm:term-structure}, and the ratio of the first to the last column.}
\label{tab:empirical-hedge}
\end{table}

\begin{figure}[t]
\centering
{\setlength{\unitlength}{\textwidth}
\begin{picture}(1,0.357895)
\put(0,0){\includegraphics[width=\textwidth]{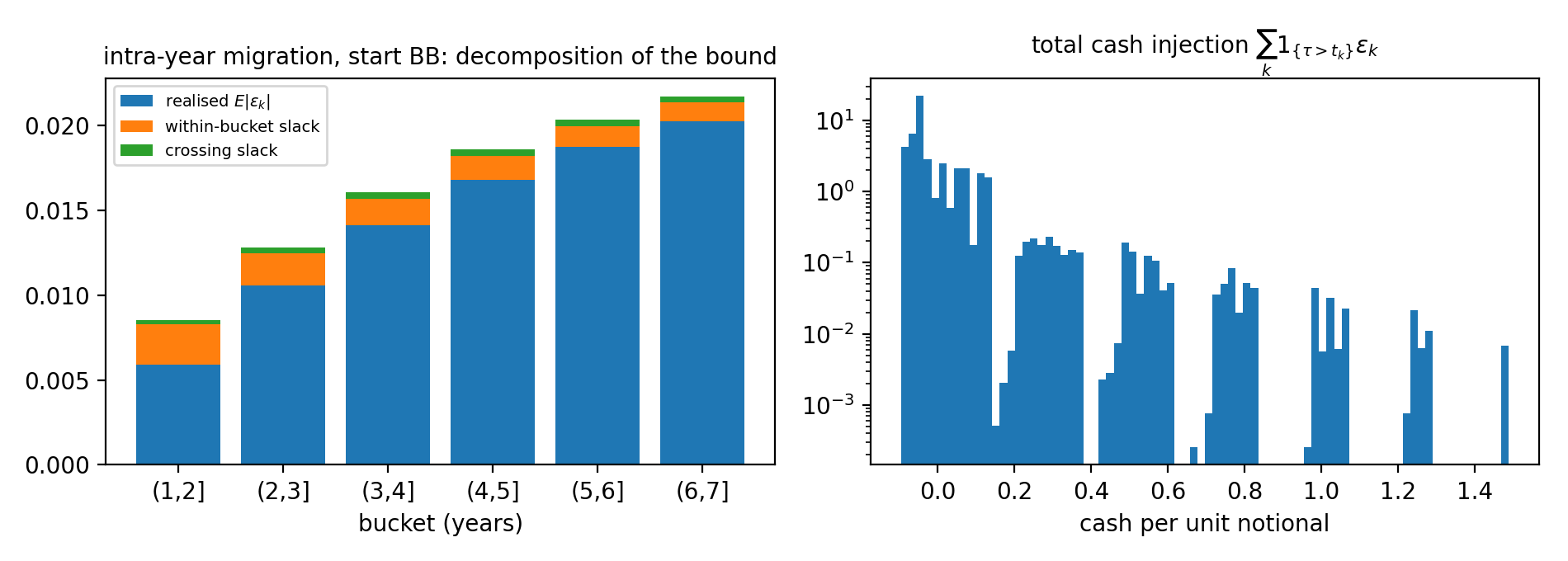}}
\put(0.55,0.3105){\color{white}\rule{0.445\textwidth}{0.0474\textwidth}}
\put(0.55,0.3105){\parbox[b][0.0474\textwidth][c]{0.445\textwidth}{\centering\small total cash residual $\sum_k\mathbf1_{\{\tau>t_k\}}\varepsilon_k$}}
\end{picture}}
\caption{Intra-year migration model, initial rating BB. Left: for each bucket, the realised expected absolute residual, the within-bucket (conditional Jensen) slack and the hazard-crossing slack, which stack to the bound of Theorem~\ref{thm:term-structure}. Right: density of the total cash residual $\sum_k\mathbf{1}_{\{\tau>t_k\}}\varepsilon_k$ on a logarithmic scale.}
\label{fig:empirical-hedge}
\end{figure}

\medskip
Table~\ref{tab:empirical-hedge} and Figure~\ref{fig:empirical-hedge} summarize the study. With reviews at the pins, the three quantities coincide to all reported digits. For the BB issuer, the Monte Carlo estimate of $\mathbb{E}[\mathbf{1}_{\{\tau>t_k\}}|\varepsilon_k|]$ obtained by simulating defaults also agrees with them within one standard error in every bucket, which confirms the equality statement of the theorem. When migrations occur inside the year, the expected absolute residual is $74\%$ to $89\%$ of the bound. The slack is dominated by the within-bucket term, because information released during the year is averaged by the conditional price at the pin, while hazard crossings contribute between $1\%$ and $2.5\%$ of the bound. The effect is strongest for B issuers, whose migrations are most frequent. For the CIR model, the ratio is essentially independent of the volatility, because the residual and the bound both scale with $\sigma$.

\medskip
In all models, the mean signed residual is of order $10^{-4}$ or smaller, that is, zero within Monte Carlo error, as calibration requires. The total cash residual is strongly right-skewed. Downgrades make the next bond cheaper and generate windfalls, while upgrades produce small losses. Its expected absolute value, $0.078$ for the intra-year BB case, is smaller than the sum $0.086$ of the bucket-wise quantities, because residuals of different buckets partly offset.

\subsection{Hedging with several instruments}\label{subsec:empirical-multi}

We repeat the study of Section~\ref{subsec:empirical-hedge} for the two rating versions of the market model, now letting both desks of Section~\ref{subsec:window} trade every bond within a liquidity window of $m$ years. All expectations that depend only on the rating at a pin are computed exactly from \eqref{eq:rating-prices}, while the distributions of total cash and the bounds of Theorem~\ref{thm:window} are estimated from $10^6$ paths. The two desks account for the liability differently. The bootstrap desk pays for rolling its position, whereas the market desk marks the liability to market. We therefore compare the expected absolute net cash residual over the life of the hedge for each desk.

\begin{table}[t]
\centering
\small
\setlength{\tabcolsep}{4pt}
\begin{tabular}{c cccc cccc}
\toprule
& \multicolumn{4}{c}{initial rating BB} & \multicolumn{4}{c}{initial rating B}\\
\cmidrule(lr){2-5}\cmidrule(lr){6-9}
$m$ & $\Sigma\,\mathbb{E}|\varepsilon^{(m)}_k|$ & $\Sigma\,\mathbb{E}|\Delta_k|$ & bootstrap & market & $\Sigma\,\mathbb{E}|\varepsilon^{(m)}_k|$ & $\Sigma\,\mathbb{E}|\Delta_k|$ & bootstrap & market\\
\midrule
1 & 0.0866 & 0.0965 & 0.0783 & 0.0786 & 0.1470 & 0.1953 & 0.1335 & 0.1339\\
2 & 0.0612 & 0.0882 & 0.0556 & 0.0294 & 0.0659 & 0.1698 & 0.0626 & 0.0311\\
3 & 0.0396 & 0.0756 & 0.0370 & 0.0035 & 0.0335 & 0.1380 & 0.0320 & 0.0003\\
4 & 0.0229 & 0.0598 & 0.0219 & 0.0000 & 0.0195 & 0.1036 & 0.0190 & 0.0000\\
5 & 0.0110 & 0.0415 & 0.0108 & 0.0000 & 0.0097 & 0.0686 & 0.0096 & 0.0000\\
6 & 0.0035 & 0.0214 & 0.0035 & 0.0000 & 0.0033 & 0.0339 & 0.0032 & 0.0000\\
\bottomrule
\end{tabular}
\caption{Hedging a seven-year zero-recovery bond with a liquidity window of $m$ years in the intra-year migration model. The first two columns of each block are the expected absolute residuals of the bootstrap desk and the pathwise bound of Theorem~\ref{thm:window}. The last two are the expected absolute total cash residual of the bootstrap desk and of the market-optimal desk.}
\label{tab:empirical-multi}
\end{table}

\begin{figure}[t]
\centering
\includegraphics[width=0.62\textwidth]{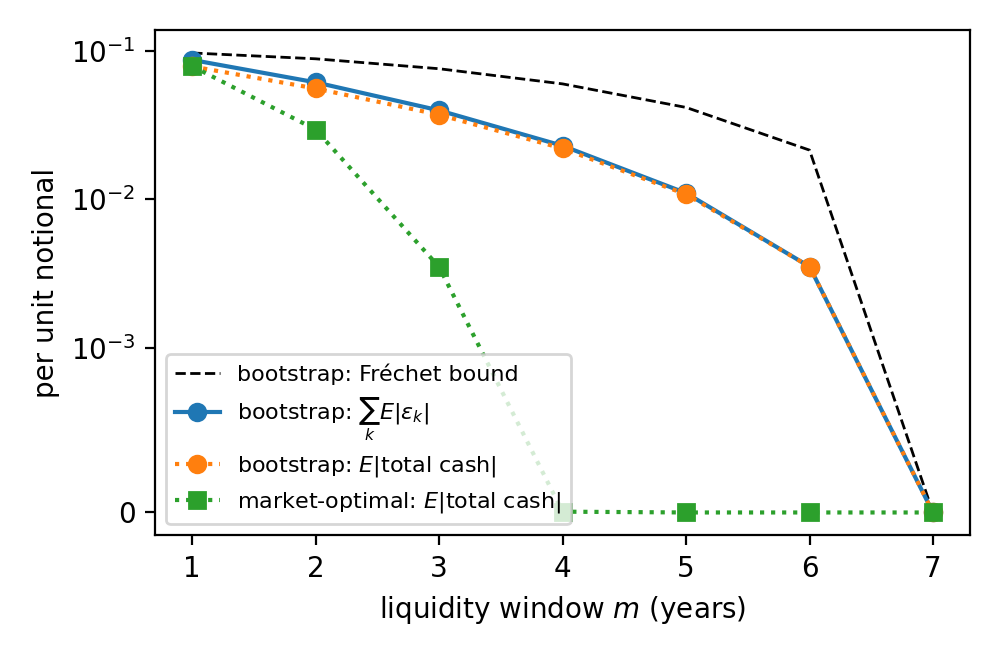}
\caption{Intra-year migration model, initial rating BB: hedging errors as functions of the liquidity window (symmetric logarithmic scale).}
\label{fig:empirical-multi}
\end{figure}

\medskip
Table~\ref{tab:empirical-multi} and Figure~\ref{fig:empirical-multi} show the intra-year migration model. The review model gives the same picture. Its expected absolute total cash for BB is $0.0825$, $0.0588$ and $0.0395$ for the bootstrap desk with $m=1,2,3$, against $0.0827$, $0.0357$ and $0.0036$ for the market desk. For $m=1$ the exact values of $\sum_k\mathbb{E}|\varepsilon_k|$ agree with Table~\ref{tab:empirical-hedge} within Monte Carlo error.

\medskip
With a single instrument, the two desks perform almost identically in total cash, $0.0783$ against $0.0786$ for BB. In this experiment, knowing the market model gives no material reduction in the reported absolute-loss criterion when only the next bond can be traded. Each additional maturity changes this picture. The market desk reduces its error by more than half at $m=2$, by more than an order of magnitude at $m=3$, and is exact to four digits from $m=4$ on. This is the near-spanning predicted by Proposition~\ref{prop:spanning}, a quantitative instance of bond-liability spanning in a Markov chain market \cite{Norberg2003}. In the review model the singular values of the Krylov matrix $[\mathbf{1},\widehat K\mathbf{1},\ldots,\widehat K^{6}\mathbf{1}]$ are approximately $6.45$, $0.61$, $0.091$, $4.3\cdot10^{-3}$, $8.1\cdot10^{-5}$, $9\cdot10^{-7}$ and $6\cdot10^{-9}$, so that four consecutive maturities span the relevant price vectors to within $3\cdot10^{-4}$ in relative terms. The bootstrap desk, in contrast, improves only gradually, roughly halving its error per additional year of liquidity. At $m=3$, using the specified market-based strategy reduces the expected absolute net cash residual relative to the bootstrap strategy by about $90\%$ for BB and $99\%$ for B, on the same hedging universe. Richer hedging universes therefore do not make the choice of model irrelevant. They make it the dominant source of error.

\medskip
As for the bounds, the pathwise quantity $\sum_k\mathbb{E}|\Delta_k|$ of Theorem~\ref{thm:window} stays within a factor of two of the realised residual for BB when $m\le3$ and tracks its decay in $m$. The ratio deteriorates for longer windows, where the residuals themselves are small. The adapted bound on the right of \eqref{eq:window-bound} is sharp for $m=1$ but increases with $m$, reaching about $0.34$ for BB at $m=4$, because it ignores the cancellation between the bond sold and the bond bought at each pin.

\section*{Acknowledgements}
F.\,M. thanks the Institut für Mathematische Statistik und Versicherungslehre, Universität Bern, for its kind hospitality during the visit in which a first draft of this paper was written.
F.\,M. acknowledges support from the ERC Advanced Grant NEITALG, grant agreement No.~101198055. 

\begin{center}
  \FundingLogos
  
  \vspace{0.5em}
  \begin{tcolorbox}\centering\small
   
    Funded by the European Union. Views and opinions expressed are, however, those of the author(s) only and do not necessarily reflect those of the European Union or the European Research Council Executive Agency. Neither the European Union nor the granting authority can be held responsible for them. This project has received funding from the European Research Council (ERC) under the European Union’s Horizon Europe research and innovation programme (grant agreement No.~101198055, project acronym NEITALG).
    
  \end{tcolorbox}
\end{center}

\bibliographystyle{abbrv}
\bibliography{aw_credit}

@article{Pflug2009,
  author  = {Pflug, Georg Ch.},
  title   = {Version-independence and nested distributions in multistage stochastic optimization},
  journal = {SIAM Journal on Optimization},
  volume  = {20},
  number  = {3},
  pages   = {1406--1420},
  year    = {2009},
  doi     = {10.1137/080718401}
}

@article{PflugPichler2012,
  author  = {Pflug, Georg Ch. and Pichler, Alois},
  title   = {A distance for multistage stochastic optimization models},
  journal = {SIAM Journal on Optimization},
  volume  = {22},
  number  = {1},
  pages   = {1--23},
  year    = {2012},
  doi     = {10.1137/110825054}
}

@book{PflugPichler2014,
  author    = {Pflug, Georg Ch. and Pichler, Alois},
  title     = {Multistage Stochastic Optimization},
  series    = {Springer Series in Operations Research and Financial Engineering},
  publisher = {Springer},
  address   = {Cham},
  year      = {2014},
  doi       = {10.1007/978-3-319-08843-3}
}

@article{BackhoffBeiglbockLinZalashko2017,
  author  = {Backhoff-Veraguas, Julio and Beiglb{\"o}ck, Mathias and Lin, Yiqing and Zalashko, Anastasiia},
  title   = {Causal transport in discrete time and applications},
  journal = {SIAM Journal on Optimization},
  volume  = {27},
  number  = {4},
  pages   = {2528--2562},
  year    = {2017},
  doi     = {10.1137/16M1080197}
}

@article{Lassalle2018,
  author  = {Lassalle, R{\'e}mi},
  title   = {Causal transport plans and their {M}onge--{K}antorovich problems},
  journal = {Stochastic Analysis and Applications},
  volume  = {36},
  number  = {3},
  pages   = {452--484},
  year    = {2018},
  doi     = {10.1080/07362994.2017.1422747}
}

@article{AcciaioBackhoffZalashko2020,
  author  = {Acciaio, Beatrice and Backhoff-Veraguas, Julio and Zalashko, Anastasiia},
  title   = {Causal optimal transport and its links to enlargement of filtrations and continuous-time stochastic optimization},
  journal = {Stochastic Processes and their Applications},
  volume  = {130},
  number  = {5},
  pages   = {2918--2953},
  year    = {2020},
  doi     = {10.1016/j.spa.2019.08.009}
}

@article{BackhoffBeiglbockEderPichler2020,
  author  = {Backhoff-Veraguas, Julio and Beiglb{\"o}ck, Mathias and Eder, Manu and Pichler, Alois},
  title   = {Fundamental properties of process distances},
  journal = {Stochastic Processes and their Applications},
  volume  = {130},
  number  = {9},
  pages   = {5575--5591},
  year    = {2020},
  doi     = {10.1016/j.spa.2020.03.017}
}

@article{BackhoffBartlBeiglbockEder2020Top,
  author  = {Backhoff-Veraguas, Julio and Bartl, Daniel and Beiglb{\"o}ck, Mathias and Eder, Manu},
  title   = {All adapted topologies are equal},
  journal = {Probability Theory and Related Fields},
  volume  = {178},
  pages   = {1125--1172},
  year    = {2020},
  doi     = {10.1007/s00440-020-00993-8}
}

@article{BackhoffBartlBeiglbockEder2020Finance,
  author  = {Backhoff-Veraguas, Julio and Bartl, Daniel and Beiglb{\"o}ck, Mathias and Eder, Manu},
  title   = {Adapted {W}asserstein distances and stability in mathematical finance},
  journal = {Finance and Stochastics},
  volume  = {24},
  number  = {3},
  pages   = {601--632},
  year    = {2020},
  doi     = {10.1007/s00780-020-00426-3}
}

@article{BackhoffBartlBeiglbockWiesel2022,
  author  = {Backhoff-Veraguas, Julio and Bartl, Daniel and Beiglb{\"o}ck, Mathias and Wiesel, Johannes},
  title   = {Estimating processes in adapted {W}asserstein distance},
  journal = {The Annals of Applied Probability},
  volume  = {32},
  number  = {1},
  pages   = {529--550},
  year    = {2022},
  doi     = {10.1214/21-AAP1687}
}

@article{BartlBeiglbockPammer2024,
  author  = {Bartl, Daniel and Beiglb{\"o}ck, Mathias and Pammer, Gudmund},
  title   = {The {W}asserstein space of stochastic processes},
  journal = {Journal of the European Mathematical Society},
  volume  = {28},
  number  = {1},
  pages   = {393--454},
  year    = {2026},
  doi     = {10.4171/JEMS/1554},
  eprint  = {2104.14245},
  archivePrefix = {arXiv}
}

@misc{BartlBeiglbockPammerSchrottZhang2025,
  author  = {Bartl, Daniel and Beiglb{\"o}ck, Mathias and Pammer, Gudmund and Schrott, Stefan and Zhang, Xin},
  title   = {The {W}asserstein space of stochastic processes in continuous time},
  year    = {2025},
  eprint  = {2501.14135},
  archivePrefix = {arXiv},
  howpublished = {arXiv:2501.14135}
}

@article{BartlWiesel2023,
  author  = {Bartl, Daniel and Wiesel, Johannes},
  title   = {Sensitivity of multiperiod optimization problems with respect to the adapted {W}asserstein distance},
  journal = {SIAM Journal on Financial Mathematics},
  volume  = {14},
  number  = {2},
  pages   = {704--720},
  year    = {2023},
  doi     = {10.1137/22M1537746}
}

@article{AcciaioBackhoffPammer2025,
  author  = {Acciaio, Beatrice and Backhoff-Veraguas, Julio and Pammer, Gudmund},
  title   = {Quantitative fundamental theorem of asset pricing},
  journal = {Mathematical Finance},
  volume  = {35},
  pages   = {636--660},
  year    = {2025},
  doi     = {10.1111/mafi.12457}
}

@article{BackhoffKallbladRobinson2025,
  author  = {Backhoff-Veraguas, Julio and K{\"a}llblad, Sigrid and Robinson, Benjamin A.},
  title   = {Adapted {W}asserstein distance between the laws of stochastic differential equations},
  journal = {Stochastic Processes and their Applications},
  volume  = {189},
  pages   = {104689},
  year    = {2025},
  doi     = {10.1016/j.spa.2025.104689},
  eprint  = {2209.03243},
  archivePrefix = {arXiv}
}

@article{BeiglbockPfluglSchrott2026,
  author  = {Beiglb{\"o}ck, Mathias and Pfl{\"u}gl, Susanne and Schrott, Stefan},
  title   = {A probabilistic view on the adapted {W}asserstein distance},
  journal = {Stochastic Processes and their Applications},
  volume  = {201},
  pages   = {105032},
  year    = {2026},
  doi     = {10.1016/j.spa.2026.105032}
}

@article{EcksteinPammer2024,
  author  = {Eckstein, Stephan and Pammer, Gudmund},
  title   = {Computational methods for adapted optimal transport},
  journal = {The Annals of Applied Probability},
  volume  = {34},
  number  = {1A},
  pages   = {675--713},
  year    = {2024}
}

@article{BionNadalTalay2019,
  author  = {Bion-Nadal, Jocelyne and Talay, Denis},
  title   = {On a {W}asserstein-type distance between solutions to stochastic differential equations},
  journal = {The Annals of Applied Probability},
  volume  = {29},
  number  = {3},
  pages   = {1609--1639},
  year    = {2019},
  doi     = {10.1214/18-AAP1423}
}

@article{ElKarouiJeanblancShreve1998,
  author  = {El Karoui, Nicole and Jeanblanc-Picqu{\'e}, Monique and Shreve, Steven E.},
  title   = {Robustness of the {B}lack and {S}choles formula},
  journal = {Mathematical Finance},
  volume  = {8},
  number  = {2},
  pages   = {93--126},
  year    = {1998},
  doi     = {10.1111/1467-9965.00047}
}

@incollection{FollmerSchweizer1991,
  author    = {F{\"o}llmer, Hans and Schweizer, Martin},
  title     = {Hedging of contingent claims under incomplete information},
  booktitle = {Applied Stochastic Analysis},
  editor    = {Davis, M. H. A. and Elliott, R. J.},
  series    = {Stochastics Monographs},
  volume    = {5},
  pages     = {389--414},
  publisher = {Gordon and Breach},
  address   = {London},
  year      = {1991}
}

@article{Cont2006,
  author  = {Cont, Rama},
  title   = {Model uncertainty and its impact on the pricing of derivative instruments},
  journal = {Mathematical Finance},
  volume  = {16},
  number  = {3},
  pages   = {519--547},
  year    = {2006},
  doi     = {10.1111/j.1467-9965.2006.00281.x}
}

@article{GlassermanXu2014,
  author  = {Glasserman, Paul and Xu, Xingbo},
  title   = {Robust risk measurement and model risk},
  journal = {Quantitative Finance},
  volume  = {14},
  number  = {1},
  pages   = {29--58},
  year    = {2014},
  doi     = {10.1080/14697688.2013.822989}
}

@article{MohajerinKuhn2018,
  author  = {Mohajerin Esfahani, Peyman and Kuhn, Daniel},
  title   = {Data-driven distributionally robust optimization using the {W}asserstein metric: performance guarantees and tractable reformulations},
  journal = {Mathematical Programming},
  volume  = {171},
  pages   = {115--166},
  year    = {2018},
  doi     = {10.1007/s10107-017-1172-1}
}

@article{BlanchetMurthy2019,
  author  = {Blanchet, Jose and Murthy, Karthyek},
  title   = {Quantifying distributional model risk via optimal transport},
  journal = {Mathematics of Operations Research},
  volume  = {44},
  number  = {2},
  pages   = {565--600},
  year    = {2019},
  doi     = {10.1287/moor.2018.0936}
}

@article{BartlDrapeauOblojWiesel2021,
  author  = {Bartl, Daniel and Drapeau, Samuel and Ob{\l}{\'o}j, Jan and Wiesel, Johannes},
  title   = {Sensitivity analysis of {W}asserstein distributionally robust optimization problems},
  journal = {Proceedings of the Royal Society A},
  volume  = {477},
  number  = {2256},
  pages   = {20210176},
  year    = {2021},
  doi     = {10.1098/rspa.2021.0176}
}

@article{Meucci2008,
  author  = {Meucci, Attilio},
  title   = {Fully flexible views: theory and practice},
  journal = {Risk},
  volume  = {21},
  number  = {10},
  pages   = {97--102},
  year    = {2008}
}

@misc{Vecer2026,
  author  = {Ve{\v{c}}e{\v{r}}, Jan},
  title   = {Geometric and arithmetic likelihood aggregation for diffusions with heterogeneous volatility},
  year    = {2026},
  eprint  = {2609.09470},
  archivePrefix = {arXiv},
  howpublished = {arXiv:2609.09470}
}

@article{JarrowLandoTurnbull1997,
  author  = {Jarrow, Robert A. and Lando, David and Turnbull, Stuart M.},
  title   = {A {M}arkov model for the term structure of credit risk spreads},
  journal = {The Review of Financial Studies},
  volume  = {10},
  number  = {2},
  pages   = {481--523},
  year    = {1997},
  doi     = {10.1093/rfs/10.2.481}
}

@article{Lando1998,
  author  = {Lando, David},
  title   = {On {C}ox processes and credit risky securities},
  journal = {Review of Derivatives Research},
  volume  = {2},
  pages   = {99--120},
  year    = {1998},
  doi     = {10.1007/BF01531332}
}

@article{DuffieSingleton1999,
  author  = {Duffie, Darrell and Singleton, Kenneth J.},
  title   = {Modeling term structures of defaultable bonds},
  journal = {The Review of Financial Studies},
  volume  = {12},
  number  = {4},
  pages   = {687--720},
  year    = {1999},
  doi     = {10.1093/rfs/12.4.687}
}

@article{DuffieLando2001,
  author  = {Duffie, Darrell and Lando, David},
  title   = {Term structures of credit spreads with incomplete accounting information},
  journal = {Econometrica},
  volume  = {69},
  number  = {3},
  pages   = {633--664},
  year    = {2001},
  doi     = {10.1111/1468-0262.00208}
}

@article{ElliottJeanblancYor2000,
  author  = {Elliott, Robert J. and Jeanblanc, Monique and Yor, Marc},
  title   = {On models of default risk},
  journal = {Mathematical Finance},
  volume  = {10},
  number  = {2},
  pages   = {179--195},
  year    = {2000},
  doi     = {10.1111/1467-9965.00088}
}

@article{JarrowProtter2004,
  author  = {Jarrow, Robert A. and Protter, Philip},
  title   = {Structural versus reduced form models: a new information based perspective},
  journal = {Journal of Investment Management},
  volume  = {2},
  number  = {2},
  pages   = {1--10},
  year    = {2004}
}

@book{BieleckiRutkowski2002,
  author    = {Bielecki, Tomasz R. and Rutkowski, Marek},
  title     = {Credit Risk: Modeling, Valuation and Hedging},
  publisher = {Springer},
  address   = {Berlin},
  year      = {2002},
  doi       = {10.1007/978-3-662-04821-4}
}

@incollection{BieleckiJeanblancRutkowski2004,
  author    = {Bielecki, Tomasz R. and Jeanblanc, Monique and Rutkowski, Marek},
  title     = {Hedging of defaultable claims},
  booktitle = {Paris--Princeton Lectures on Mathematical Finance 2003},
  series    = {Lecture Notes in Mathematics},
  volume    = {1847},
  pages     = {1--132},
  publisher = {Springer},
  address   = {Berlin},
  year      = {2004},
  doi       = {10.1007/978-3-540-44468-8_1}
}

@book{Schonbucher2003,
  author    = {Sch{\"o}nbucher, Philipp J.},
  title     = {Credit Derivatives Pricing Models: Models, Pricing and Implementation},
  publisher = {Wiley},
  address   = {Chichester},
  year      = {2003}
}

@book{Lando2004,
  author    = {Lando, David},
  title     = {Credit Risk Modeling: Theory and Applications},
  publisher = {Princeton University Press},
  address   = {Princeton},
  year      = {2004}
}

@book{JeanblancYorChesney2009,
  author    = {Jeanblanc, Monique and Yor, Marc and Chesney, Marc},
  title     = {Mathematical Methods for Financial Markets},
  publisher = {Springer},
  address   = {London},
  year      = {2009},
  doi       = {10.1007/978-1-84628-737-4}
}

@book{OKane2008,
  author    = {O'Kane, Dominic},
  title     = {Modelling Single-name and Multi-name Credit Derivatives},
  publisher = {Wiley},
  address   = {Chichester},
  year      = {2008}
}

@article{FreyRoeslerLu2019,
  author  = {Frey, R{\"u}diger and R{\"o}sler, Lars and Lu, Dan},
  title   = {Corporate security prices in structural credit risk models with incomplete information},
  journal = {Mathematical Finance},
  volume  = {29},
  number  = {1},
  pages   = {84--116},
  year    = {2019},
  doi     = {10.1111/mafi.12176}
}

@article{KalkbrenerPackham2026,
  author  = {Kalkbrener, Michael and Packham, Natalie},
  title   = {A {M}arkov approach to credit rating migration conditional on economic states},
  journal = {Canadian Journal of Statistics},
  volume  = {54},
  pages   = {e70039},
  year    = {2026},
  doi     = {10.1002/cjs.70039},
  eprint  = {2403.14868},
  archivePrefix = {arXiv}
}

@article{LandoSkodeberg2002,
  author  = {Lando, David and Sk{\o}deberg, Torben M.},
  title   = {Analyzing rating transitions and rating drift with continuous observations},
  journal = {Journal of Banking \& Finance},
  volume  = {26},
  number  = {2--3},
  pages   = {423--444},
  year    = {2002},
  doi     = {10.1016/S0378-4266(01)00228-X}
}

@article{JafrySchuermann2004,
  author  = {Jafry, Yusuf and Schuermann, Til},
  title   = {Measurement, estimation and comparison of credit migration matrices},
  journal = {Journal of Banking \& Finance},
  volume  = {28},
  number  = {11},
  pages   = {2603--2639},
  year    = {2004},
  doi     = {10.1016/j.jbankfin.2004.06.004}
}

@article{FrydmanSchuermann2008,
  author  = {Frydman, Halina and Schuermann, Til},
  title   = {Credit rating dynamics and {M}arkov mixture models},
  journal = {Journal of Banking \& Finance},
  volume  = {32},
  number  = {6},
  pages   = {1062--1075},
  year    = {2008},
  doi     = {10.1016/j.jbankfin.2007.09.013}
}

@article{MichielonKhedherSpreij2024,
  author  = {Michielon, Matteo and Khedher, Asma and Spreij, Peter},
  title   = {Proxying credit curves via {W}asserstein distances},
  journal = {Annals of Operations Research},
  volume  = {336},
  pages   = {1351--1367},
  year    = {2024},
  doi     = {10.1007/s10479-022-04552-3}
}

@article{MarshallOlkin1967,
  author  = {Marshall, Albert W. and Olkin, Ingram},
  title   = {A multivariate exponential distribution},
  journal = {Journal of the American Statistical Association},
  volume  = {62},
  number  = {317},
  pages   = {30--44},
  year    = {1967},
  doi     = {10.1080/01621459.1967.10482885}
}

@book{Bremaud1981,
  author    = {Br{\'e}maud, Pierre},
  title     = {Point Processes and Queues: Martingale Dynamics},
  publisher = {Springer},
  address   = {New York},
  year      = {1981}
}

@book{JacodShiryaev2003,
  author    = {Jacod, Jean and Shiryaev, Albert N.},
  title     = {Limit Theorems for Stochastic Processes},
  edition   = {2nd},
  series    = {Grundlehren der mathematischen Wissenschaften},
  volume    = {288},
  publisher = {Springer},
  address   = {Berlin},
  year      = {2003}
}

@article{BremaudYor1978,
  author  = {Br{\'e}maud, Pierre and Yor, Marc},
  title   = {Changes of filtrations and of probability measures},
  journal = {Zeitschrift f{\"u}r Wahrscheinlichkeitstheorie und verwandte Gebiete},
  volume  = {45},
  pages   = {269--295},
  year    = {1978},
  doi     = {10.1007/BF00537538}
}

@book{Lindvall1992,
  author    = {Lindvall, Torgny},
  title     = {Lectures on the Coupling Method},
  publisher = {Wiley},
  address   = {New York},
  year      = {1992}
}

@article{BurdzyKendall2000,
  author  = {Burdzy, Krzysztof and Kendall, Wilfrid S.},
  title   = {Efficient {M}arkovian couplings: examples and counterexamples},
  journal = {The Annals of Applied Probability},
  volume  = {10},
  number  = {2},
  pages   = {362--409},
  year    = {2000},
  doi     = {10.1214/aoap/1019487348}
}

@article{ConnorJacka2008,
  author  = {Connor, Stephen and Jacka, Saul},
  title   = {Optimal co-adapted coupling for the symmetric random walk on the hypercube},
  journal = {Journal of Applied Probability},
  volume  = {45},
  number  = {3},
  pages   = {703--713},
  year    = {2008},
  doi     = {10.1239/jap/1222441824}
}

@article{KeilsonKester1977,
  author  = {Keilson, Julian and Kester, Adri},
  title   = {Monotone matrices and monotone {M}arkov processes},
  journal = {Stochastic Processes and their Applications},
  volume  = {5},
  number  = {3},
  pages   = {231--241},
  year    = {1977},
  doi     = {10.1016/0304-4149(77)90033-3}
}

@article{Lorentz1953,
  author  = {Lorentz, G. G.},
  title   = {An inequality for rearrangements},
  journal = {The American Mathematical Monthly},
  volume  = {60},
  number  = {3},
  pages   = {176--179},
  year    = {1953}
}

@article{CambanisSimonsStout1976,
  author  = {Cambanis, Stamatis and Simons, Gordon and Stout, William},
  title   = {Inequalities for ${E}k({X},{Y})$ when the marginals are fixed},
  journal = {Zeitschrift f{\"u}r Wahrscheinlichkeitstheorie und verwandte Gebiete},
  volume  = {36},
  pages   = {285--294},
  year    = {1976},
  doi     = {10.1007/BF00532695}
}

@article{BurkardKlinzRudolf1996,
  author  = {Burkard, Rainer E. and Klinz, Bettina and Rudolf, R{\"u}diger},
  title   = {Perspectives of {M}onge properties in optimization},
  journal = {Discrete Applied Mathematics},
  volume  = {70},
  number  = {2},
  pages   = {95--161},
  year    = {1996},
  doi     = {10.1016/0166-218X(95)00103-X}
}

@techreport{SP2025,
  author      = {Serino, Nicole and Kraemer, Nick W.},
  title       = {Default, Transition, and Recovery: 2024 Annual Global Corporate Default and Rating Transition Study},
  institution = {S\&P Global Ratings},
  type        = {RatingsDirect report},
  year        = {2025},
  month       = mar,
  note        = {Data from S\&P Global Market Intelligence's CreditPro}
}

@article{GunasingamWong2025,
  author  = {Gunasingam, Madhu and Wong, Ting-Kam Leonard},
  title   = {Adapted optimal transport between {G}aussian processes in discrete time},
  journal = {Electronic Communications in Probability},
  year    = {2025},
  eprint  = {2404.06625},
  archivePrefix = {arXiv}
}

@misc{GunasingamWong2026,
  author  = {Gunasingam, Madhu and Wong, Ting-Kam Leonard},
  title   = {Adapted optimal transport between filtered {G}aussian processes},
  year    = {2026},
  eprint  = {2604.22159},
  archivePrefix = {arXiv},
  howpublished = {arXiv:2604.22159}
}

@misc{GunasingamMattesiniWieselWong2026,
  author  = {Gunasingam, Madhu and Mattesini, Francesco and Wiesel, Johannes and Wong, Ting-Kam Leonard},
  title   = {Adapted {W}asserstein barycenters of {G}aussian processes},
  year    = {2026},
  eprint  = {2604.22453},
  archivePrefix = {arXiv},
  howpublished = {arXiv:2604.22453}
}

@article{AcciaioHouPammer2025,
  author  = {Acciaio, Beatrice and Hou, Songyan and Pammer, Gudmund},
  title   = {Entropic adapted {W}asserstein distance on {G}aussians},
  journal = {Electronic Communications in Probability},
  volume  = {30},
  pages   = {1--14},
  year    = {2025},
  eprint  = {2412.18794},
  archivePrefix = {arXiv}
}

@misc{AcciaioBartlGrassHouPammer2026,
  author  = {Acciaio, Beatrice and Bartl, Daniel and Grass, Anne and Hou, Songyan and Pammer, Gudmund},
  title   = {The geometry of the adapted {B}ures--{W}asserstein space},
  year    = {2026},
  eprint  = {2602.00623},
  archivePrefix = {arXiv},
  howpublished = {arXiv:2602.00623}
}

@article{Ruschendorf1985,
  author  = {R{\"u}schendorf, Ludger},
  title   = {The {W}asserstein distance and approximation theorems},
  journal = {Zeitschrift f{\"u}r Wahrscheinlichkeitstheorie und verwandte Gebiete},
  volume  = {70},
  number  = {1},
  pages   = {117--129},
  year    = {1985},
  doi     = {10.1007/BF00532240}
}

@article{OConnorMcGoffNobel2022,
  author  = {O'Connor, Kevin and McGoff, Kevin and Nobel, Andrew B.},
  title   = {Optimal transport for stationary {M}arkov chains via policy iteration},
  journal = {Journal of Machine Learning Research},
  volume  = {23},
  number  = {45},
  pages   = {1--52},
  year    = {2022}
}

@article{GlanzerPflugPichler2019,
  author  = {Glanzer, Martin and Pflug, Georg Ch. and Pichler, Alois},
  title   = {Incorporating statistical model error into the calculation of acceptability prices of contingent claims},
  journal = {Mathematical Programming},
  volume  = {174},
  number  = {1--2},
  pages   = {499--524},
  year    = {2019},
  doi     = {10.1007/s10107-018-1352-7}
}

@article{BangiaEtAl2002,
  author  = {Bangia, Anil and Diebold, Francis X. and Kronimus, Andr{\'e} and Schagen, Christian and Schuermann, Til},
  title   = {Ratings migration and the business cycle, with application to credit portfolio stress testing},
  journal = {Journal of Banking \& Finance},
  volume  = {26},
  number  = {2--3},
  pages   = {445--474},
  year    = {2002},
}

@article{Norberg2003,
  author  = {Norberg, Ragnar},
  title   = {The {M}arkov chain market},
  journal = {ASTIN Bulletin},
  volume  = {33},
  number  = {2},
  pages   = {265--287},
  year    = {2003},
}

@article{BenTalHochman1972,
  author  = {Ben-Tal, Aharon and Hochman, Eithan},
  title   = {More bounds on the expectation of a convex function of a random variable},
  journal = {Journal of Applied Probability},
  volume  = {9},
  number  = {4},
  pages   = {803--812},
  year    = {1972},
}

@article{PostekEtAl2018,
  author  = {Postek, Krzysztof and Ben-Tal, Aharon and den Hertog, Dick and Melenberg, Bertrand},
  title   = {Robust optimization with ambiguous stochastic constraints under mean and dispersion information},
  journal = {Operations Research},
  volume  = {66},
  number  = {3},
  pages   = {814--833},
  year    = {2018}
}

@article{Vershik1994,
  author  = {Vershik, Anatoly M.},
  title   = {The theory of decreasing sequences of measurable partitions},
  journal = {St. Petersburg Mathematical Journal},
  volume  = {6},
  number  = {4},
  pages   = {705--761},
  year    = {1995},
  note    = {Translated from Algebra i Analiz 6 (1994)}
}

@article{DallAglio1956,
  author  = {Dall'Aglio, Giorgio},
  title   = {Sugli estremi dei momenti delle funzioni di ripartizione doppia},
  journal = {Annali della Scuola Normale Superiore di Pisa, Classe di Scienze, Serie 3},
  volume  = {10},
  number  = {1--2},
  pages   = {35--74},
  year    = {1956}
}

\end{document}